\documentclass[12pt,a4paper]{article}
\pdfoutput=1
\usepackage[utf8]{inputenc}
\usepackage[T1]{fontenc}
\usepackage[margin=2.0cm]{geometry}
\usepackage[pdfstartview={FitH},bookmarks=false,linktoc=page,colorlinks=true,linkcolor=blue,citecolor=blue,urlcolor=blue]{hyperref}
\usepackage[numbered]{bookmark}
\usepackage{mathtools}
\usepackage{amssymb}
\usepackage{graphicx}
\usepackage[font=small,labelfont=bf]{caption}
\usepackage{authblk}
\usepackage{cite}
\usepackage{dsfont}
\usepackage[dvipsnames]{xcolor} 
\usepackage[titles]{tocloft}
\usepackage{float}
\usepackage{subcaption}
\usepackage{color}
\usepackage{tikz}
\usepackage{ifthen}
\usepackage[warn]{textcomp}
\numberwithin{equation}{section}
\allowdisplaybreaks
\usepackage{multirow}
\usepackage{cancel}
\usepackage{soul}

\usetikzlibrary{patterns}

\newcommand{\dbar}{{\mkern3mu\mathchar'26\mkern-12mu d}}

\usepackage{comment}

\tikzset{ntp/.style={circle, thin, minimum size=2mm, inner sep=0,
fill=white,#1}}

\usepackage{caption}
\usepackage{subcaption}

\begin{document}
\title{\vspace{2cm}\textbf{Black Holes and Thermogeometric Optimization}\vspace{1cm}}

\author[a]{V. Avramov}
\author[a,b]{H. Dimov}
\author[a]{M. Radomirov}
\author[a,c]{R. C. Rashkov}
\author[a]{T. Vetsov}

\affil[a]{\textit{Department of Theoretical Physics, Sofia University,}\authorcr\textit{5 J. Bourchier Blvd., 1164 Sofia, Bulgaria}}
	
\affil[b]{\textit{The Bogoliubov Laboratory of Theoretical Physics, JINR,}\authorcr\textit{141980 Dubna,
		Moscow region, Russia}
}
 
\affil[c]{\textit{Institute for Theoretical Physics, Vienna University of Technology,}
  \authorcr\textit{Wiedner Hauptstr. 8–10, 1040 Vienna, Austria}
\vspace{10pt}\texttt{v.avramov,h\_dimov,radomirov,rash,vetsov@phys.uni-sofia.bg}

\vspace{-30pt}\texttt{}\vspace{0.0cm}}
\date{}
\maketitle

\begin{abstract}
We suggest a finite-time geometric optimization framework to analyze thermal fluctuations and optimal processes in black holes. Our approach implement geodesics in thermodynamic space to define optimal pathways between equilibrium and non-equilibrium states. Since thermodynamic metrics need not be positive-definite, the method ensures a positive thermodynamic length by incorporating simple scale factor into the metric. 
We show that the scale factor is sensitive to phase transitions in entropy representation, addressing a key gap in Hessian thermodynamic geometry. Additionally, we link the scale factor to the sign of thermodynamic curvature, connecting it to the information geometry governing optimal processes. 
Our results indicate that optimal fluctuations can drive the evaporation of Schwarzschild and Kerr black holes, which may significantly differ from Hawking radiation. We also explore optimal accretion-driven processes supported by an external inflow of energy.
\end{abstract}

\thispagestyle{empty}
\tableofcontents

\section{Introduction}

One of the defining features of black holes is their intricate relationship with the laws of thermodynamics. Ever since the pioneering works in black hole thermodynamics \cite{Bekenstein:1972tm, Bekenstein:1973ur, Bekenstein:1974ax, Bekenstein:1975tw, Hawking:1971tu, Hawking:1975vcx, Hawking:1976de, Bardeen:1973gs, Davies:1977bgr, Davies_1978} 
 there has been numerous modifications and generalizations, incorporating various effects. Additionally, recent observational advancements, such as the Event Horizon Telescope (EHT) collaboration \cite{EventHorizonTelescope:2019dse, EventHorizonTelescope:2021dqv,EventHorizonTelescope:2022wkp} and the gravitational wave observations \cite{LIGOScientific:2016aoc, LIGOScientific:2016vlm}, have significantly improved our understanding of black holes and future high-resolution experiments are anticipated to be crucial in further advancing our knowledge on these phenomena.

An important problem in black hole thermodynamics is understanding how the states of such stellar objects change under certain conditions. Phenomena like matter accretion \cite{Luminet:1979nyg, Page:1974he, Thorne:1974ve, Gyulchev:2019tvk, Gyulchev:2021dvt}, black hole collisions \cite{LIGOScientific:2016aoc, LIGOScientific:2016vlm}, gravitational wave emissions \cite{1982TaylorPulse}, and Hawking radiation  \cite{Hawking:1975vcx}, directly impact the macro physics of black holes. Understanding these events is crucial not only for contemporary physics, but also for the potential future exploitation of black holes as energy sources. For instance, the Penrose energy extraction from a rotating black hole is one such possibility \cite{penrose1971extraction}. Therefore, a natural question is whether other processes can force black holes to change their thermodynamic state. 

A positive answer can be found in Thermodynamic geometry \cite{Weinhold:1975a, Ruppeiner:1983zz, Ruppeiner:1995ss}, which analyzes spontaneous or excited fluctuations within a thermodynamic system. As thermal systems, black holes fall under this framework. The principal strength of the geometric approach to thermodynamics lies in its emphasis on the relationships between elements, such as distance and curvature, facilitating the natural extraction of essential and unique characteristics of the system under study. Thermodynamic geometry was introduced by F. Weinhold \cite{Weinhold:1975a} and later developed further by G. Ruppeiner \cite{Ruppeiner:1983zz, Ruppeiner:1995ss}. Weinhold demonstrated that the laws of equilibrium thermodynamics could be described within an abstract metric space. In his approach the Hessian of the internal energy with respect to the system's extensive parameters serves as a metric on the space of macro states. Conversely, Ruppeiner's approach was grounded in fluctuation theory, employing entropy as the thermodynamic potential, where the Hessian of the entropy is used to determine the probability of fluctuations between different macrostates. It was subsequently shown that both metrics are conformally related, with temperature acting as the conformal factor \cite{Ruppeiner:1995ss, salamon1984relation, mrugala1984equivalence}.

Although the Hessian approach offers a straightforward route to defining geometric structures, the resulting geometry depends on the choice of thermodynamic potential -- an undesirable feature if the geometry is intended to reflect intrinsic physical properties. To address this limitation, an alternative framework, known as Geometrothermodynamics (GTD), was introduced by H. Quevedo and collaborators \cite{Quevedo:2007ws, Quevedo:2007mj, Quevedo:2017tgz, pineda2019physical}. GTD is specifically constructed to produce metrics that are invariant under Legendre transformations, thereby ensuring consistency across different thermodynamic potentials. This approach involves formulating the theory in an extended thermodynamic phase space equipped with a contact structure. Nevertheless, the formalism does not single out a unique Legendre-invariant metric, instead, it identifies three distinct families of metrics that satisfy the invariance conditions. A recent study \cite{Quevedo:2023vip} showed that by combining the GTD framework with the homogeneous\footnote{ Systems that satisfy the Euler identity.} or quasi-homogeneous\footnote{Systems that satisfy a generalized Euler identity.} properties of the thermodynamic system, the free parameters of these metrics can be fixed, allowing for their unambiguous construction. Various works have already applied these and similar methods to study thermodynamic properties of gravitational systems \cite{Ruppeiner:2007hr, Ruppeiner:2008kd, Ruppeiner:2013yca, Ruppeiner:2018pgn, Mansoori:2013pna, Mansoori:2014oia, Mansoori:2016jer, HosseiniMansoori:2019jcs, Mahmoudi:2023uxr, Quevedo:2007ws, Quevedo:2007mj, Quevedo:2017tgz, pineda2019physical, Quevedo:2023vip, Dimov:2019pkk, Dimov:2019fxp, Vetsov:2018dte}.  

The metric approach to the space of macrostates can be naturally extended to incorporate time as an ordering parameter, thus introducing the {geometric} framework of finite-time thermodynamics {(FTT)} \cite{salamon1977finite, andresen1977extremals, andresen1977optimization, salamon1980significance, salamon1980minimum, salamon1983thermodynamic, andresen1984thermodynamics, andresen1983availability, salamon1985length, andresen2011current, andresen1996finite, berry2022finite}. {This framework offers an alternative or complementary perspective to more traditional formulations of FTT, such as those presented in \cite{CurzonAhlborn1975, Bejan1995, Cheng2019}.}
The goal is to transition between two arbitrary states with minimal effort by following the shortest thermodynamic paths, which represent optimal processes. These paths are governed by the thermodynamic length \cite{2007PhRvL99j0602C, cafaro2022thermodynamic}, a concept that has significantly influenced thermodynamic optimization and control theory across diverse systems \cite{2007PhRvL99j0602C, cafaro2022thermodynamic,avramov2023thermodynamic, Bravetti:2015xsp,2020Entrp221076A,ANDRESEN1996647,2019Quant3197S, Gruber:2016xui, Gruber:2016mqb}. 
To evaluate the relevant parameters such as  relaxation time and  thermodynamic speed, it suffices to establish a positive thermodynamic distance between the initial and final states.

In Ruppeiner's fluctuation theory the thermodynamic length governs the probability of fluctuations between states. The general principle is that the probability of transitioning from one state to another is related to the thermodynamic distance between these states in a way that larger distances assume smaller probabilities \cite{Ruppeiner:1983zz, Ruppeiner:1995ss}. In this context, the positive-definiteness of the thermodynamic metric is linked to the system's stability, enabling these metrics to describe transitions between equilibrium states. However, in the non-equilibrium regime, the Hessian of the energy (or the negative Hessian of the entropy) is not a positive-definite matrix. To ensure a positive thermodynamic length between non-equilibrium states, it becomes necessary to modify these metrics. The latter is implemented by incorporating a simple scaling factor in front of the Hessian, which subsequently connects to the type of information geometry employed to define the optimal processes. 

The finite-time thermodynamics framework has been utilized in recent studies to explore various properties of black holes \cite{avramov2023thermodynamic, Bravetti:2015xsp, 2020Entrp221076A, ANDRESEN1996647, 2019Quant3197S, Gruber:2016xui, Gruber:2016mqb}. Building on these efforts, the current investigation seeks to further extend these concepts in black hole physics, with particular emphasis on the key distinctions discussed in Section \ref{secTGO}, where a suitable optimization algorithm is developed.

The paper is organized as follows. In Section \ref{secTGO}, we introduce the finite-time thermogeometric optimization method (TGO) for investigating thermal fluctuations and optimal processes in black holes. This method consists of two key components. First, it builds on thermodynamic geometry and Ruppeiner's fluctuation theory \cite{Ruppeiner:1983zz, Ruppeiner:1995ss}, where the Hessian matrix of a relevant thermodynamic potential defines the metric in the space of macro states. Second, it utilizes the concept that geodesics in this space can be used to identify optimal processes in thermal systems \cite{salamon1977finite, andresen1977extremals, andresen1977optimization, salamon1980significance, salamon1980minimum, salamon1983thermodynamic, andresen1984thermodynamics, andresen1983availability, salamon1985length, andresen2011current, andresen1996finite, berry2022finite, Bravetti:2015xsp, Gruber:2016xui, Gruber:2016mqb, avramov2023thermodynamic,2007PhRvL99j0602C, cafaro2022thermodynamic,
2020Entrp221076A,ANDRESEN1996647,2019Quant3197S}. Additionally, TGO generalizes previous propositions by incorporating a simple scale factor into the Hessian metric. This factor can be linked to the type of information geometry and proves to be sensitive to the location of Davies phase transition curves in entropy representation.

In Section \ref{secSchwTGO}, we apply TGO to study optimal thermal evaporation in  Schwarzschild black hole. To preserve key geometric features, we treat the Schwarzschild state space as the zero angular momentum limit of the larger Kerr state space. Our findings show that the Schwarzschild black hole can fully evaporate due to fluctuations. We also compare these optimal fluctuations to an evaporation caused by a Hawking radiation model \cite{page1976thermal, page2013jcap, Nian:2019buz, Arevalo:2024kmo}. 

In Section \ref{secKerrTGO}, we examine fluctuation-driven evaporation and accretion-driven processes for Kerr black hole in both entropy and energy representations. The thermodynamic geodesic equations governing these processes are highly nonlinear and their investigation require numerical treatment. 
For an optimal evaporation in entropy representation, we demonstrate that the black hole's spin consistently decreases, ending when the specific spin reaches a Davies phase transition point. At this point all optimal processes terminate, since classical description breaks down. For accretion-driven processes, the black hole's energy increases until its spin encounters a phase transition. In both scenarios, the sign of the metric scale factor effectively detects the location of the Davies point.  

An intriguing feature of our (optimal) fluctuation model in entropy representation is that it precludes the complete evaporation of a Kerr black hole. All optimal pathways for the black hole's specific spin terminate at a Davies phase transition point before full evaporation can occur. On the other hand, in energy representation, the optimal evaporation of the Kerr black hole is allowed at rates that are either faster or slower than Hawking radiation evaporation. Furthermore, in energy representation, the thermodynamic length, nor the metric scale factor, can be used to detect the Davies critical points. This is related to the fact that the thermodynamic curvature of the Kerr black hole is zero in this representation.
 
Finally, in Section \ref{secConcl}, we provide a brief summary of our results.


\section{Thermogeometric optimization}\label{secTGO}

Our approach to thermal fluctuations and optimal processes is based on the concept that geodesics in the space of macrostates can be utilized to determine optimal paths between states in any thermal system. We introduce a straightforward algorithm for constructing these optimal protocols, we call Thermogeometric Optimization (TGO), which involves five steps:

\begin{enumerate}
    \item Choosing the thermodynamic representation.
    \item Analyzing the thermodynamic (in)stability. 
    \item Choosing the thermodynamic metric.
    \item Analyzing optimal processes (geodesics) on the space of states.
    \item Extracting thermodynamic length, speed, relaxation time etc.
\end{enumerate}

In the following we provide a concise overview of the steps involved in the algorithm.

\subsection{Thermodynamic representation}

The first step involves selecting the appropriate thermodynamic potential, which determines the thermodynamic representation (ensemble). This choice is dictated by the control parameters available in the system. For instance, in the context of energy  representation, the first law of thermodynamics can be expressed as follows:
\begin{equation}\label{eqFirstLaw}
d E=\sum\limits_{a=1}^n I_a dE^a=\vec I. d\vec E,
\end{equation}
where $I^a$ represent the set of intensive parameters and $ E^a $ are the natural energy extensive control variables. The  relations between the extensive and the intensive parameters are imposed by  the equations of state (EoS). For example, in the energy representation one has\footnote{In thermodynamics, partial derivatives are generally referred to as Nambu brackets, which naturally account for any Jacobians when there is a change of variables. We present Nambu brackets in App. \ref{appA}.}:
\begin{equation}
I_a=\frac{\partial E(\vec E)}{\partial E^a}\bigg|_{E^1,...,\hat E^a,...,E^n},
\end{equation}
where the hatted quantities are the ones being varied and thus removed from the list of constant variables. The expression $ E = E(E^a) $ defines the energy fundamental relation, representing an $ n $-dimensional hypersurface embedded in $ (n+1) $-dimensional space spanned by $ (E^1, \ldots, E^n, E) $. The definition of such space is justified by the fact that the first law of thermodynamics (\ref{eqFirstLaw}) is written as a product of generalized forces $ \vec{I} $ and differentials of generalized coordinates  $ \vec{E} $, analogous to the work defined in classical mechanics.

In many practical scenarios, control parameters may consist of a mix of intensive and extensive variables, necessitating an appropriate change of potential. All energy or entropy based potentials (such as free energies or free entropies) result from the Legendre transformation of the initial potential (either $E$ or $S$) along some or all of the corresponding natural directions. For instance, all free energies $\Phi_{(a)}$,  defined by a Legendre transformation along one of the energy natural parameters, say $E_a$, are given by (no summation over $a$)\footnote{As an example, performing a Legendre transformation $\mathcal{L}_S E$ of the energy $E$ with respect to entropy $S$ yields the Helmholtz potential: $\Phi_{(S)}(T,...) = \mathcal{L}_{S} E = E - TS \equiv F$. It naturally depends on the temperature $T$ and thus defines the corresponding Helmholtz thermodynamic representation (the canonical ensemble).}:
\begin{align}
&\Phi_{(a)}(I_a,E_1,...,\hat E_a,..,E_n)=\mathcal{L}_{E^a}E=E-I_a E^a.
\end{align}
Consequently, potentials derived by a Legendre transformation along two directions are written by (no summation over $a, b$):
\begin{align}
\Phi_{(ab)}(I_a,I_b, E_1,...,\hat E_a,..,\hat E_b,...,E_n)=\mathcal{L}_{E^a,E^b}E=E-I_a E^a-I_b E^b,\quad a \neq b.
\end{align}
Evidently, the sequence can be extended further until one takes a Legendre transformation along all energy natural parameters. Constructing similar sequence of transformations of the entropy one finds the family of Massieu-Planck potentials.

Finally, one has to consider the homogeneity of the thermodynamic potentials. Note that in standard thermodynamics all potentials are homogeneous functions of degree one. For instance, under scaling $\lambda > 0$, the energy potential behaves as follows $
    E(\lambda E^1, \lambda E^2, ..., \lambda E^n) = \lambda^1 E(E^1, E^2, ..., E^n)$.
This characteristic unavoidably renders the Legendre transformation simultaneously along all parameters  trivial. However, this does not hold true in black hole physics, where potentials exhibit generalized quasi-homogeneity of degree $r$ and type $(r_1, ..., r_n)$, \cite{Belgiorno:2002}, e.g. $
    E(\lambda^{r_1} E^1, \lambda^{r_2} E^2, ..., \lambda^{r_n} E^n) = \lambda^r E(E^1, E^2, ..., E^n)$.
According to the generalized Euler's homogeneity theorem, this leads to
\begin{equation}
rE=\sum\limits_{a=1}^{n} r_a I_a E^a,
\end{equation}
which in black hole physics is know as the Smarr relation \cite{smarr1973mass}.

\subsection{Thermodynamic (in)stability}

The purpose of analyzing thermodynamic (in)stability is to gain insight into the (non)equilibrium landscape of possible states of the system. Anything beyond stable regions inherently exists in a non-equilibrium state, resulting in continuous change of system's parameters.  In the equilibrium scenario the Hessian metrics are positive-definite, while in the non-equilibrium case they are negative definite, since certain heat capacities are negative. Nevertheless, the TGO algorithm operates effectively in both scenarios, requiring only the existence of a positive thermodynamic length. In the non-equilibrium case this is always assured by the sign of a simple  metric scale $\epsilon$ multiplying all components of the Hessian.

There are several standard criteria for assessing global and local thermodynamic stability. Global stability can be determined by applying either the Sylvester criterion or the eigenvalue criterion, \cite{Avramov:2023eif, Avramov:2024hys}. Both rely on the fundamental requirement that energy and entropy are strictly convex or concave functions at equilibrium\footnote{Energy is strictly convex function with a minimum at equilibrium, while entropy is strictly concave with a maximum at equilibrium.}. According to the theory of such functions, the Hessian matrix of a convex (concave) function must be strictly positive (negative) definite, meaning all its eigenvalues should be strictly positive (negative). 

Let $\varepsilon_k$ represent the eigenvalues of the Hessian of the energy, and $s_k$ denote the eigenvalues of the Hessian of the entropy. The eigenvalue criterion states that the system reaches global thermodynamic equilibrium if $\epsilon_k > 0$ or $s_k < 0$ for all $k = 1, \dots, n$. Alternatively, let $\Delta_k^{(E)}$ or $\Delta_k^{(S)}$ represent the determinants of the principal minors of the Hessian of the energy or entropy at level\footnote{If $k=n$, then $\Delta_n$ represents the determinant of the entire Hessian matrix. For $k<n$, $\Delta_k$ is constructed by removing $n-k$ number of rows and columns.} $k$. According to the Sylvester criterion, the system achieves global thermodynamic stability if (no summation over $k$): $\Delta_k^{(E)} > 0$ or $(-1)^k \Delta_k^{(S)} > 0$.

An inconclusive outcome based on these criteria occurs when some eigenvalues are zero or when some determinants of the principal minors are also zero. This situation is observed, for instance, in the stability analysis of an ideal gas, where the determinant of the Hessian of the entropy is zero, leading to one or more zero eigenvalues. In such cases, the system's local thermodynamic stability can be assessed by examining the properties of local heat capacities or other relevant thermodynamic response functions. { 
In classical thermodynamics, it is well established that positive heat capacities are required to maintain local thermodynamic equilibrium. This principle was applied in the pioneering work of Davies \cite{Davies:1977bgr}, who demonstrated that Schwarzschild and Kerr-Newman black holes are thermodynamically unstable due to their negative heat capacities -- implying they should radiate. While heat capacities may suffice to assess the local thermodynamic stability of standard black hole solutions in general relativity, this is generally not sufficient for black holes in modified theories of gravity. In such cases, the full set of thermodynamic response functions must be examined\footnote{See for instance \cite{avramov2023thermodynamic}, where the authors found that a particular black hole solution is globally thermodynamically unstable, despite all heat capacities being positive. This indicates local stability with respect to certain processes, but not all -- suggesting that some other thermodynamic response coefficients must be negative.}.
}

{Here we focus only on the properties of the local heat capacities.} The extensive quantity $C$, representing the total heat capacity of the system, is defined as follows:
\begin{equation}\label{eqTotalCapacity}
    C=\frac{\dbar Q}{dT}=T\frac{\partial S}{\partial T}.
\end{equation}
Since heat $\dbar Q$ depends on the nature of the process (hence the inexact differential $\dbar$), the heat capacity $C$ also varies accordingly. Therefore for different processes $C$ defines distinct thermodynamic responses. This prompts a broader definition of heat capacity, which specifies the particular process used to alter the system's state. Generally, $C_{x^1,x^2,...,x^{n-1}}$ denotes the heat capacity at a fixed set of thermodynamic parameters $(x^1,x^2,...,x^{n-1})$. If it depends on a number of variables, say $(y^1,y^2,...,y^{n})$, then one can write \cite{Mansoori:2014oia,  avramov2023thermodynamic}:
\begin{equation}
C_{x^1,x^2,...,x^{n-1}}(y^1,y^2,...,y^{n})=T \frac{\partial S}{\partial T}\bigg|_{x^1,x^2,...,x^{n-1}}=T \frac{\{S,x^1,x^2,...,x^{n-1}\}_{y^1,y^2,...,y^n}}{\{T,x^1,x^2,...,x^{n-1}\}_{y^1,y^2,...,y^n}},
\end{equation}
where the Nambu brackets $\{ \cdot \}$, employed above, extend the concept of Poisson brackets to encompass three or more independent variables (see App. \ref{appA}). Additionally, the set of constant parameters $(x^1,x^2,...,x^{n-1})$ may comprise a mixture of intensive and extensive variables. Moreover, all relevant state quantities become functions of the parameters $(y^1,y^2,...,y^{n})$. In such a scenario, we refer to $(y^1,...,y^n)$ as defining the coordinates of the state space, although they may not be natural for the chosen thermodynamic representation.

Local thermodynamic equilibrium arises from quasi-equilibrium\footnote{
Quasi-equilibrium refers to a regime where a system evolves slowly enough that each local region remains nearly in thermodynamic equilibrium, despite small spatial or temporal gradients in variables like temperature or pressure. This permits the local application of equilibrium thermodynamics, with well-defined quantities that vary smoothly in space and time. The approximation breaks down when macroscopic changes occur on timescales comparable to or shorter than the local relaxation time.} among different segments of the system, permitting sufficiently small gradients of parameters. Associating local stability with the positivity of specific heat capacities is tied to the components of the Hessian matrix, where fulfilling the generic stability conditions always demands $C>0$. Therefore, if even one of the heat capacities is negative, the system becomes thermodynamically unstable. This is particularly evident in simple systems (see, for instance, \cite{bazarov1964thermodynamics, callen2006thermodynamics, greiner2012thermodynamics, swendsen2020introduction, blundell2010concepts}). Hence, it can be argued that the classical criterion for local thermodynamic stability, within a given process with fixed control parameters ($x^1,x^2,...,x^{n-1}$), is defined by
$C_{x^1,x^2,...,x^{n-1}}>0$.

Finally, one notes that heat capacities are crucial for identifying phase transitions within the system. Specifically, a divergence or change in the sign of a heat capacity indicates the onset of a phase transition and the breakdown of the equilibrium thermodynamic description. Paul Davies first highlighted this concept for black holes, noting that the divergence of the heat capacity in the Kerr-Newman black hole signifies a second-order phase transition \cite{Davies:1977bgr}.

 \subsection{Thermodynamic metric}

The thermodynamic metric formalism uses the Hessian of a specific thermodynamic potential. In entropy representation, this is captured by the Ruppeiner thermodynamic metric defined by the Hessian of the entropy with respect to its natural parameters \cite{Ruppeiner:1983zz, Ruppeiner:1995ss}:
 \begin{equation}\label{eqRupMetGen}
 ds^2_{(R)}=\sum_{a,b}g_{ab}^{(R)}(\vec S) dS^a dS^b=\epsilon\sum_{a,b}\frac{\partial^2 S(\vec S)}{\partial S^a\partial S^b}  dS^a dS^b,\quad \epsilon\in\mathbb{R}.
 \end{equation}
The scale factor $\epsilon$ assures positive-definite thermodynamic length with respect to this metric. In this case, the line element $ds^2$
measures the dissipative work needed to drive the system between nearby states\footnote{Therefore, following the interpretation of the square of the thermodynamic length, $\mathcal{L}^2$, as the minimum energy required to change the system's state, we can adjust the value of $\epsilon$ accordingly. Hence, in this context, we conjecture that for black holes $\epsilon$ is uniquely determined when $\mathcal{L}^2$, which governs the evaporation process, corresponds to the black hole's initial energy. For more details see Sec. \ref{secOESCHBH}. Furthermore, in Sec. \ref{secKGSrep} we show that $\epsilon$ is also related to the sign of the thermodynamic curvature, which, according to Ruppeiner's interpretation, is related to the type of interactions of the underlying statistical model \cite{Ruppeiner:2010}. Later we also show that the sign of $\epsilon$ is vital to maintain real {and positive} thermodynamic length. 
}. The original value of the scale factor in Ruppeiner's fluctuation theory is $\epsilon=-1$:
 \begin{equation}\label{eqRupMetGen1}
	ds^2_{(R)}=\sum_{a,b}g_{ab}^{(R)}(\vec S) dS^a dS^b=-\sum_{a,b}\frac{\partial^2 S(\vec S)}{\partial S^a\partial S^b}  dS^a dS^b.
\end{equation}

The inclusion of the scale factor in (\ref{eqRupMetGen}) is an important key difference that distinguishes our approach from previous studies, \cite{Bravetti:2015xsp, Gruber:2016xui, Gruber:2016mqb, avramov2023thermodynamic}. {To show its importance let us consider how it affects the probability for a fluctuation or a process to occur\footnote{In fact, if we consider small fluctuations of the thermodynamic potential around each equilibrium point, the second moment of the fluctuation turns out to be directly related to the components of the corresponding Hessian metric.}. In Ruppeiner’s original fluctuation theory the minus sign in the metric (\ref{eqRupMetGen1}) arises from fundamental principles of thermodynamics and statistical mechanics. Entropy $S$ is maximized in equilibrium, so its Hessian is negative-definite for stable systems.
	The minus sign in the metric ensures  $g_{ab}$ is positive-definite, as required for a Riemannian metric. The probability of thermodynamic fluctuation is then given by Einstein’s formula \cite{Ruppeiner:1983zz, Ruppeiner:1995ss} (sums over $a$ and $b$ are assumed):
	\begin{equation}
		p\propto e^{\Delta S}\approx e^{-\frac{1}{2}  g_{a b} \Delta S^a \Delta S^b}=e^{\frac{1}{2}\frac{\partial^2 S}{\partial S^a\partial S^b} \Delta S^a \Delta S^b},
	\end{equation}
	where $\Delta S^a$ are the fluctuations of the entropy natural parameters. This formula suggests that without the minus sign in the metric, fluctuations would be exponentially enhanced leading to an unphysical situations for stable systems. Therefore, the minus sign in $g_{ab}$ ensures it also defines a real and positive distance in state space.
	
	However, if we replace the minus sign in Ruppeiner’s metric (\ref{eqRupMetGen1}) by a real number $\epsilon$, then several important things change. First, if the scale $\epsilon<0$ we retain Ruppeiner’s original fluctuation theory, but the probability for a process or a fluctuation to occur is now scaled by this factor and still retaining positive distance in state space:
	\begin{equation}\label{eqProbFlucTh}
		p\propto e^{-\frac{\epsilon}{2} \frac{\partial^2 S(\vec S)}{\partial S^a\partial S^b} \Delta S^a \Delta S^b}.
	\end{equation}
	This is very important, when considering stable systems which are not described by homogeneity one\footnote{According to the standard thermodynamics potentials, such as internal energy and entropy, should be a homogeneous functions of their natural parameters of order 1, e.g. $E(\lambda S, \lambda V,\lambda N)=\lambda^1 E(S,V,N)$, where $\lambda$ is some positive scale. According to the generalized Euler homogeneity theorem black holes' potentials may exhibit different degree of homogeneity, which is expressed  in their Smarr relations. } thermodynamics, such as the majority of black holes. The interparticle interactions of such systems may have stronger or weaker strength than systems from the standard thermodynamics, hence a simple way to account for this strength would be via the scale factor $\epsilon$, as shown in Section \ref{secKGSrep}. The latter would agree by the interpretation of the Ruppeiner's curvature as a measure of the strength of interactions. In the rescaled version of the metric (\ref{eqRupMetGen}) the curvature scales proportional to $\epsilon^{-1}$, e.g. in Eq. (\ref{eqRupScCurve}) for the Kerr black hole.
	
	Furthermore, since in the non-equilibrium case the Hessian of the entropy may become positive-definite, then one can adjust $\epsilon>0$ in such a way that a positive-definite distance can be obtained and the probability (\ref{eqProbFlucTh}) for a fluctuation  will still be exponentially suppressed. }
	
Another useful Hessian metric appears in energy representation, where we encounter Weinhold's thermodynamic metric \cite{Weinhold:1975a}:
\begin{equation}\label{eqWeinhMetricdif}
	ds^2_{(W)}=\sum_{a,b}g_{ab}^{(W)}(\vec E) dE^a dE^b=\epsilon\sum_{a,b}\frac{\partial^2 E(\vec E)}{\partial E^a\partial E^b}  dE^a dE^b.
\end{equation}
Here $E$ is the energy potential. Both metrics are  conformally related by the temperature $T$ being the  conformal factor, i.e. 	$ds^2_{(W)}=T ds^2_{(R)}$. 
	
Despite their apparent  simplicity, Hessian metrics exhibit a notable drawback: the physical properties do not remain invariant under changes in the potential. This limitation was addressed by H. Quevedo and collaborators \cite{Quevedo:2007mj, pineda2019physical, Quevedo:2007ws, Quevedo:2017tgz}, who introduced Geometrothermodynamics (GTD) approach, which employs a Legendre-invariant extension of these metrics. {However, the more complex structure of GTD metrics lacks the clear probabilistic interpretation associated with Hessian metrics, making their physical meaning less transparent. Nonetheless, recent studies \cite{pineda2019physical, Quevedo:2023ypd} have demonstrated that the components of GTD metrics can also be interpreted as second moments of fluctuations of a thermodynamic potential, thereby offering new insights into the physical significance of the geometrothermodynamic framework.}

\subsection{Thermodynamic length and finite-time thermodynamics}

The concept of length is natural in a metric space. In thermodynamic sense, paths between states in a given space of macro states correspond to specific thermodynamic processes. Therefore, the length of these paths should be related to the characteristics of the given process. While there are many types of thermodynamic transformations or protocols, one often prefers the so-called optimal processes. These are protocols that extremize important system properties, such as dissipation, energy loss, entropy production, etc, \cite{salamon1977finite, andresen1977extremals, andresen1977optimization, salamon1980significance, salamon1980minimum, salamon1983thermodynamic, andresen1984thermodynamics, andresen1983availability, salamon1985length, andresen2011current, andresen1996finite, berry2022finite}. 

In this context, the paths connecting two states are the shortest paths, or geodesics. The length of these paths, known as thermodynamic length \cite{2007PhRvL99j0602C, cafaro2022thermodynamic}, quantifies the distance between two (non)equilibrium states. In fluctuation theory, thermodynamic length measures the number of fluctuations associated with the change of states \cite{Ruppeiner:1983zz, Ruppeiner:1995ss}. Let $\Phi$ represents a generic thermodynamic potential and $\vec \Phi=(\Phi^1,...,\Phi^n)$ are its natural parameters spanning the thermodynamic space. Given a metric $ g_{ab}(\vec \Phi)$ in this space, the functional of the thermodynamic length along a path $\gamma$ between two points is defined by
\begin{equation}\label{eqGeodAction}
{\cal L}[\gamma] = \int_\gamma \sqrt{g_{ab}(\vec \Phi) d\Phi^a d\Phi^b}.
\end{equation}

One can rewrite this functional in terms of a specific paths, parametrized by some affine parameter $t$. Let the path $\gamma$ is parameterized by $t$ (not necessarily time). Then all natural parameters are parametrization dependent functions $\Phi^a=\Phi^a(t)$ and hence the length functional becomes:
\begin{equation}\label{eqGeodAction2}
	{\cal L}[\gamma(t)] = \int \nolimits_{0}^{\tau} \sqrt {{g_{ab}}\big(\vec \Phi(t)\big)\dot \Phi^a(t) \dot \Phi^b(t)}\, dt.
\end{equation}
Here the dots represent derivatives with respect to $t$ and $\tau$ is the final value of $t$.

The key difference between the two definitions (\ref{eqGeodAction}) and (\ref{eqGeodAction2}) of the thermodynamic length is that $\mathcal{L}[\gamma]$ does not necessarily yield the optimal thermodynamic distance, whereas $\mathcal{L}[\gamma(t)]$, when evaluated on a geodesic profile for $\Phi^a(t)$, is optimal\footnote{A simple example, demonstrating the difference between the two definition of the length, is given by the uniform rotation of a point particle around a circle with a radius $R$. In polar coordinates: $x=\rho \cos\varphi$ and $y=\rho \sin\varphi$, the metric on the disk is $ds^2=d\rho^2+\rho^2 d\varphi^2$. For a circle $\rho=R$ and $0\leq \varphi\leq 2 \pi$, Eq. (\ref{eqGeodAction}) yields $\mathcal{L}=2 \pi R$, which is the circumference of the circle. However, Eq. (\ref{eqGeodAction2}) gives the law of motion $\mathcal{L}=\omega R \tau=v \tau$, which follows from the solutions $\rho(t)=R$ and $\varphi(t)=\varphi_0+\omega t$ to the corresponding geodesic equations.}. If one considers the affine parameter $t$ as time, then  (\ref{eqGeodAction2}) is a particular (geometric) way to introduce finite-time processes in thermodynamics.

One way to interpret the thermodynamic length\footnote{Note that the units of thermodynamic length vary depending on the thermodynamic representation.} is that the square of its extreme values represent the minimum work required to drive a system from one state to another. If the process is carried out reversibly\footnote{In a reversible process, the system passes through a series of equilibrium states and remains in thermal equilibrium with its surroundings at each stage.} or quasi-reversibly -- slowly enough to avoid entropy generation -- then the square of the thermodynamic length provides a lower bound on dissipation, which is a consequence of the Cauchy-Schwarz inequality,
\begin{equation}\label{eqGDivergence}
	{\mathcal {J}} = \tau \int_0^\tau {g_{ab}}(\vec \Phi)\dot \Phi^a \dot \Phi^b dt\ge {{\cal L}^2}.
\end{equation}
Here, the thermodynamic divergence $\mathcal{J}$ measures the efficiency of the quasi-static protocol used in the transformation. Hence $\mathcal{L}$ characterizes the geometric path that {minimizes the dissipation} or maximizes the efficiency of a given  finite-time thermodynamic process, \cite{berry2022finite}.

\subsection{Geodesics on the space of states and optimal processes}

The final step in our optimization algorithm is to derive the corresponding thermodynamic geodesic equations, whose solutions extremize the thermodynamic length. A straightforward variation of (\ref{eqGeodAction2}) with respect to the fields $\Phi^a(t)$ leads to:
\begin{equation}\label{eqGeodesicGeneric}
        \ddot \Phi^c(t)+\Gamma^c_{ab}\big(\hat g,\vec \Phi\big) \dot \Phi^a(t)\dot \Phi^b(t)=0,
\end{equation}
where $\Gamma^c_{ab}(\hat g,\vec \Phi)$ are the Christoffel symbols defined by the derivatives of the thermodynamic metric with respect to the parameters $\Phi^a$:
\begin{equation}\label{eqChristofel}
\Gamma^{c}_{ab}=\frac{1}{2}g^{c d}(\partial_a g_{db}+\partial_b g_{da}-\partial_d g_{ab}).
\end{equation}
The solutions of (\ref{eqGeodesicGeneric}) provide the geodesic profiles of the control parameter that define the optimal finite-time thermodynamic process.

Note also that, since the factor $\epsilon$  is an overall scale in the metric (\ref{eqRupMetGen}) or (\ref{eqWeinhMetricdif}), the geodesic equations and, consequently, the geodesic profiles are independent of $\epsilon$. However, the thermodynamic length scales with $\sqrt{\epsilon}$, making this parameter essential to ensure a real-valued thermodynamic distance between states. Furthermore, in Section \ref{secOPSrep} we will show that this property is also crucial for detecting the Davies phase transition points in the Kerr black hole within entropy representation. However, as shown in Section \ref{secOPErep}, this does not hold in energy representation, since the thermodynamic curvature is zero and the metric scale cannot be associated to the specific type of information geometry.

\section{Optimal processes for Schwarzschild black hole}\label{secSchwTGO}

We apply the TGO procedure for the Schwarzschild solution within the framework of Thermodynamic geometry, using Hessian metrics in the space of states. In order to preserve important geometric structures in one-dimensional thermodynamic space\footnote{Such as avoiding the lack of intrinsic thermodynamic curvature in one-dimensional spaces.} we consider the Schwarzschild black hole as the $J\to 0$ limit of the Kerr solution. In this case, we derive explicit analytic expressions for the thermodynamic length and the evaporation lifespan of the black hole in each thermodynamic representation.

\subsection{Thermodynamics of the Schwarzschild black hole}

Let us briefly review the thermodynamic properties of the Schwarzschild black hole in energy, entropy and Helmoltz free energy representations, correspondingly. All quantities are expressed in SI units.

\subsubsection{The Schwarzschild solution}
 The spherically symmetric Schwarzschild black hole is described by the metric:
\begin{equation}
    ds^2=-\bigg(1-\frac{R_S}{r}\bigg)c^2dt^2+\bigg(1-\frac{R_S}{r}\bigg)^{-1}dr^2+r^2 \big( d\theta^2 +\sin^2\theta d\varphi^2 \big),
\end{equation}
where $R_S={2 G M}/{c^2}$ is the Schwarzschild gravitational radius of the event horizon, $M$ is the mass of the black hole, $G$ is the gravitational constant and $c$ is the speed of light. 

The corresponding black hole thermodynamics is described by three parameters: entropy $S$, energy $E=M c^2$ and temperature $T$, satisfying the first law of thermodynamics:
\begin{equation}
    dE=T dS,\quad S=\frac{k A}{4 l_p^2},\quad A=4 \pi R_S^2,\quad T=\frac{\kappa}{2 \pi}=\frac{c^3 \hbar }{8 \pi  G k M}.
\end{equation}
Here $A$ is the area of the event horizon, $l_p$ is the Planck length, $\kappa$ is the surface gravity, $k$ is the Boltzmann constant, and $\hbar$ is the Dirac constant. Since there are only three thermodynamic parameters the Schwarzschild black hole has three natural thermodynamic representations.  

\subsubsection{Entropy representation}

The entropy representation is determined by the fundamental relation $S=S(E)$ on the event horizon of the black hole\footnote{Conversion to the Planck system of units is presented in Appendix \ref{appPlUnits}.}:
\begin{align}\label{eqSTSchw}
S(E)=\lambda E^2, \quad\frac{1}{T}=\frac{dS}{ d E}=2 \lambda E, \quad \lambda =\frac{4 \pi k G}{\hbar c^5},\quad dS=\frac{1}{T} dE,
\end{align}
where $\lambda=4.52\times 10^{-41}\, (\text{JK})^{-1}$ is a physical scale factor.
The state space, $ S = S(E) $, forms a one-dimensional curve parameterized by the energy $ E $, which is embedded in the two-dimensional $(E, S)$ plane. 

\subsubsection{Energy representation}

The energy representation is defined by the relation $E=E(S)$, representing a one-dimensional curve parameterized by $ S $ and embedded in the $(S, E)$ plane. The relevant relations in this representation are given by: 
\begin{equation}
   E(S)=\eta  \sqrt{S},\quad T(S)=\frac{dE}{dS}=\frac{\eta }{2 \sqrt{S}}, \quad\lambda\eta ^2 =1,\quad dE=T dS,
\end{equation}
where for later convenience we have introduced a second scale parameter $\eta$.
\subsubsection{Free energy representation}
The Helmholtz free energy ensemble, $F=F(T)$, represents a curve in $(T, F)$ space, parameterized by the Hawking temperature $T$:
\begin{equation}\label{eqHelmholtzSchw}
   F(T)=E-T S= \frac{\eta ^2}{4 T},\quad E(T)=\frac{\eta ^2}{2 T},\quad S(T)=\frac{\eta ^2}{4 T^2}.
\end{equation}
This representation follows by a Legendre transformation of the energy of the black hole along the entropy direction.

\subsubsection{Thermodynamic instability}

The Schwarzschild black hole is unstable both locally and globally from a classical thermodynamic perspective. This is a consequence of its Hessian of the energy/entropy not being positive/negative definite, and the fact that its heat capacity is always negative:
\begin{equation}
  \frac{d^2 E}{dS^2}=-\frac{\eta }{4 S^{3/2}}<0,\quad \frac{d^2 S}{dE^2}=2\lambda>0,\quad  C=T \frac{dS}{dT}=-\frac{\eta ^2}{2 T^2}<0.
\end{equation}
Therefore, an isolated  Schwarzschild black hole should continuously radiate energy, for instance in terms of Hawking radiation. In  what follows, our goal is to  show that the Schwarzschild black hole may also radiate energy due to (optimal) thermal fluctuations on the event horizon.

\subsection{Optimal evaporation in entropy representation}\label{secHTDMER}

According to Ruppeiner's fluctuation theory \cite{Ruppeiner:1983zz, Ruppeiner:1995ss}, the probability of fluctuations between macro states is proportional to the thermodynamic length connecting them. A geodesic path in this space represents the optimal process used to achieve this change. Using an affine parametrization of the path, a proper time parameter can be introduced, enabling the process to be described by its  length, speed, and relaxation time. Here we apply this framework to study the optimal fluctuation induced evaporation of the Schwarzschild black hole solution in entropy representation.

\subsubsection{Geodesic profile of the energy}

In thermodynamic geometry, the Ruppeiner metric is defined by the Hessian of the entropy. For the Schwarzschild black hole, the space of states in entropy representation is one-dimensional, described by the curve $ S = S(E) $, where $ E $ is the energy coordinate. As a result, the thermodynamic metric reduces to a scalar function with a single component:
\begin{equation}\label{eqHessE}
g(E)=\epsilon\frac{d^2 S }{dE^2}=2 \epsilon\lambda.
\end{equation}
This situation is not well-suited for our analysis, as one-dimensional space lacks intrinsic curvature. To address this limitation, we embed the Schwarzschild parameter space $ (E, S(E)) $ within the higher-dimensional Kerr state space $ (E, J, S(E, J)) $, and subsequently take the limit $ J \to 0 $. Since the limit is regular\cite{Avramov:2023eif}, this approach allows for smooth transition from Kerr to Schwarzschild state space while retaining the relevant geometric structures. Therefore, Eqs. (\ref{eqKerrTGOSrepE}) and (\ref{eqKerrTGOSrepJ}) result in a single geodesic equation for the energy profile:
\begin{equation}\label{eqOnedGeodSchwHess}
    \ddot E(t)=0,\quad E(0)=E_0,\quad \dot E(0)=\dot E_0,
\end{equation}
where $E_0$ is the initial energy of the black hole and $\dot E_0=\dot E(0)$ is the initial rate of energy change.
The optimal path connecting the initial state with energy $ E(0)=E_0 $ to a state with energy $E(\tau)= E_\tau $, is just a straight line:
\begin{align}\label{eqGeodEprofSchw}
 E(t)= E_0+\dot E_0 t.
\end{align}
If the initial rate of change is positive, $ \dot{E}_0 > 0 $, the black hole is increasing its energy (mass).  This situation corresponds to an accretion-driven optimal process, where the black hole is fed on by an external source of matter or energy. Conversely, if the rate of change is negative, $ \dot{E}_0 = -|\dot{E}_0| < 0 $, the black hole's energy will decrease over time, indicating evaporation through fluctuations. Let us focus on the evaporation process.

\subsubsection{Relaxation time and thermodynamic length}

Let $\tau$ be the relaxation time required to transition from state with energy $E_0$ to a state with energy $E_\tau<E_0$. Using the geodesic profile (\ref{eqGeodEprofSchw}) one can extract $\tau$ by imposing $E(\tau)=E_\tau$:
\begin{equation}\label{eqEvapTSchchSrep}
    \tau=\frac{E_0-E_{\tau}}{|\dot E_0|}, \quad E_0>E_\tau.
\end{equation}
We observe that $ \tau $ is proportional to the absolute change of the energy $\Delta E = E_0 - E_\tau > 0 $ of the black hole, divided by the initial rate $|\dot E_0|$ of the process. Hence, considering fixed $E_0$ and $E_\tau$, the duration of the process will be affected only by the initial rate of change of the energy. Notice that $\tau$ remains unaffected by $\lambda$ and $\epsilon$. This is also true for the profile function (\ref{eqGeodEprofSchw}). However, once the process starts, the probability of fluctuating to the state $ E_\tau $ will be determined by the thermodynamic length at $ \tau $, which depends on both $\lambda$ and $\epsilon$. In the $J\to 0$ limit, the thermodynamic length in entropy representation yields\footnote{Note that by inserting $\tau$ from (\ref{eqEvapTSchchSrep}) one converts the result of (\ref{eqGeodAction2}) to (\ref{eqGeodAction}). In the latter one has an integral from $E_0$ to $E_\tau$ and $\sqrt{2 \epsilon \lambda dE^2}=\sqrt{2 \epsilon \lambda}\,|dE|=-\sqrt{2 \epsilon \lambda}\,dE$, since $dE<0$ for an evaporation.}:
\begin{equation}\label{eqTDLSchwHess}
\mathcal{L}(\tau)=\int\limits_0^\tau\sqrt{g_{ab}(\vec S) \dot S^a(t) \dot S^b(t)} \,dt=\sqrt{2 \epsilon\lambda  }\,|\dot E_0|  \,\tau=v\tau=\sqrt{2\epsilon \lambda  }(E_0-E_{\tau}),
\end{equation}
where $g_{ab}$ is defined in Eq. (\ref{eqKerrTDEntropy}), $S^a=(E,J)$, and $v=\sqrt{2 \epsilon\lambda  }\,|\dot E_0|$ is the thermodynamic speed of the process. 
For $ E(t) $ to represent a valid geodesic path in the state space, its associated thermodynamic length must be positive\footnote{In entropy representation, the units of $\mathcal{L}$ are $(\text{J/K})^{1/2}$. Its square, $\mathcal{L}^2$, has units of entropy $(\text{J/K})$, traditionally interpreted as the minimal entropy produced during a transformation of the system's state.}, which can be ensured by setting $ \epsilon >0 $. It is important to note that the final expression for $\mathcal{L}$ depends on the energy difference $(E_0 - E_\tau)$ between the initial and final states. As a result, the thermodynamic length increases for a larger energy difference, implying that it is more likely for the black hole to fluctuate to a nearby state $ E_\tau \approx E_0 $ rather than to a distant one for which $ E_\tau \ll E_0 $.

\subsubsection{Optimal evaporation of the Schwarzschild black hole}\label{secOESCHBH}

Consequently, one can look for an optimal path in parameter space with a proper length that can spontaneously evaporate the black hole. By setting $ E(\tau_{evap}) = E_\tau = 0 $, we can estimate the fluctuation evaporation time $\tau_{evap}$ and its associated  length:
\begin{equation}\label{eqEvapTsch}
\tau_{evap}=\frac{E_0}{|\dot E_0|}, \quad \mathcal{L}_{evap}=\sqrt{2\epsilon\lambda  }E_0.
\end{equation}
The expression for the length suggests that larger black holes are less likely to spontaneously evaporate, whereas the probability increases for smaller black holes. On the other hand, the expression for the evaporation time, caused by fluctuations, is straightforward but lacks context, unless compared to a known process. For this purpose, let us compare it to the evaporation occurring during the thermal Hawking radiation emitted by the black hole. In this case, we can use the Stefan-Boltzmann power law of black body radiation.  

By definition the power radiated by a Schwarzschild black hole is given by:
\begin{equation}
    P=-\frac{dE}{dt}=\sigma \varepsilon A T^4=\frac{\alpha}{E^2},\quad \alpha=\frac{\hbar c^{10} }{15360 \pi   G^2},\quad \sigma=\frac{\pi ^2 k^4}{60 \hbar ^3 c^2 },
\end{equation}
where $A=4 \pi R_S^2$ is the area of the event horizon, $\varepsilon=1$ is the gray-body factor for the black hole, and $\sigma$ is the Stefan-Boltzmann constant. This equation is separable in terms of $t$ and $E$, hence one can obtain the energy profile $E_H(t)$ of the Hawking radiation: 
\begin{equation}\label{eqHawkEvapESchw}
    E_H(t)=(E_0^3-3\alpha t)^{1/3},\quad \dot E_{H,0}=\dot E_H(0)=-\frac{\alpha }{E_0^2},
\end{equation}
where the index $H$ stands for Hawking.
Full evaporation occurs when $E_H(t=\tau_{H,\,evap})=0$, thus the Hawking evaporation time of the Schwarzschild black hole yields
\begin{equation}\label{eqHevapTSch}
\tau_{H,\,evap}=\frac{1}{3 \alpha}E_0^3 .
\end{equation}

Now, let us assume that the fluctuation evaporation time (\ref{eqEvapTsch}) is equal to the Hawking evaporation time\footnote{This implies that thermal fluctuations could imitate Hawking radiation.} (\ref{eqHevapTSch}). In this case, one can find the initial rate of energy change $\dot E_0$ required to excite the fluctuation process:
\begin{equation}
    \dot E_0=-\frac{3\alpha}{E_0^2}.
\end{equation}
Note that this rate differs 3 times from the Hawking initial rate $\dot E_{H,0}$ in (\ref{eqHawkEvapESchw}).
For a solar-mass black hole, $E_0\equiv E_\odot=M_\odot c^2= 1.8 \times 10^{47} \, \text{J}$, one has tremendously slow evaporation taking more than the current age of the Universe:
\begin{equation}\label{eqSSSchIC}
    \tau_{evap}=6.74\times 10^{74}\,\text{s}= 2.14\times 10^{58} \,\text{byr},
\end{equation}
where $1\,\text{byr}\approx 3.16\times 10^{16} \,\text{s}$.
The corresponding initial Hawking and fluctuation rates are:
\begin{equation}
  \dot E_{H,0}=-8.9\times 10^{-29}\,\text{J/s}=-2.8\times  10^{-12}\,\text{J/byr},  \quad \dot E_0=3 \dot E_{H,0}.
\end{equation}
In this case, the value of the square of the thermodynamic length for the evaporation of a solar-mass Schwarzschild black hole is
\begin{equation}
\mathcal{L}^2_{evap}=\epsilon \,3\times 10^{54} \,\text{J/K},
\end{equation}
which can also be interpret as the minimal entropy produced during this process. Note that it depends on $\epsilon$. Assuming that the square of the thermodynamic length is at least equal to the initial entropy of the black hole, $\mathcal{L}^2_{evap} = 2\epsilon \lambda E_0^2 \equiv S_0 = \lambda E_0^2$, we can fix $\epsilon = 1/2$. Since this choice guarantees that the minimal entropy produced during the evaporation process is at least equal to the black hole's initial entropy, it follows that $\epsilon < 1/2$ should not be permitted if full evaporation is achievable through optimal fluctuations.

\begin{figure}[H]
\centering
  \begin{subfigure}{0.43\textwidth}
  	\includegraphics[width=1.0\linewidth]{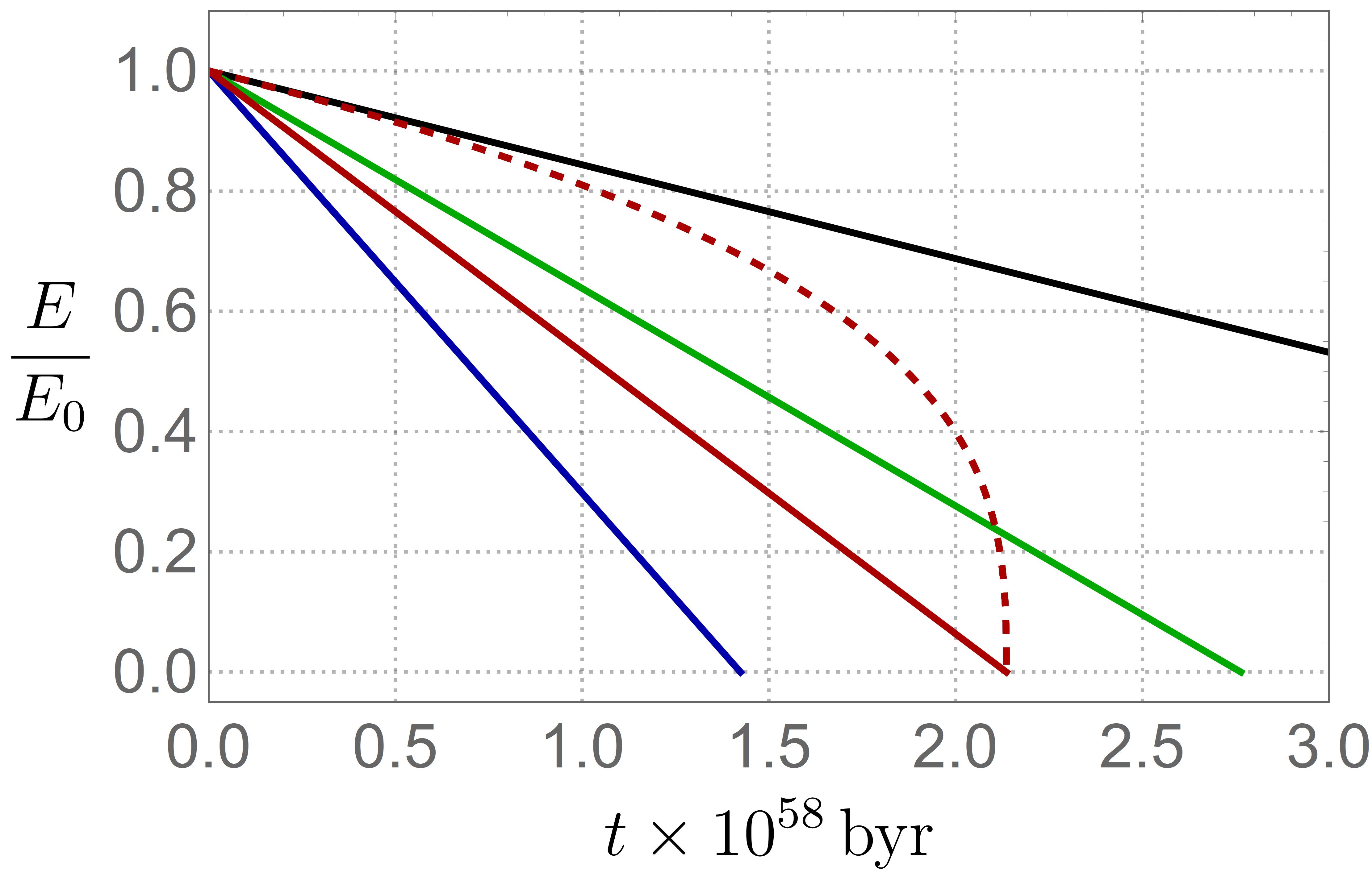}
  	\caption{Evaporation in entropy representation.}\label{figEHcomp}
  	\end{subfigure}
  	\hspace{0.7 cm}
  	\begin{subfigure}{0.43\textwidth}\includegraphics[width=1.0\linewidth]{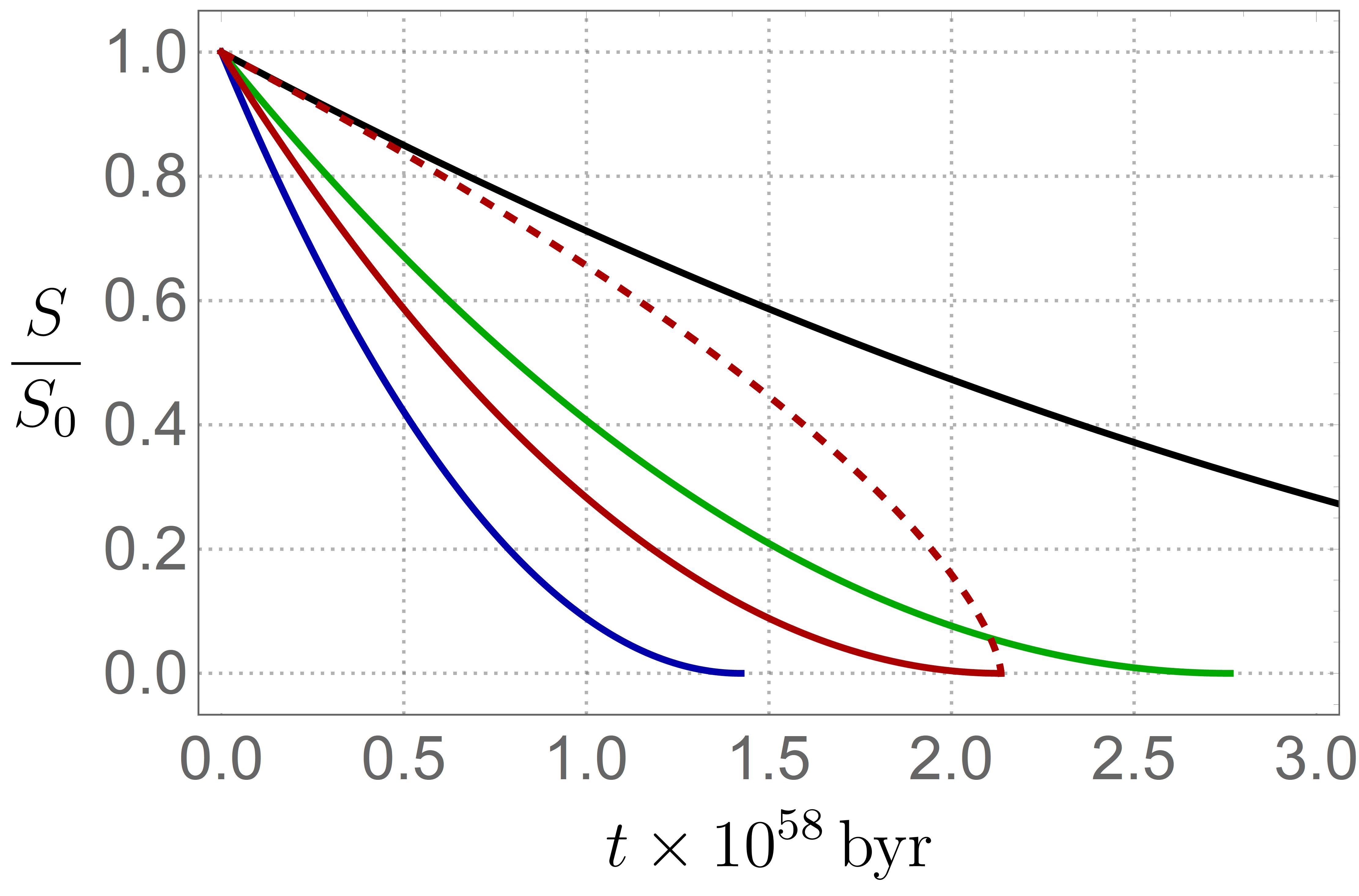}
  \caption{Evaporation in energy representation.}\label{figSHcomp}
  \end{subfigure}
  \hspace{1.5 cm}
      \caption{The evaporation profiles of  energy and entropy of a solar-mass Schwarzschild black hole are shown via two processes: \textbf{(a)} fluctuations (optimal processes) represented by solid curves, and \textbf{(b)} the Hawking evaporation process shown by dashed red curves. Both profiles are normalized using the energy and entropy of a solar-mass Schwarzschild black hole (see App. \ref{appNDF1}). 
}\label{figEScomSchwz}
\end{figure}

A second scenario is to consider fluctuations with equal to Hawking initial rates of change of the energy, $\dot E_0=\dot E_{H,0}$. In this case, the fluctuation evaporation time is about three times larger than the Hawking evaporation time. 
A comparison between the Hawking energy evaporation profile (\ref{eqHawkEvapESchw}) and the energy fluctuation profile (\ref{eqGeodEprofSchw}) is shown on Fig. \ref{figEHcomp}. The red dashed and the red solid curves represent scenarios where the fluctuation evaporation time matches the Hawking evaporation time. In this case, the initial fluctuation evaporation rate is about three times larger than the initial Hawking evaporation rate. The black solid curves depict fluctuations with initial evaporation rates equal to Hawking's rates. In this case, the fluctuation evaporation time is about three times larger than the Hawking evaporation time. The blue curves indicate faster optimal evaporation than Hawking, leading to complete evaporation earlier. In contrast, the green curves initially evaporate more quickly than Hawking, but take longer to fully evaporate. Notably, evaporation via thermal fluctuations (solid  curves) is different from the Hawking evaporation (the dashed red curve). In general, the optimal processes on the event horizon of the black hole do not need to follow Stefan-Boltzmann power law of black body radiation.

\subsubsection{The role of the CMB for isolated black holes}

Let us now briefly comment on the impact of the Cosmic Microwave Background (CMB) on black hole's evaporation. Notably, isolated stellar black holes with masses exceeding 0.75\% of the Earth's mass are cooler than the CMB, which has a temperature of approximately $ T \approx 2.725\,\text{K} $. These black holes will absorb energy from their surroundings, causing them to grow rather than evaporate. 
However, the processes governed by Eq. (\ref{eqGeodEprofSchw}) can occur spontaneously. Once initiated, there exists a non-zero probability for the black hole to emit energy or completely evaporate via thermodynamic fluctuations, irrespective of the CMB's presence. However, the probability for the latter process to occur is negligible due to the large evaporation thermodynamic length, as shown in the previous section.

Conversely, for Planck-scale black holes\footnote{A black hole with Planck energy has a gravitational radius $R_S = 2 l_p $, where $l_p $ is the Planck length.}, with $E_0 \equiv E_p = 1.96 \times 10^9\,\text{J} $, spontaneous evaporation is far more likely due to their small thermodynamic length:  
\[
\tau_{\text{evap}} = 9 \times 10^{-40}\,\text{s}, \quad \dot{E}_0 = -2 \times 10^{48}\,\text{J/s}, \quad T_0 = 5.7 \times 10^{30}\,\text{K}, \quad \mathcal{L}_{\text{evap}}^2 = \epsilon \, 3 \times 10^{-22}\,\text{J/K}.
\]
Essentially, such rapid evaporation can also be realized through the standard Hawking radiation.

\subsubsection{Thermodynamic curvature}

Let us also consider the thermodynamic curvature. Since we are embedding the Schwarzschild state space into the higher-dimensional Kerr state space, it acquires a nonzero thermodynamic curvature, given by the $J\to 0$ limit of Eq. (\ref{eqRupScCurve}), $\mathcal{R} = (\epsilon \lambda E^2)^{-1}$. Based on its interpretation, for $\epsilon = 1/2$, one has $\mathcal{R} > 0$, indicating that the optimal evaporation fluctuations of the Schwarzschild black hole in entropy representation arise from the repulsive nature of interactions between the quantum bits on the event horizon.

\subsection{Optimal evaporation in energy representation}

The corresponding entropy equation in energy representation follows by taking the $J\to 0$ limit of equation (\ref{eqGeodEnerKerr}):
\begin{equation}\label{eqHessErep}
    \ddot S(t)-\frac{3 }{4 }\frac{\dot S^2(t)}{S(t)}=0,\quad S(0)=S_0,\quad \dot S(0)=-|\dot S_0|<0,
\end{equation}
where we already assumed a decreasing entropy for an evaporating black hole\footnote{The inverse process of an optimally growing black hole can also be considered with $\dot S(0)=\dot S_0>0$}. The solution  leads to the following entropy profile:
\begin{equation}\label{eqSprofHessSchw}
    S(t)= \frac{(4 S_0-|\dot S_0| t)^4}{256 S_0^3}.
\end{equation}
The relaxation time from $S_0$ to $S_\tau$, and the evaporation time at $S_\tau=0$, are given by
\begin{equation}\label{eqStimeHess}
 \tau=\frac{4 S_0^{3/4} \left(\sqrt[4]{S_0}-\sqrt[4]{S_\tau}\right)}{|\dot S_0|},\quad\tau_{evap}=\frac{4 S_0}{|\dot S_0|}.
\end{equation}
The associated thermodynamic length is:
\begin{equation}
    \mathcal{L}(\tau)=\frac{ \sqrt{-\epsilon\eta } \,|\dot S_0| }{2 S_0^{3/4}}\,\tau
    =v \tau=2\sqrt{-\epsilon\eta }\big(\sqrt[4]{S_0}-\sqrt[4]{S_\tau}\big),\quad \mathcal{L}_{evap}=2\sqrt{-\epsilon\eta }\,S_0^{1/4},
\end{equation}
which is real and positive if $\epsilon<0$. Note that in energy representation the units of the square of the thermodynamic length $\mathcal{L}^2$ are Joules. Therefore, one can interpret $\mathcal{L}^2$ as the minimal energy (or work) required to transform the state of the system. In this case, considering evaporation, the black hole energy provides the minimal available energy for the fluctuations, thus $\mathcal{L}^2_{evap}=-4\epsilon\eta\,S_0^{1/2}=-4\epsilon\eta\,\sqrt{\lambda} E_0\equiv E_0$, hence $\epsilon=-1/4$. This choice for $\epsilon$ ensures that the energy available for the fluctuations driving the black hole's optimal evaporation is at least equal to the black hole's initial energy. For $\epsilon < -1/4$, the energy required to evaporate the black hole exceeds the initial energy of the black hole itself. As a result, additional energy would be needed, and the fluctuations cannot occur spontaneously. However, the analysis above remains valid also in this case.

Using Eqs. (\ref{eqSTSchw}) and (\ref{eqHawkEvapESchw}), we can find the Hawking entropy of the black hole\footnote{This is the time evolution of the entropy of the bare black hole and not the entropy of the Hawking radiation itself as seen by an asymptotic observer. }:
\begin{equation}\label{eqHEPSSchwz}
    S_H(t)=\lambda E_H^2(t)=\lambda (E_0^3-3\alpha t)^{2/3}=\lambda (\eta^3 S_0^{3/2}-3\alpha t)^{2/3},\quad \dot S_{H,0}=\dot S_H(0)=-\frac{2 \alpha }{\eta^3\sqrt{S_0}}.
\end{equation}
Hawking evaporation time is then given by $S(t=\tau_{H,\,evap})=0$:
\begin{equation}\label{eqHevapTimeS}
 \tau_{H,\,evap}= \frac{\eta ^3 }{3 \alpha }S_0^{3/2}.
\end{equation}
Equating (\ref{eqHevapTimeS}) and (\ref{eqStimeHess}), one finds the initial entropy fluctuation rate of change\footnote{A solar-mass Schwarzschild black hole has initially $S_0=1.5\times 10^{54} \,\text{J/K}$ and $|\dot S_0|=8.7 \times 10^{-21} \,\text{J/(Ks)}$.}:
\begin{equation}
    \dot S_0=-\frac{12 \alpha}{\eta^3 \sqrt{S_0}}.
\end{equation}
Note that it differs six times from the Hawking initial rate $\dot S_{H,0}$ in (\ref{eqHEPSSchwz}). A comparison between the Hawking entropy evaporation profile (\ref{eqHEPSSchwz}) and the entropy fluctuation profile (\ref{eqSprofHessSchw}) is shown on Fig. \ref{figSHcomp}.

\subsection{Optimal evaporation in Helmholtz representation}

One could also write the corresponding geodesic equation for the temperature profile $T(t)$ in canonical ensemble:
\begin{equation}\label{eqTempSchGeod}
    \ddot T(t)-\frac{3 }{2}\frac{\dot T^2(t)}{T(t)}=0,\quad T(0)=T_0,\quad \dot T(0)=\dot T_0>0.
\end{equation}
The rate of change for the temperature should be positive to account for the increasing temperature of the black hole when its mass-energy decreases. The solution to \eqref{eqTempSchGeod} yields
\begin{equation}\label{eqHelhTDlHess}
    T(t)= \frac{4 T_0^3}{(\dot T_0 t-2 T_0)^2}.
\end{equation}
The temperature $T(t)$ is an increasing function of time from $t=0$ to $t=2 T_0/\dot T_0$, where it diverges. The latter correspond to the optimal evaporation time. The relaxation time $\tau$ between two states is defined by $T(\tau)=T_\tau$, while the evaporation time is defined when the temperature becomes infinitely high, hence:
\begin{equation}
    \tau = \frac{2 T_0 \left(\sqrt{T_\tau}-\sqrt{T_0}\right)}{\dot T_0 \sqrt{T_\tau}},\quad \tau_{evap}=\frac{2 T_0}{\dot T_0}.
\end{equation}
The thermodynamic length is positive for $\epsilon>0$:
\begin{equation}
    \mathcal{L}(\tau)=\frac{\sqrt{\epsilon}\,\eta \dot T_0}{\sqrt{2}\, T_0^{3/2}} 
    \tau
    =v \tau=\frac{\sqrt{2\epsilon} \,\eta\left(\sqrt{T_\tau}-\sqrt{T_0}\right)}{ \sqrt{T_0 T_\tau}}, \quad\mathcal{L}_{evap}=\frac{\sqrt{2\epsilon}\,\eta}{\sqrt{T_0}}.
\end{equation}
The units of $\mathcal{L}^2$ are  in Joules and hence it sets lower bound on the energy used to transform the state of the black hole. In case of an evaporation, setting the lower bound to be the initial energy of the black hole, $\mathcal{L}^2_{evap}=2 \epsilon \eta^2 T_0^{-1}=4 \epsilon \eta^2\lambda E_0\equiv E_0$, one finds $\epsilon=1/4$.

\subsection{Short summary}
Let us briefly summarize the results for the Schwarzschild black hole. We found that there is small, but nonzero probability for optimal thermal fluctuations to completely evaporate the black hole across three different thermodynamic representations. Depending on the initial rates of change of the parameters, the optimal fluctuation-driven evaporation process can either be faster or slower compared to Hawking radiation. In Fig. \ref{figEScomSchwz}, we demonstrated that the optimal evaporation process, when initiated with the same rates of change as the Hawking profiles, results in slower evaporation. However, increasing the initial rates of change allows for a faster evaporation.

\section{Optimal processes for Kerr black hole}\label{secKerrTGO}

In this section, we investigate optimal thermal fluctuations and processes on the space states of the Kerr black hole solution. Our numerical results indicate that, under specific initial conditions, fluctuation-driven processes can lead the black hole to stop rotating, completely evaporate, or approach a Davies phase transition point.  We also show that optimal accretion-driven processes can also significantly alter the state of the black hole.

Importantly, we demonstrate that the thermodynamic length, combined with the sign of the metric scale $\epsilon$, provide effective tools for identifying Davies phase transition points in entropy representation.  

\subsection{Thermodynamics of the Kerr black hole}

Let us briefly review the thermodynamic properties of the Kerr black hole in energy and entropy representations.

\subsubsection{The Kerr solution}

The Kerr metric in Boyer-Lindquist coordinates $(t, r, \theta, \phi)$ is given by:
\begin{align}
\nonumber ds^2 = -\left(1 - \frac{R_S r}{\Sigma}\right)c^2 dt^2 - \frac{2R_S a r\sin^2\theta}{\Sigma} c \, dt \, d\phi + \frac{\Sigma}{\Delta}dr^2 + \Sigma \, d\theta^2 
\\[5pt]
    + \left(r^2 + a^2 + \frac{R_S a^2 r\sin^2\theta}{\Sigma}\right)\sin^2\theta \, d\phi^2,
\end{align}
where we have used the standard notations:
\begin{equation}
  \Sigma = r^2 + a^2\cos^2\theta,\quad
\Delta = r^2 - R_S r + a^2,\quad a = \frac{J}{Mc},\quad a_*=\frac{2 a}{R_S},\quad R_S=\frac{2 G M}{c^2}.
\end{equation}
Here $M$ is the mass of the black hole, $J$ is the angular momentum, $a$ is the spin parameter, and $ -1< a_*< 1$ is the specific (dimensionless) spin parameter. The positions of the event horizon $r_+$ and the internal Cauchy horizon $r_-$ are given by the roots of $\Delta=0$:
\begin{equation}\label{eqRpmKerr}
    r_{\pm}= \frac{1}{2} \Big(R_S\pm \sqrt{R_S^2-4 a^2}\Big).
\end{equation}
Consequently, Kerr's thermodynamics is determined by\footnote{It is easy to reproduce the expressions for $T$ and $\Omega$ by setting $R_S=r_- + r_+$ and $a=\sqrt{r_- r_+}$ and then using the standard relations to the rotating Killing vector of the Kerr black hole.}:
\begin{equation}\label{eqKerrTD}
 S=\frac{k A}{4 l_p^2}, \quad A=4 \pi (r_{+}^2+a^2),\quad T=\frac{\hbar c}{4 \pi  k r_+}\frac{r_+ - r_-}{r_+ + r_-}, \quad \Omega=\frac{c r_-}{a (r_+ +r_-)}.
\end{equation}

We are now ready to consider the relevant thermodynamic representations of the Kerr black hole solution.

\subsubsection{Entropy representation}

Inserting $r_+$ from Eq. (\ref{eqRpmKerr}) into $S$ from  Eq. (\ref{eqKerrTD}) one  finds\footnote{All quantities are expressed in SI units: $S\,[\text{J/K}]$, $E\,[\text{J}]$, $J\,[\text{J\,s}]$, $T\,[\text{K}]$, and $\Omega\,[\text{s}^{-1}]$. }:
\begin{align}\label{eqKerrEntr}
    &S(E,J) = \frac{\lambda}{2} \big( E^2+\sqrt{ E^4 -  \zeta^2 J^2} \big),
    \\[5pt]\label{eqKerrTemp}
    &T(E,J) = \bigg(\frac{\partial S}{\partial E} \bigg)^{\!\!-1}\bigg|_J=\frac{\sqrt{E^4-\zeta ^2 J^2}}{\lambda E  \big(E^2+\sqrt{E^4-\zeta ^2 J^2}\big)},
    \\[5pt]
    &\Omega (E,J) = -T\frac{\partial S}{\partial J} \bigg|_E=\frac{\zeta ^2 J}{2 E\big( E^2+ \sqrt{E^4-\zeta ^2 J^2}\big)},
    \\[5pt]
    &dS=\frac{1}{T} dE-\frac{\Omega}{T} dJ,\quad \lambda =\frac{4 \pi k G}{\hbar c^5},\quad \zeta=\frac{c^{5}}{G},\quad a_*=\frac{2 a}{R_S}=\frac{\zeta J}{E^2},
\end{align}
where\footnote{The parameter $\lambda$ can be expressed in terms of the Planck energy $E_p$ and the Planck temperature $T_p$: $\lambda=\frac{4 \pi}{E_p T_p}$. The parameter $\zeta$ is called the Planck power: $\zeta=\frac{E_p}{t_p}$, where $t_p$ is the Planck time.} $\lambda= 4.52\times 10^{-41}\, (\text{JK})^{-1}$,  $\zeta= 3.64\times 10^{52} \,\text{W}$, and $ -1< a_*< 1$ is the specific spin parameter of the Kerr black hole\footnote{Conversion to the Planck system of units is presented in App. \ref{appPlUnits}.}. The non-extremality condition $T>0$ $(r_+>r_-)$ leads to:
\begin{equation}\label{eqKerrExsistS}
    E^2>\zeta |J|,
\end{equation}
The extremal case $r_-=r_+$ ($a_*=\pm 1$) is classically forbidden due to the third law of thermodynamics.

\subsubsection{Energy representation}

The  first law of thermodynamics in energy representation is $dE=T dS+\Omega dJ$, where the relevant quantities in ($S,J$) space are given by:
\begin{align}\label{eqEnRepKerr}
&E(S,J)=\frac{\sqrt{\zeta ^2 \lambda ^2 J^2+4 S^2}}{2 \sqrt{\lambda S}},
\\[5pt]
& T(S,J)=
\frac{\partial E}{\partial S}\bigg|_J=\frac{4 S^2-\zeta ^2 \lambda ^2 J^2}{4 S \sqrt{\lambda S(\zeta ^2 \lambda ^2 J^2  +4  S^2)}},
\\[5pt]
& \Omega(S,J)=\frac{\partial E}{\partial J}\bigg|_S=\frac{  \lambda ^{3/2}\zeta ^2 J}{2 \sqrt{S(\zeta ^2  \lambda ^2 J^2+4 S^2)}}.
\end{align}
In this case, the non-extremality condition $T>0$ leads to
\begin{equation}\label{eqKerrExsistE}
   2 S> \zeta  \lambda |J| .
\end{equation}
In addition, the specific spin parameter $a_*$ in $(S,J)$ space yields:
\begin{equation}
a_*=\frac{4\zeta\lambda J S}{\zeta^2 \lambda^2 J^2+4 S^2}, \quad -1< a_*<1.
\end{equation}

\subsubsection{Thermodynamic instability of Kerr black hole}

The thermodynamic instability of the Kerr black hole has been thoroughly analyzed in \cite{Avramov:2023eif}.  Due to the fact that the system is globally unstable the Kerr black hole can radiate energy. As a consequence, the inherent thermodynamic instability creates a favorable environment for thermodynamic fluctuations. These fluctuations may arise spontaneously or be driven by accretion processes. In either scenario, the optimal trajectories link two non-equilibrium states of the system. Once the initial fluctuations subside, the black hole either initiates a new cascade of fluctuations or continues its evolution along conventional pathways.

\subsection{Comments on Hawking evaporation profiles for Kerr black hole}

Let us consider the Stefan-Boltzmann power law for Kerr:
\begin{equation}\label{eqHKerrSBlaw}
    \dot E(t)=-\sigma A T^4=-\frac{f(a_*)}{E^2(t)},\quad f(a_*)=\frac{b \big(1-a_*^2 (t)\big)^2}{\big(\sqrt{1-a_*^2 (t)}+1\big)^3},\quad  a_*(t)=\frac{\zeta  J(t)}{E^2(t)},\quad b=\frac{2 \pi ^3 k^4}{15\hbar ^3 \zeta ^2 \lambda ^4 }.
\end{equation}
In order to solve these equations one needs to set the profile of one of the functions, e.g. the specific spin parameter $a_*(t)$. The latter is achieved in the following model \cite{page1976thermal, page2013jcap, Nian:2019buz, Arevalo:2024kmo}:
\begin{equation}
    \gamma=-E^3(t) \frac{d \ln X(t)}{dt},
\end{equation}
where $\gamma$ is some constant and $X(t)$ represents the parameter that needs to be
considered for time evolution. Applying this model to the energy and the angular momentum we find:
\begin{equation}\label{eq_EJ_model}
    \dot E(t) E^2(t)=-\gamma_1,\quad \zeta \dot J(t) E(t)=-\gamma_2 a_*(t),
\end{equation}
where $\gamma_{1,2}$ are constants. By comparing the first equation with (\ref{eqHKerrSBlaw}), we obtain $\gamma_1 = f(a_*)$, leading to an algebraic equation for $a_*(t)$. This implies that $a_*(t)$ is time-independent constant, denoted $a_*$, which allows both equations in \eqref{eq_EJ_model} to be solved. As a result, the Hawking profiles for the energy and the angular momentum of Kerr black hole are:
\begin{equation}\label{eqEJprofsKerH}
    E_H(t)= \big( E_0^3-3 \gamma _1 t \big)^{1/3},\quad J_H(t)= J_0-\frac{a_* \gamma _2} {2  \zeta \gamma _1}\left[E_0^2-\left(E_0^3-3 \gamma _1 t\right)^{2/3}\right].
\end{equation}
The profile for the energy becomes equal to the Schwarzschild one from Eq.(\ref{eqHawkEvapESchw}), if $a_*=0$, which leads to  $\gamma_1=\alpha$. The extremal case $a_*=1$ leads to $\gamma_1=0$. This shows that we only need to adjust the values of $a_*$ and $\gamma_2$. The Hawking evaporation time follows from $E_H(t)=0$:
\begin{equation}\label{eqKerrEvapTime}
    \tau_{H,evap}=\frac{E_0^3}{3 \gamma _1},\quad \gamma_1>0.
\end{equation}
Inserting this in  $J_H(t)=0$, we find  $\gamma_2$ for a complete Kerr black hole evaporation via Hawking radiation:
\begin{equation}
    \gamma _2=\frac{2 \zeta  J_0}{a_* E_0^2}\gamma _1,\quad \gamma_2 >0.
\end{equation}

Finally, inserting the Hawking profiles from (\ref{eqEJprofsKerH}) into (\ref{eqKerrEntr}), we can model the time evolution of the entropy on the event horizon of the Kerr black hole, if it emits Hawking radiation:
\begin{equation}\label{eqSHKerrProf}
    S_H(t)=\frac{\lambda}{2} \bigg( E_H^2(t)+\sqrt{ E_H^4(t) -  \zeta^2 J^2_H(t)} \bigg).
\end{equation}

\begin{figure}[H]
	\centering
	\begin{subfigure}{0.45\textwidth}
		\includegraphics[width=1.0\linewidth]{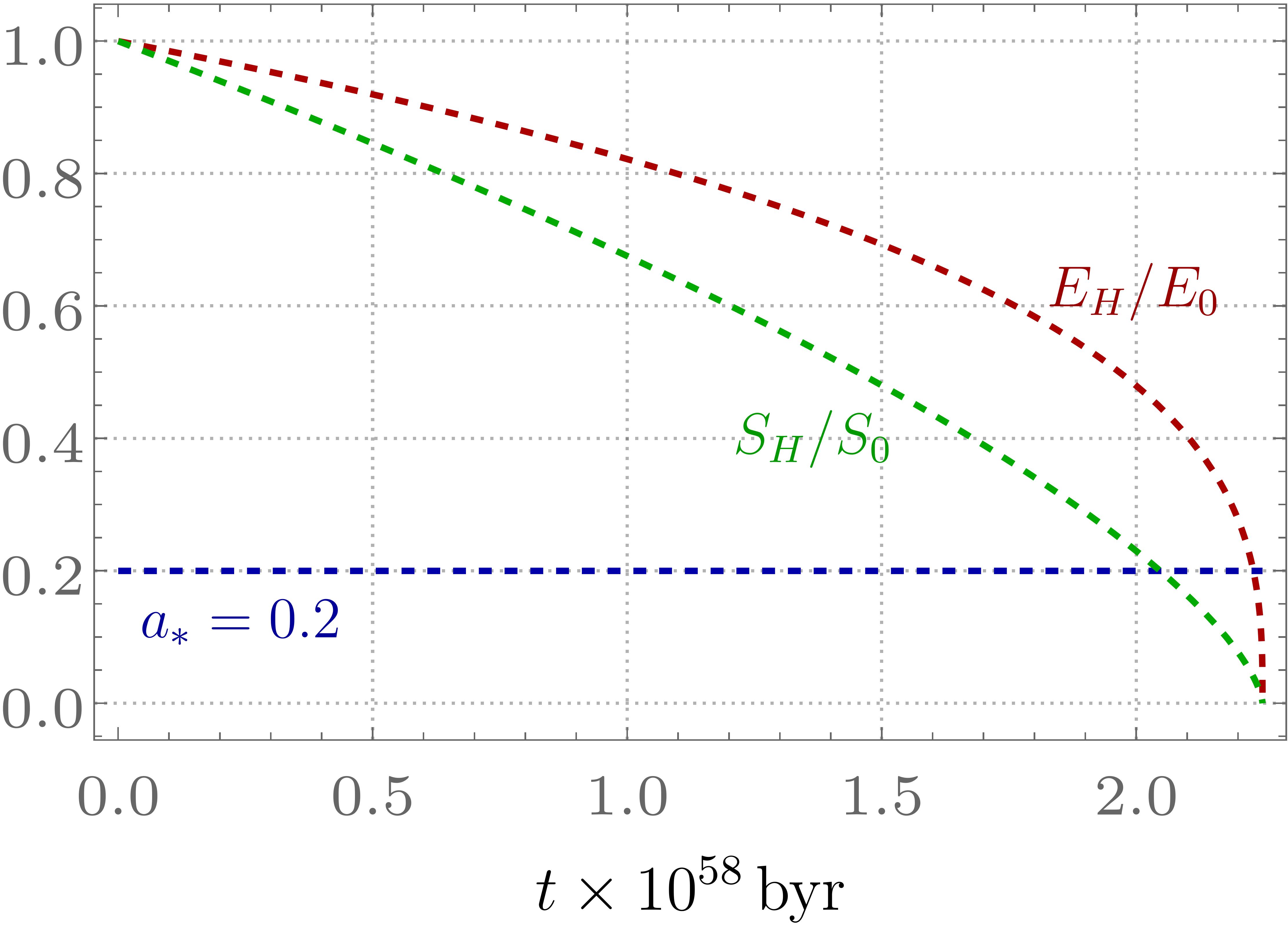}
		\caption{Hawking evaporation of Kerr at $a_*=0.2$.}\label{figHevapKE}
	\end{subfigure}
	\hspace{0.8 cm}
	\begin{subfigure}{0.45\textwidth}\includegraphics[width=1.0\linewidth]{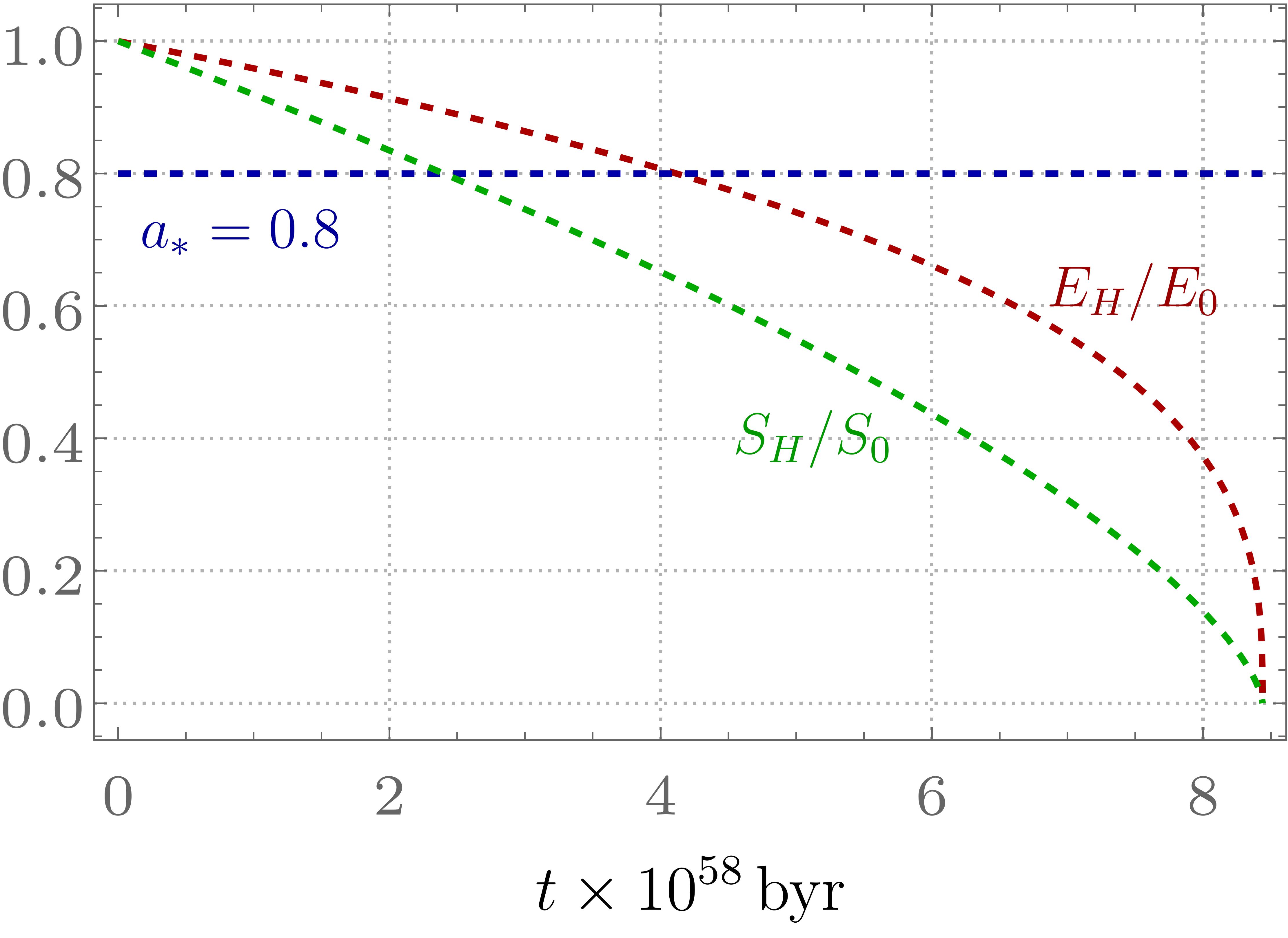}
		\caption{Hawking evaporation of Kerr at $a_*=0.8$.}\label{figHevapKS}
	\end{subfigure}
	\hspace{1.5 cm}
	\caption{Hawking evaporation profiles (\ref{eqEJprofsKerH}) and (\ref{eqSHKerrProf}) for a solar-mass Kerr black hole at different specific spins. The numeric data for these profiles is provided in App. \ref{appfigKerr0208}. 
	}\label{figKerr0208}
\end{figure}

Consequently, it is easy to find the initial Hawking rates of change\footnote{Typical initial Hawking energy, entropy and angular momentum rates of change for a solar-mass Kerr black hole $(a_*=0.6)$ are: $|\dot E_{H,0}|\sim 10^{-29}\,\text{J/s}$, $|\dot S_{H,0}|\approx 10^{-22}\,\text{J/(Ks)}$ and $|\dot J_{H,0}|\approx 10^{-34}\,\text{J}$ (see App. \ref{appfigKerr0208}).} at $t=0$ for this model, simply by differentiating the profiles with respect to time:
\begin{equation}\label{eqHERESJ}
    \dot E_{H,0}=-\frac{\gamma_1}{E_0^2},
    \quad \dot J_{H,0}=-\frac{ \gamma_2 a_*}{ \zeta E_0},\quad \dot S_{H,0}= \frac{4 \zeta  \lambda ^{5/2} J_0 \sqrt{S_0} \left(\gamma _1 \zeta \lambda J_0 -a_*  \gamma _2 S_0\right)}{\left(\zeta ^2 \lambda ^2 J_0^2-4 S_0^2\right) \sqrt{\zeta ^2 \lambda ^2J_0^2 +4 S_0^2}}.
\end{equation}
Note that all of these rates are negative due to the evaporation. As an illustration, the Hawking evaporation profiles (\ref{eqEJprofsKerH}) and (\ref{eqSHKerrProf}) for a solar-mass Kerr black hole at different specific spins are depicted on Fig. \ref{figKerr0208}. 

In addition, Hawking profiles and the optimal thermogeometric ones have to obey certain restrictions. For example, one has to take into account the occurrence of possible Davies phase transition curves. We will  comment on this issue in the following subsection.  

\subsection{Comments on phase transitions and Hessian metrics}

In the early development of black hole thermodynamics Paul Davies suggested that phase transitions in black holes occur at points where their heat capacities diverge \cite{Davies:1977bgr, Davies_1978}. Later, within the framework of Thermodynamic geometry, it was observed that the curvature of Hessian metrics might not capture all critical points of the system. For example, thermodynamic curvature (\ref{eqRupScCurve}) is sensitive only to the extremal case (\ref{eqKerrExsistS}), where it diverges, but it does not detect the Davies curves defined by the divergences of the heat capacity $C_J(E,J)$ (see App. \ref{appA}), which are given by
\begin{equation}\label{eqDaviesCS}
     {\zeta  J}=\pm E^2 \sqrt{2 \sqrt{3}-3}.
\end{equation}
A similar situation occurs in the energy representation, where the thermodynamic curvature of the Weinhold metric for the Kerr black hole is zero. However, the divergences of the heat capacity $C_J$ in the $(S,J)$ space are described by the following Davies curves:
\begin{equation}\label{eqDaviesCE}
   \sqrt{3} \,\lambda \zeta   J =\pm  2 S \sqrt{2 \sqrt{3}-3} .
\end{equation}
Both conditions (\ref{eqDaviesCS}) and (\ref{eqDaviesCE}) can be expressed in a single equation for the specific spin:
\begin{equation}\label{eqDavies_a}
    \tilde a_*=\pm\sqrt{2 \sqrt{3}-3}\approx \pm0.681,
\end{equation}
where we used tilde to indicate that it is a single number. It is independent of the mass or the angular momentum and represents an universal characteristic for all Kerr black holes.

A key question now is what happens to a Kerr black hole when its specific spin crosses the Davies point at $ \tilde a_* = 0.861 $. In the Hawking model, which we used for comparison, the specific spin is designed to remain constant throughout the evolution of the Hawking profiles, so it may never crosses a Davies point. However, in our TGO model, the spin parameter varies over time, allowing for the possibility of crossing the Davies point.

When considering optimal processes in entropy representation, our findings indicate that the sign of the scale parameter $\epsilon$ changes depending on whether its initial spin is above or below the Davies critical point. Consequently, the process cannot initiate unless the correct sign of the $\epsilon$ parameter is considered.

On the other hand, in  energy representation, the thermodynamic length and the metric scale factor cannot be used to detect the Davies critical points. This is  rooted in the fact that the thermodynamic curvature of the Kerr black hole in this representation is zero and the metric scale cannot be associated with the information geometry on the space of states.

\subsection{Optimal processes in entropy representation}\label{secOPSrep}

We investigate evaporation-driven and accretion-driven optimal processes of the Kerr black hole within the entropy representation.

\subsubsection{Geodesic equations on the space of states}\label{secKGSrep}

Ruppeiner's thermodynamic metric for  Kerr is defined by the Hessian of the entropy:
\begin{align}\label{eqKerrTDEntropy}
&\hat g^{(R)}=\epsilon \hat 
 H_S=\epsilon \!\left(\!\!
\begin{array}{cc}
 \frac{\partial^2 S}{\partial E^2}\big|_J  \!\!& \!\!\frac{\partial^2 S}{\partial E \partial J}\\[5pt]
 \frac{\partial^2 S}{\partial E \partial J} \!\!& \!\! \frac{\partial^2 S}{\partial J^2}\big|_E\\
\end{array}
\!\!\right)
 =\epsilon\lambda\! \left(\!\!
\begin{array}{cc}
 1 +\frac{ E^2\left(E^4-3 \zeta ^2 J^2\right)}{\left(E^4-\zeta ^2 J^2\right)^{3/2}}  \!& \! \frac{\zeta ^2 J E^3}{\left(E^4-\zeta ^2 J^2\right)^{3/2}} \\[7pt]
 \frac{\zeta ^2 J E^3  }{\left(E^4-\zeta ^2 J^2\right)^{3/2}} \!& \!-\frac{\zeta ^2 E^4 }{2 \left(E^4-\zeta ^2 J^2\right)^{3/2}} \\
\end{array}
\!\!\right).
\end{align}
The sign of the scale $\epsilon$ determines the sign of the scalar thermodynamic curvature\footnote{The units of the curvature in entropy representation are inverse entropy: $\mathcal{R} \,[\text{K/J}]$.}:
\begin{equation}\label{eqRupScCurve}
\mathcal{R}^{(R)}=\frac{2 E^2-\sqrt{E^4-\zeta ^2 J^2}}{\epsilon \lambda  E^2   \sqrt{E^4-\zeta ^2 J^2}}.
\end{equation}
Assuming the non-extremality condition (\ref{eqKerrExsistS}) the thermodynamic curvature is positive if $\epsilon>0$ and negative if $\epsilon<0$.  According to the standard interpretation \cite{Ruppeiner:2010}, positive curvature characterizes an elliptic information geometry, which corresponds to dominant repulsive interactions among the microscopic degrees of freedom on the event horizon. Conversely, negative curvature represents a hyperbolic information geometry, associated with attractive interactions. In the context of optimal processes, the  sign of $\epsilon$ is fixed by the requirement of a positive-definite thermodynamic length. Consequently, the existence of a specific thermodynamic process is closely tied to the type of information geometry defined on the state space.
The corresponding system of geodesic equations (\ref{eqGeodesicGeneric}) for $E(t)$ and $J(t)$ is given by:
\begin{align}\label{eqKerrTGOSrepE}
 &\ddot E- \frac{\zeta ^2 J E^2 }{U^4 \big(U^2+E^2\big)}\dot J \dot E+\frac{\zeta ^2 E^3}{2 U^4 \big(U^2+E^2\big)}\dot J^2=0,
\\[5pt]\label{eqKerrTGOSrepJ}
   &\ddot J-\frac{2 E^2( U^2-E^2)-3 \zeta ^2 J^2}{2 J U^4}\dot J^2 +\frac{2 E^2(U^2-2E^2)-4 \zeta ^2 J^2}{E U^4}\dot J \dot E
    +\frac{3 J \left(E^4+\zeta ^2 J^2\right)}{E^2 U^4}\dot E^2=0,
\end{align}
where $U^2=\sqrt{E^4-\zeta ^2 J^2}$. 
This highly nonlinear system\footnote{Note that this system of thermodynamic geodesic equations do not allow processes with certain parameters held constant, besides for the trivial values. This indicates that all relevant quantities must be regarded as dynamical over time.  This is in contrast to the processes investigated in \cite{Gruber:2016xui}.} can be solved numerically for some appropriate initial and boundary conditions of the parameters. In the following we show this for stellar-sized Kerr black holes.

\subsubsection{Stellar-sized Kerr black holes}

Let us consider a solar-mass Kerr black hole. We can model its optimal processes within the TGO framework. When a fluctuation-driven evaporation occurs,  we can compare the time evolution of the parameters of the black hole to the Hawking evaporation profiles (\ref{eqEJprofsKerH}) and (\ref{eqSHKerrProf}).  The resulting behavior is shown in Figure \ref{fignEnJ}. The case of an optimal accretion-driven process is depicted in Fig. \ref{figKerrposEJ}. 

To quantify the evolution of a parameter $X$, we introduce the index of relative change. For instance, if $X=E$ is the energy of the black hole, the index of relative change during the process is defined by:
\begin{equation}
     \delta E = \frac{E_\tau - E_0}{E_0}.
\end{equation}
It represents the fraction of the black hole’s initial energy that has been accumulated, converted or lost during the process. This parameter does not  represent the amount of energy that can be extracted from the black hole, since our analysis does not account for the specific nature of energy conversion. To explore this effect further, one could examine black holes operating as heat engines (see for instance \cite{Johnson:2014yja, Setare:2015yra, Johnson:2015fva, Johnson:2016pfa, Chakraborty:2016ssb, Hennigar:2017apu, Chakraborty:2017weq, Johnson:2018amj, Johnson:2019olt, DiMarco:2022yhp} and references therein), which  is an interesting study on its own. The relative index of energy change is negative when the black hole is losing energy and positive when it is gaining energy.

\subsubsection{Evaporation-driven profiles}\label{secEDPERep}

Thermodynamic fluctuations on the event horizon of the black hole could trigger a process of an evaporation where the black hole is forced to lose energy or decrease its spin. Although the TGO approach does not tell us how the black hole loses energy in such process, it gives us an estimate of the how much energy is lost or transformed in the transition from one state to another. The probability of such process is governed by the value of the thermodynamic length between the initial and the final state. Longer paths are less likely to occur than shorter paths. In what follows, we show that the thermodynamic length and the relaxation time depend on the initial rates of change, which trigger the fluctuations. Once started, the optimal paths of evolution are  determined by the geodesic profiles of the relevant parameters. In other words, the black hole relaxes along the path of least resistance.

\begin{figure}[H]
	\centering
	\begin{subfigure}{0.43\textwidth}
		\includegraphics[width=1.0\linewidth]{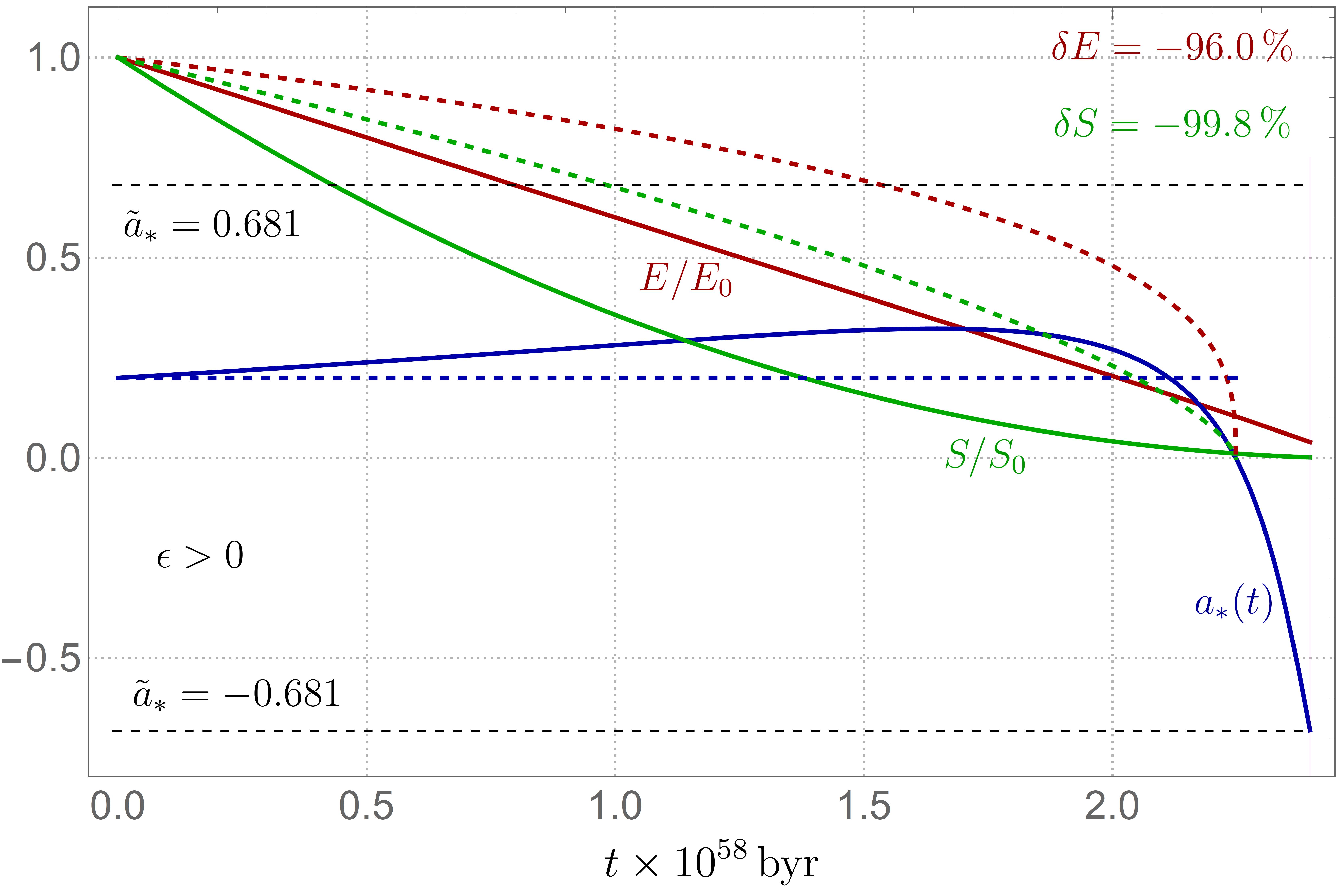}
		\caption{\scriptsize{$\dot E_0=2.7\, \dot E_{H,0}$, $\dot J_0=1.5 \,\dot J_{H,0}$, $\dot S_0=2.72 \,\dot S_{H,0}$.}\vspace{10pt}}\label{fig_a=0.2a}
	\end{subfigure}
	\hspace{0.8 cm}
	\begin{subfigure}{0.43\textwidth}\includegraphics[width=1.0\linewidth]{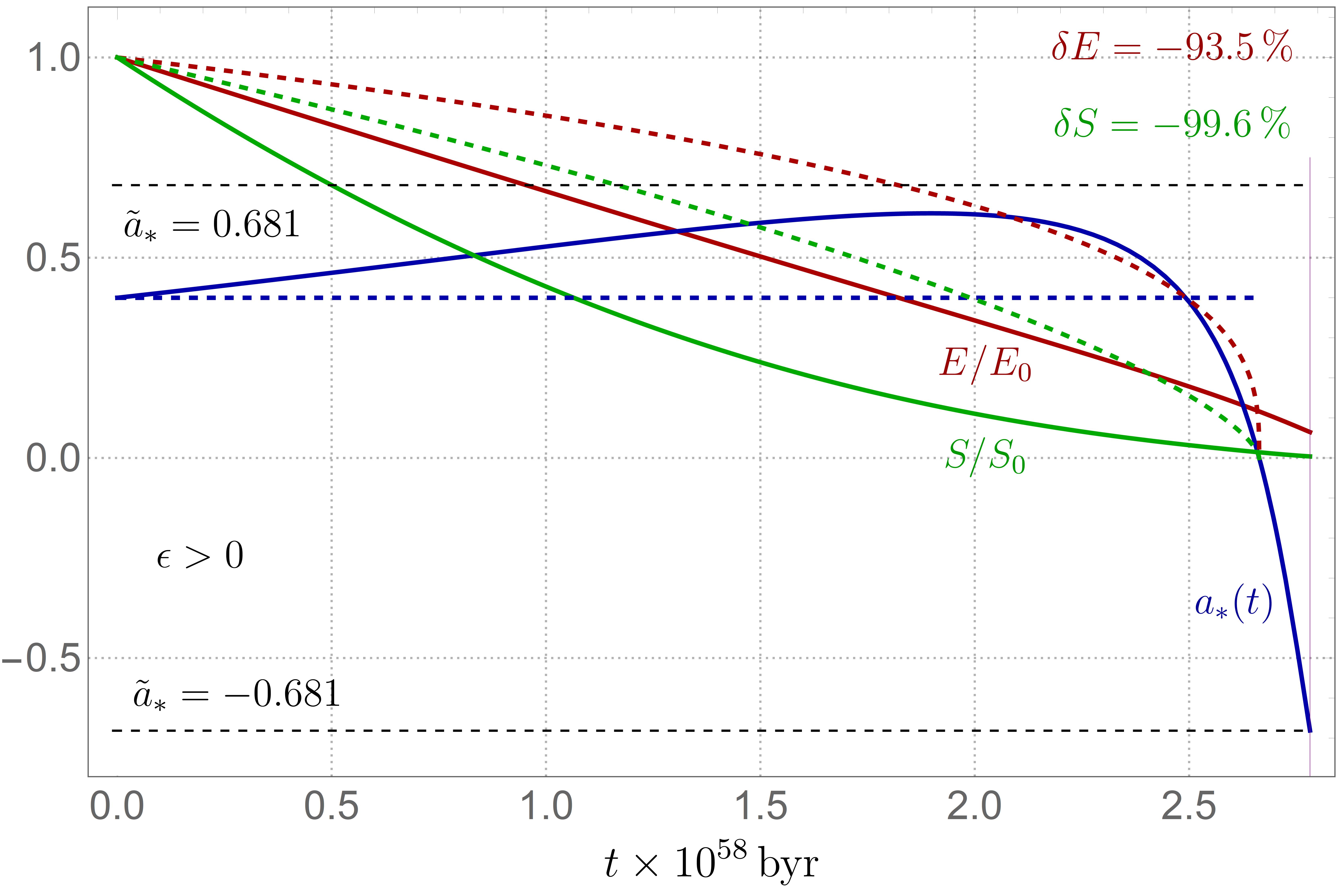}
		\caption{\scriptsize{$\dot E_0=2.7\, \dot E_{H,0}$, $\dot J_0=1.51 \dot J_{H,0}$, $\dot S_0=2.81 \dot S_{H,0}$.}\vspace{10pt}}\label{fig_a=0.4a}
	\end{subfigure}
	\hspace{0.8 cm}
	\begin{subfigure}{0.43\textwidth}\includegraphics[width=1.0\linewidth]{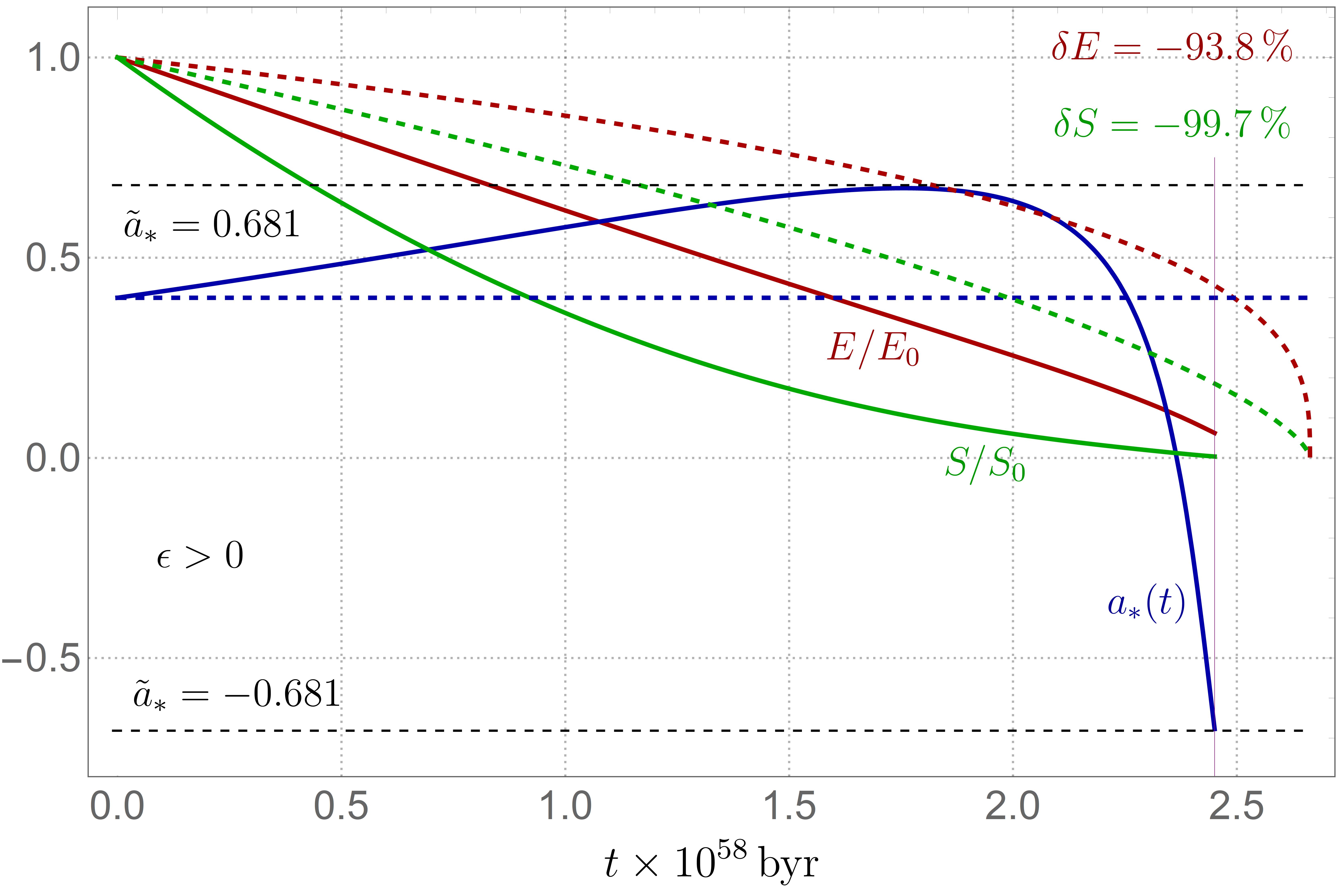}
		\caption{\scriptsize{$\dot E_0=3.1 \,\dot E_{H,0}$, $\dot J_0=1.51\, \dot J_{H,0}$, $\dot S_0=3.24 \dot S_{H,0}$.}\vspace{10pt}}\label{fig_a=0.4b}
	\end{subfigure}
	\hspace{0.8 cm}
		\begin{subfigure}{0.43\textwidth}\includegraphics[width=1.0\linewidth]{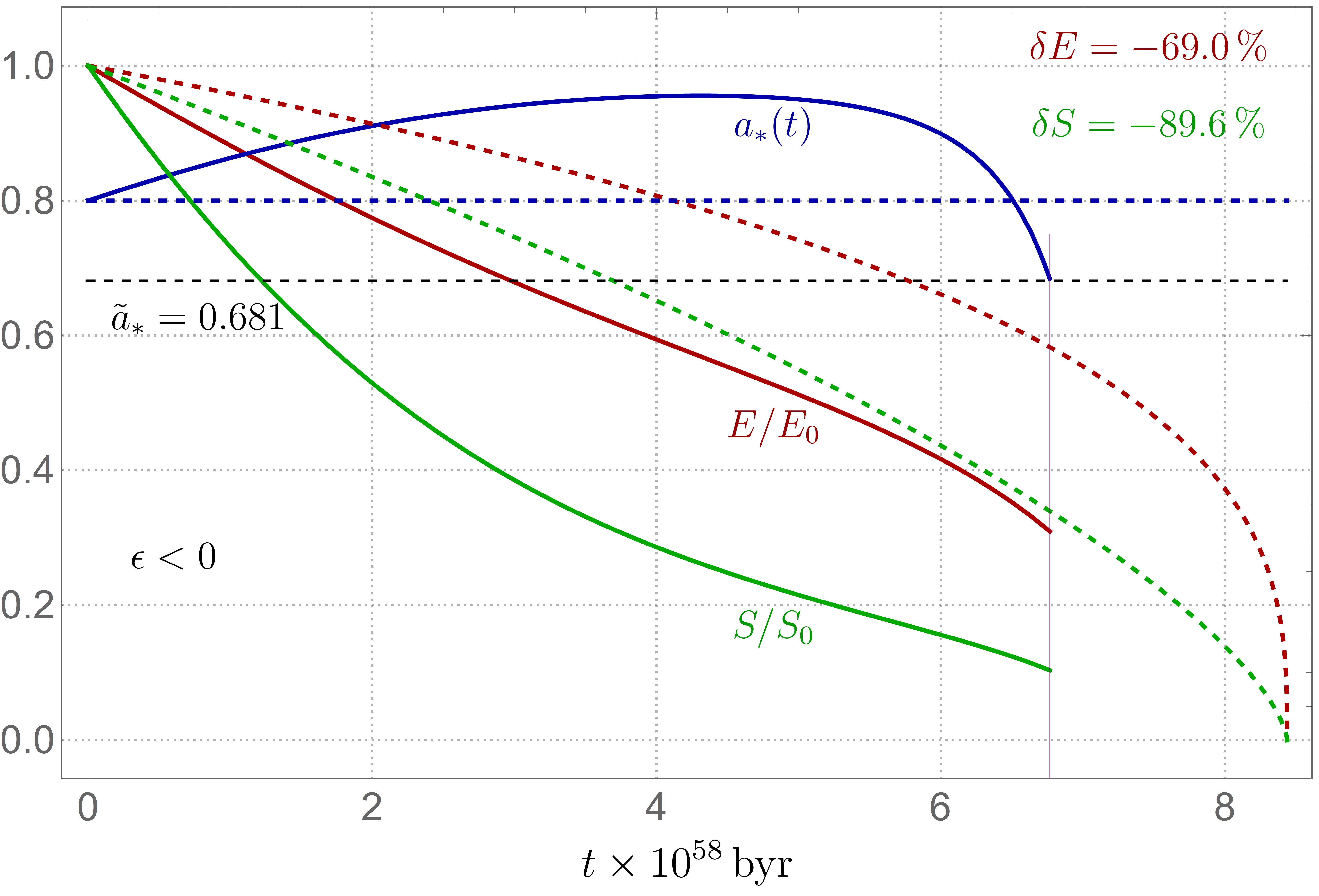}
		\caption{\scriptsize{$\dot E_0=3.1 \,\dot E_{H,0}$, $\dot J_0=2.0\, \dot J_{H,0}$, $\dot S_0=3.83 \,\dot S_{H,0}$.}\vspace{10pt}}\label{fig_a=0.8a}
	\end{subfigure}
\\	\hspace{0.296 cm}
		\begin{subfigure}{0.43\textwidth}\includegraphics[width=1.0\linewidth]{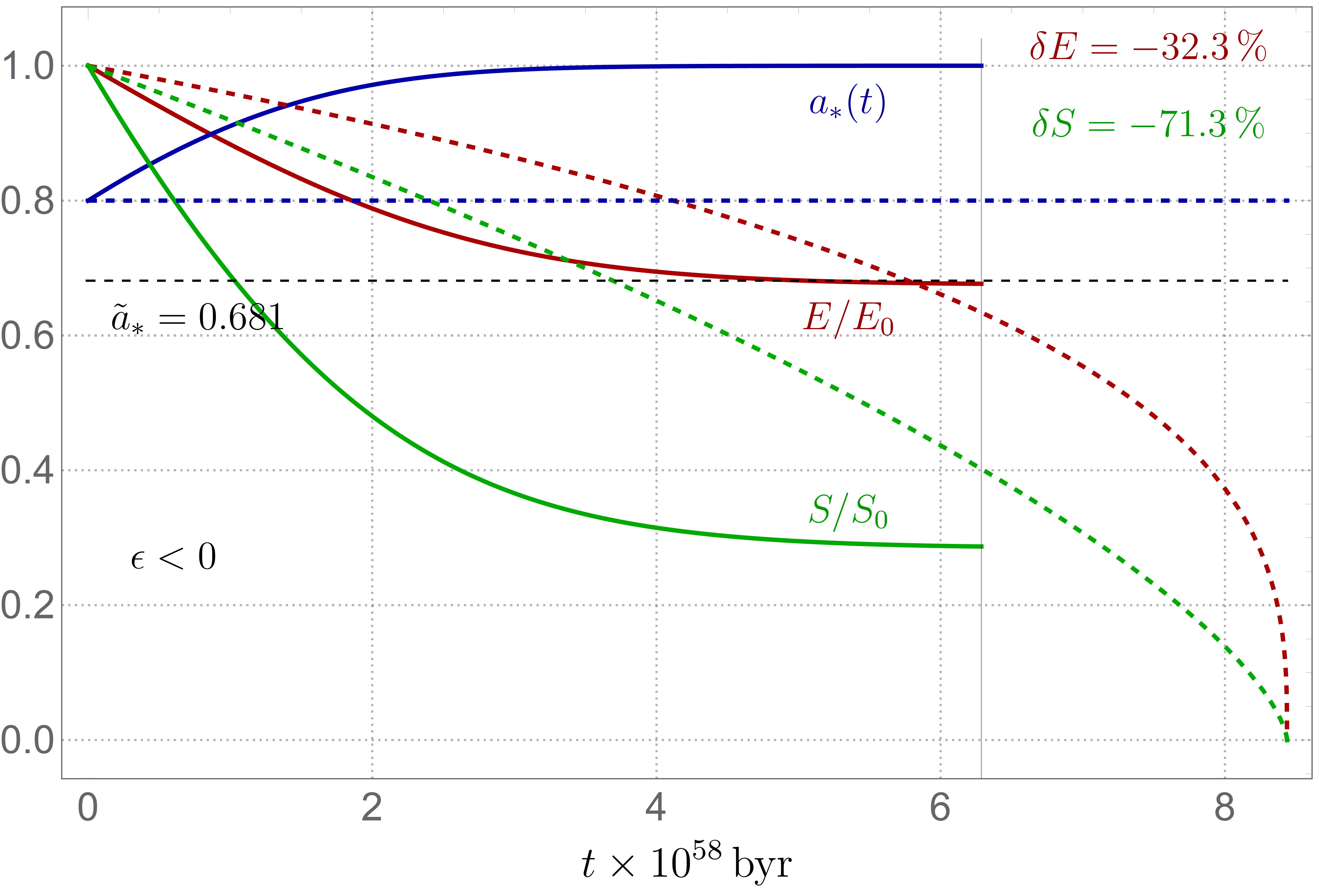}
		\caption{\scriptsize{$\dot E_0=3.1 \,\dot E_{H,0}$, $\dot J_0=1.0 \,\dot J_{H,0}$, $\dot S_0=4.5 \,\dot S_{H,0}$.}\vspace{10pt}}\label{fig_a=0.8b}
	\end{subfigure}
	\hspace{0.8 cm}
		\begin{subfigure}{0.43\textwidth}\includegraphics[width=1.0\linewidth]{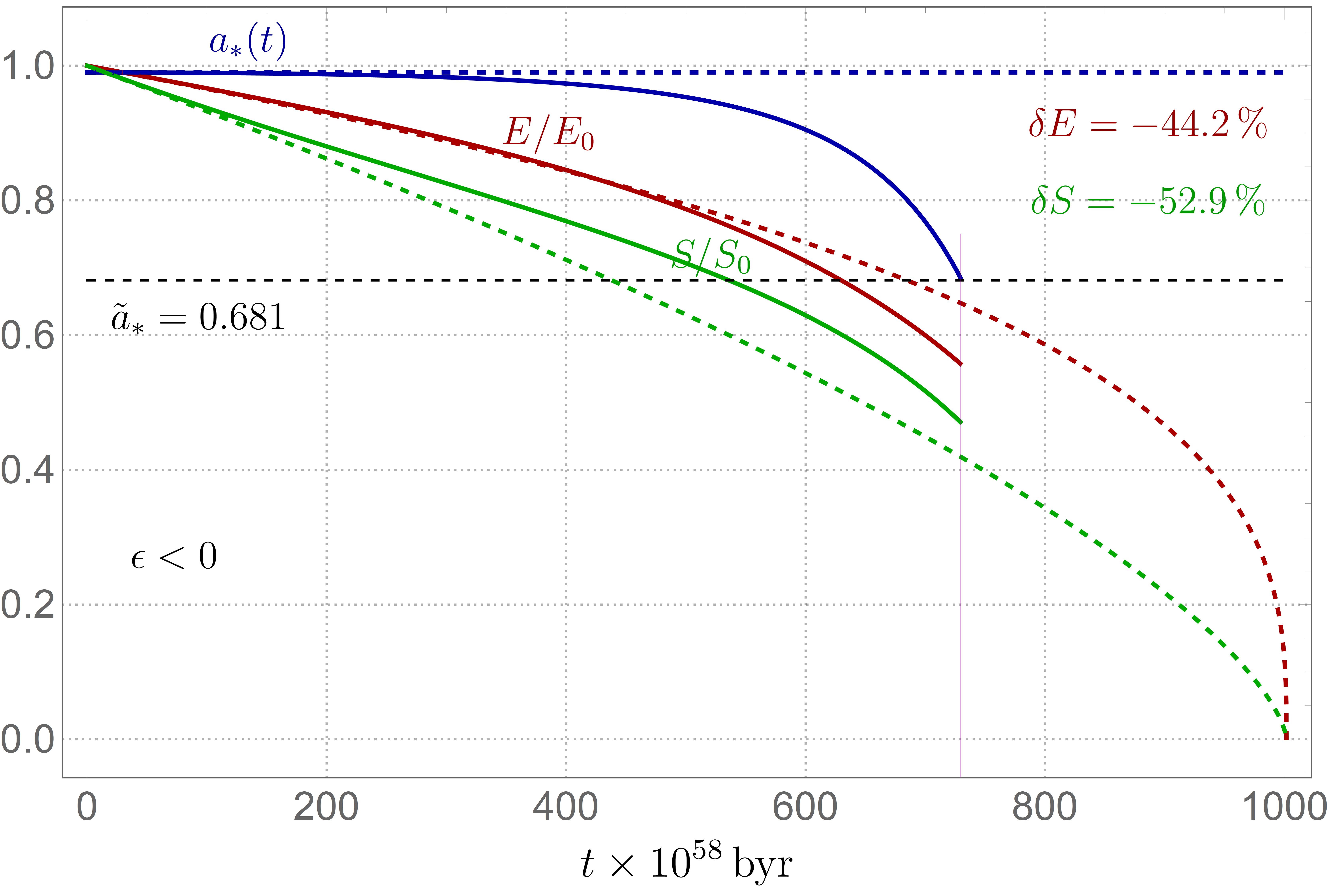}
		\caption{\scriptsize{$\dot E_0=1.0 \,\dot E_{H,0}$, $\dot J_0=1.0 \,\dot J_{H,0}$, $\dot S_0=1.0\, \dot S_{H,0}$.}\vspace{10pt}}\label{fig_a=0.99}
	\end{subfigure}
	\hspace{0.8 cm}
	\caption{The optimal fluctuation-driven evaporation profiles of energy (solid red), entropy (solid green), and angular momentum (solid blue) for a solar-mass Kerr black hole in entropy representation. The corresponding Hawking evaporation profiles are indicated by dashed curves, with the Hawking evaporation rates all negative and evaluated according to (\ref{eqHERESJ}). Their values for $a_*=0.2$, $a_*=0.4$, $a_*=0.8$, and $a_*=0.99$ are listed in Appendix \ref{appfigKerr0208}. The time scale is measured in $10^{58}$ billions of years. The information geometry is elliptic ($\epsilon > 0$) for initial spins below the Davies point ($\tilde a_* = 0.681$) and hyperbolic ($\epsilon < 0$) for initial spins above this threshold. The thin purple vertical line denotes the end time of the numerical solutions to the thermodynamic geodesic equations.}\label{fignEnJ}
\end{figure}

Figure \ref{fig_a=0.2a} illustrates the optimal evaporation profiles for the energy (solid red), entropy (solid green), and angular momentum (solid blue) of a solar mass Kerr black hole with an initial spin $a_{*0} = 0.2$. The initial rates of change are chosen so that the black hole's spin reaches zero precisely at the Hawking evaporation time\footnote{The initial rate of change of angular momentum is negative, $\dot{J}_0 = 1.5 \, \dot{J}_{H,0} < 0$. However, the spin $a_*(t)$ increases to a certain maximum during the process, before decreasing toward the negative Davies point. This behavior is consistent across other profiles.}. For comparison, the Hawking evaporation profiles are shown with corresponding colored dashed curves. During this process, the black hole loses approximately $96\%$ of its initial energy and $99\%$ of its initial entropy, under the assumption that Hawking radiation effects are neglected during fluctuations. Notably, the  spin profile, $a_*(t)$, passes through the Schwarzschild state, allowing the black hole to reverse its spin\footnote{This is allowed, since the $J\to 0$ limit of the Kerr black hole is a regular limit.}. There is no indication in the TGO method that the black hole's evolution should stop. The spin continues to evolve until it reaches the negative Davies curve at $\tilde a_*=-0.681$, where the thermodynamic length becomes complex, preventing further fluctuations. At this stage, the black hole undergoes a second-order phase transition, marking the end of the optimal fluctuation-driven evolution.

Figure \ref{fig_a=0.4a} depicts an optimal evaporation with an initial spin $ a_{*0} = 0.4 $. The evolution of the fluctuations qualitatively resembles that shown in Figure \ref{fig_a=0.2a}. All fluctuations effectively ceasing after the spin reaches the negative Davies point, where the thermodynamic length becomes complex.

Figure \ref{fig_a=0.4b} depicts a scenario starting at $ a_{*0} = 0.4 $, The chosen initial rates lead the spin of the black hole to increase its value close to the positive Davies curve and then  decrease towards zero. The result leads to the vanishing of the spin occurring earlier than the Hawking evaporation time. All fluctuations cease when the spin reaches the negative Davies point, where the thermodynamic length of the process becomes complex.

Figure \ref{fig_a=0.8a} illustrates the situation starting at $ a_{*0} = 0.8 $, which is above the Davies curve. Here, the spin increases its value close to the extremal limit before starting to decrease as it approaches the Davies curve from above. At that point the thermodynamic length becomes complex and effectively terminates all further fluctuations. Notably, the entire process occurs for $\epsilon < 0$, which contrasts with the previous cases where the initial spin parameter was below the critical value of $\tilde{a}_* = 0.681$. 

In Fig. \ref{fig_a=0.8b} the fluctuations drive the spin  to increase from $a_{*0}=0.8$  toward the extremal value $a_*=1$. At some point the black hole becomes extremal and all optimal fluctuations stop.  

In Fig, \ref{fig_a=0.99} the spin starts from a near-extremal Kerr black hole initially at $ a_* = 0.99 $, which relaxes at the Davies phase transition at $\tilde{a}_* = 0.681$. Once again, all fluctuations cease after that point, since the thermodynamic length becomes a complex number.

The change in the information geometry of the state space of a solar-mass Kerr black hole, as indicated by the complex values of the thermodynamic length at the Davies critical point $\tilde{a}_* = 0.681$, or by the sign of the metric scale factor\footnote{Depending on whether the initial spin is above or below the critical Davies point.} $\epsilon$ -- demonstrate that our thermogeometric approach effectively identifies critical points and provides meaningful insights into the nature of the Davies phase transition in the entropy representation. Notably, while this effect is present in entropy representation, it does not occur in  energy representation, as discussed in Section \ref{secOPErep}. This discrepancy arises since in energy representation the thermodynamic curvature for the Kerr black hole is zero, preventing any association of the metric scale factor $\epsilon$ with the information geometry in this context.

\subsubsection{Accretion-driven profiles}

If there is an external inflow of matter or energy into the black hole, two physical types of accretion-driven optimal processes can be considered: one involving an increase in mass and spin, and another involving a decrease in the spin of the black hole.  It is important to note that these optimal profiles are not mere fluctuations but rather driven processes resulting from interactions with the surrounding environment. 

\begin{figure}[H]
	\centering
	\begin{subfigure}{0.45\textwidth}
		\includegraphics[width=1.0\linewidth]{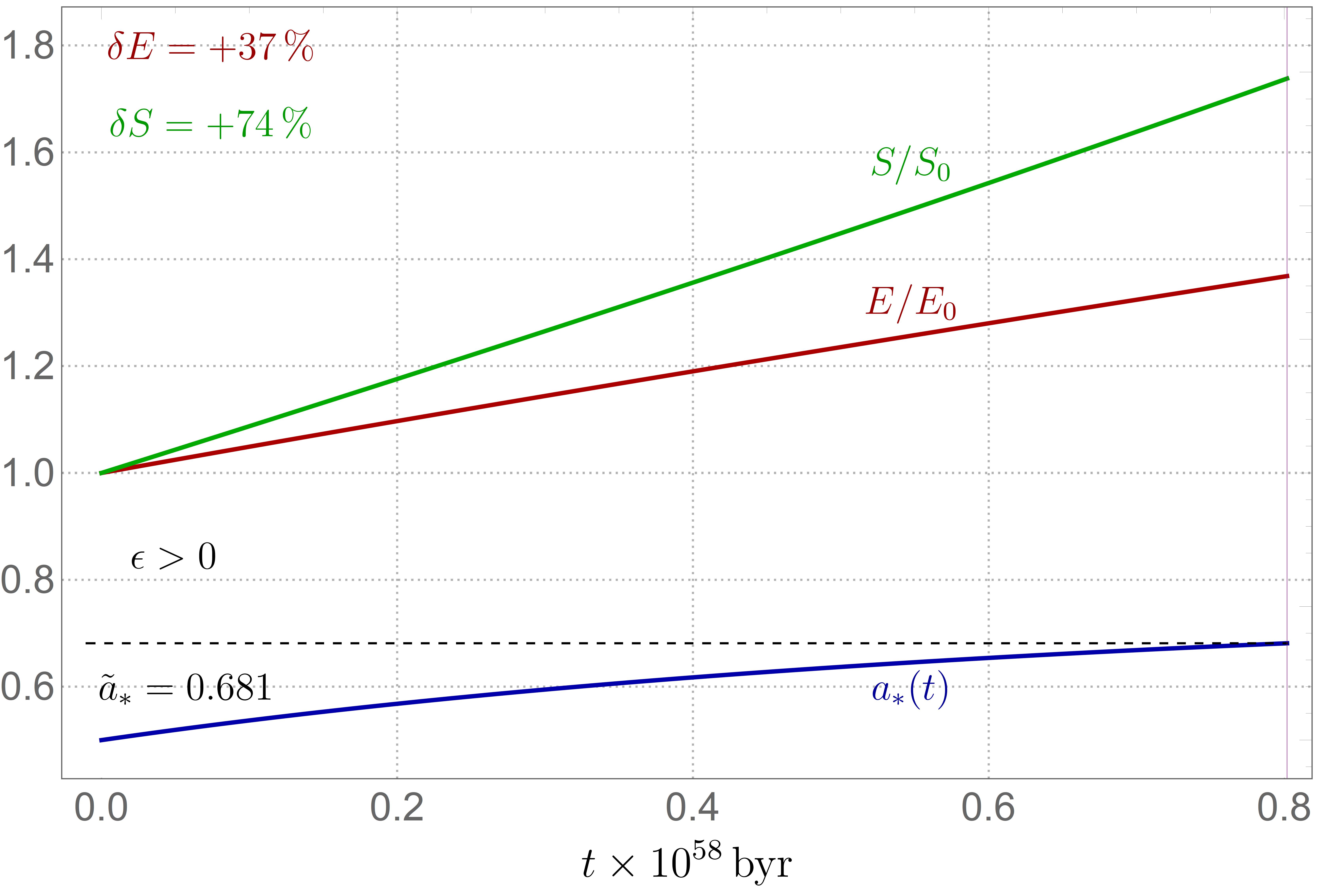}
		\caption{Initial spin at $a_{*0}=0.5$.}\label{fig4a}
	\end{subfigure}
	\hspace{0.8 cm}
	\begin{subfigure}{0.45\textwidth}\includegraphics[width=1.0\linewidth]{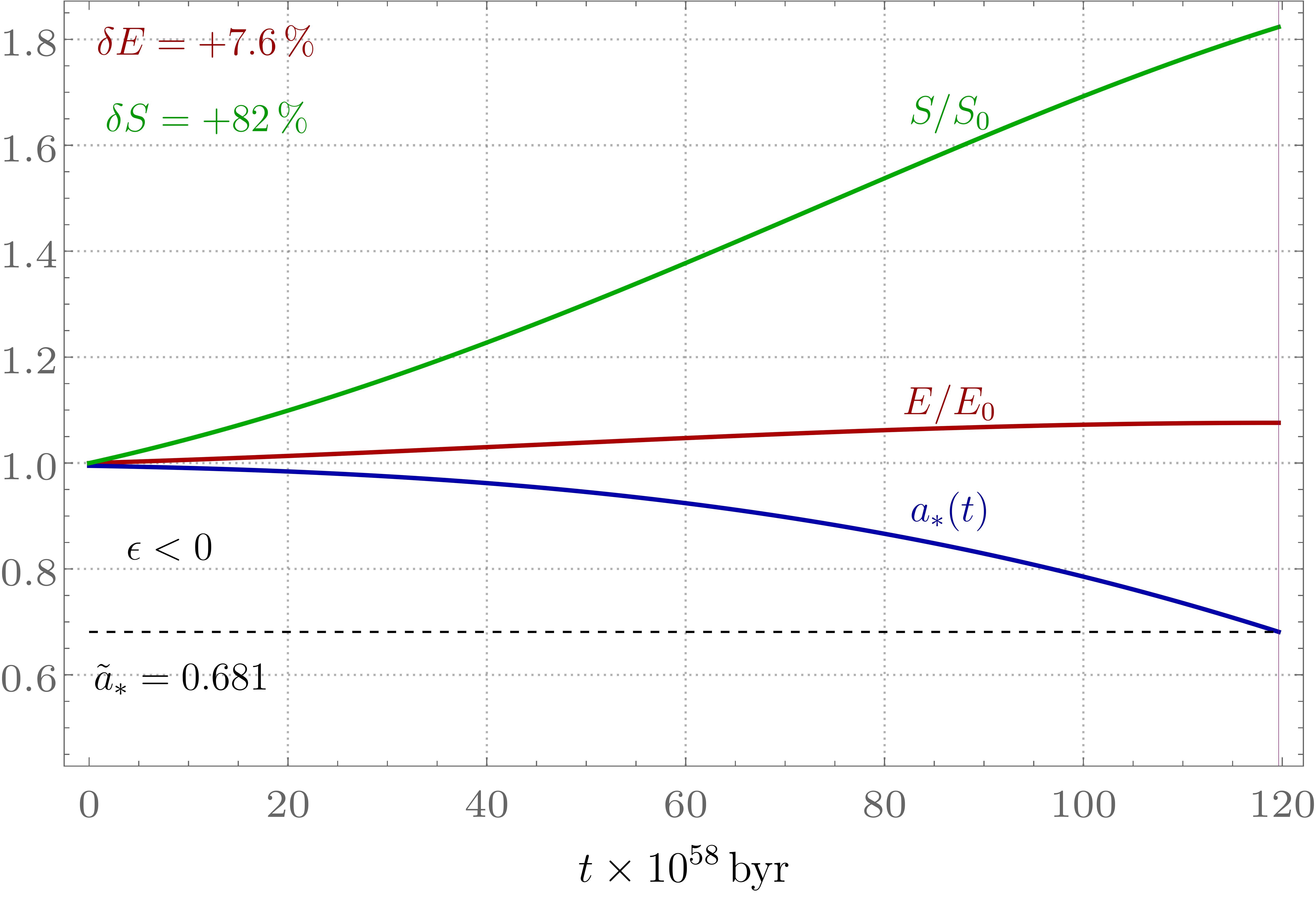}
		\caption{Initial spin at $a_{*0}=0.995$.}\label{figHevapKS4}
	\end{subfigure}
	\hspace{1.5 cm}
	\caption{ The data for these profiles is: \textbf{(a)} $\dot E_0=-4.6 \,\dot E_{H,0}$, $\dot J_0=-8.3 \,\dot J_{H,0}$, $\dot S_0=-4.0 \,\dot S_{H,0}$; \textbf{(b)} $\dot E_0=-6.1 \,\dot E_{H,0}$, $\dot J_0=-4.3 \,\dot J_{H,0}$, $\dot S_0=-22.3 \,\dot S_{H,0}$. The Hawking evaporation rates are negative and evaluated according to (\ref{eqHERESJ}). Their values for $a_*=0.5$ and $a_*=0.995$ are listed in Appendix \ref{appfigKerr0208}.}\label{figKerrposEJ}
\end{figure}

In Figure \ref{figKerrposEJ}, we analyze optimal matter accretion processes with small positive rates of energy and angular momentum. Over time, the spin of the black hole reaches the Davies curve from below or from above.

Figure \ref{fig4a} illustrates an optimal accretion process for a Kerr black hole initially spinning at $a_{*0}=0.5$. As the angular momentum increases, the spin gradually rises until it reaches the Davies critical value, $\tilde{a}_* = 0.681$. At this point, the optimal process concludes\footnote{This does not imply the end of accretion itself but signifies that the black hole may no longer accumulate matter optimally.} due to a phase transition experienced by the black hole. It is notable, that at the critical point the optimal process is terminated by an event location imposed by hand in our numerical procedure. This is due to the fact that the thermodynamic length continued to be real valued through this point and beyond and it failed to detect it.  Nevertheless, the sign of the metric scale $\epsilon$ still changes below or above this curve. 

Figure \ref{figHevapKS4} presents another optimal accretion process, beginning with a positive angular momentum rate of change, $\dot{J}_0 = -4.3\,\dot{J}_{H,0} > 0$, and initial spin at $a_{*0}=0.995$. In this scenario, the spin profile, $a_*(t)$, decreases toward the Davies point, where the thermodynamic length becomes complex and the process terminates.

\subsection{Optimal processes in energy representation}\label{secOPErep}

Although the analysis in energy representation mirrors the one in entropy representation, there are notable differences. Most significantly, the thermodynamic curvature is zero, meaning that the Davies phase transition points cannot be interpreted as a change in the type of information geometry. Consequently, the sign of the metric scale  does not relate to the type of information geometry, but merely ensures that the thermodynamic length remains positive along the optimal process. 

Another key difference is that, in entropy representation, complete evaporation of the black hole has not been found prior to the spin reaching the Davies points. In contrast, in energy representation, we obtain full optimal evaporation profiles of the Kerr black hole.

\subsubsection{Geodesic equations on the space of states}

The explicit form of the Weinhold metric for Kerr in energy representation is given by the Hessian of the energy with respect to the entropy an the angular momentum of the black hole:
\begin{align}
\hat g^{(W)}=\epsilon \hat H_E=\epsilon \!\left(\!\!
\begin{array}{cc}
 \frac{\partial^2 E}{\partial S^2}\big|_J \!\!& \!\! \frac{\partial^2 E}{\partial S \partial J}\\[5pt]
 \frac{\partial^2 E}{\partial S \partial J} \!\!& \!\! \frac{\partial^2 E}{\partial J^2}\big|_S\\
\end{array}
\!\!\right)
=\epsilon\! \left(\!\!
\begin{array}{cc}
 \frac{3 \zeta ^2 \lambda ^2 J^2\left(\zeta ^2 \lambda ^2J^2 +8 S^2\right)-16 S^4}{8 \sqrt{\lambda } S^{5/2} \left(\zeta ^2 \lambda ^2 J^2+4 S^2\right)^{3/2}} \!& \!-\frac{\zeta ^2 \lambda ^{3/2} J\left(\zeta ^2  \lambda ^2 J^2+12 S^2\right)}{4 S^{3/2}\left(\zeta ^2 \lambda ^2 J^2 +4 S^2\right)^{3/2}} \\[5pt]
  -\frac{\zeta ^2 \lambda ^{3/2} J\left(\zeta ^2  \lambda ^2 J^2+12 S^2\right)}{4 S^{3/2}\left(\zeta ^2 \lambda ^2 J^2 +4 S^2\right)^{3/2}}\!& \! \frac{2 \zeta ^2 \lambda ^{3/2} S^{3/2}}{\left(\zeta ^2 \lambda ^2 J^2 +4 S^2\right)^{3/2}} \\
\end{array}
\!\!\right)\!.
\end{align}
Its thermodynamic curvature is zero, suggesting that the system is  governed purely by random thermal fluctuations with no attractive or repulsive forces between the bits on the horizon (flat thermodynamic geometry)\footnote{See for example \cite{Mahmoudi:2023uxr,Quevedo:2007mj} and references therein.}. The corresponding system of thermodynamic geodesic equations in $(S,J)$ space is highly nonlinear and is presented in Appendix \ref{appKTDGEER}. Its numerical investigation is shown on Figures \ref{fignSnJ} and \ref{figKerrposSJ} for a solar-mass Kerr black hole with initial energy $E_{0}=E_\odot = 1.8 \times 10^{47} \,\text{J}$ and different initial spins.

\subsubsection{Evaporation-driven profiles}

\begin{figure}[H]
	\centering
	\begin{subfigure}{0.45\textwidth}
		\includegraphics[width=1.0\linewidth]{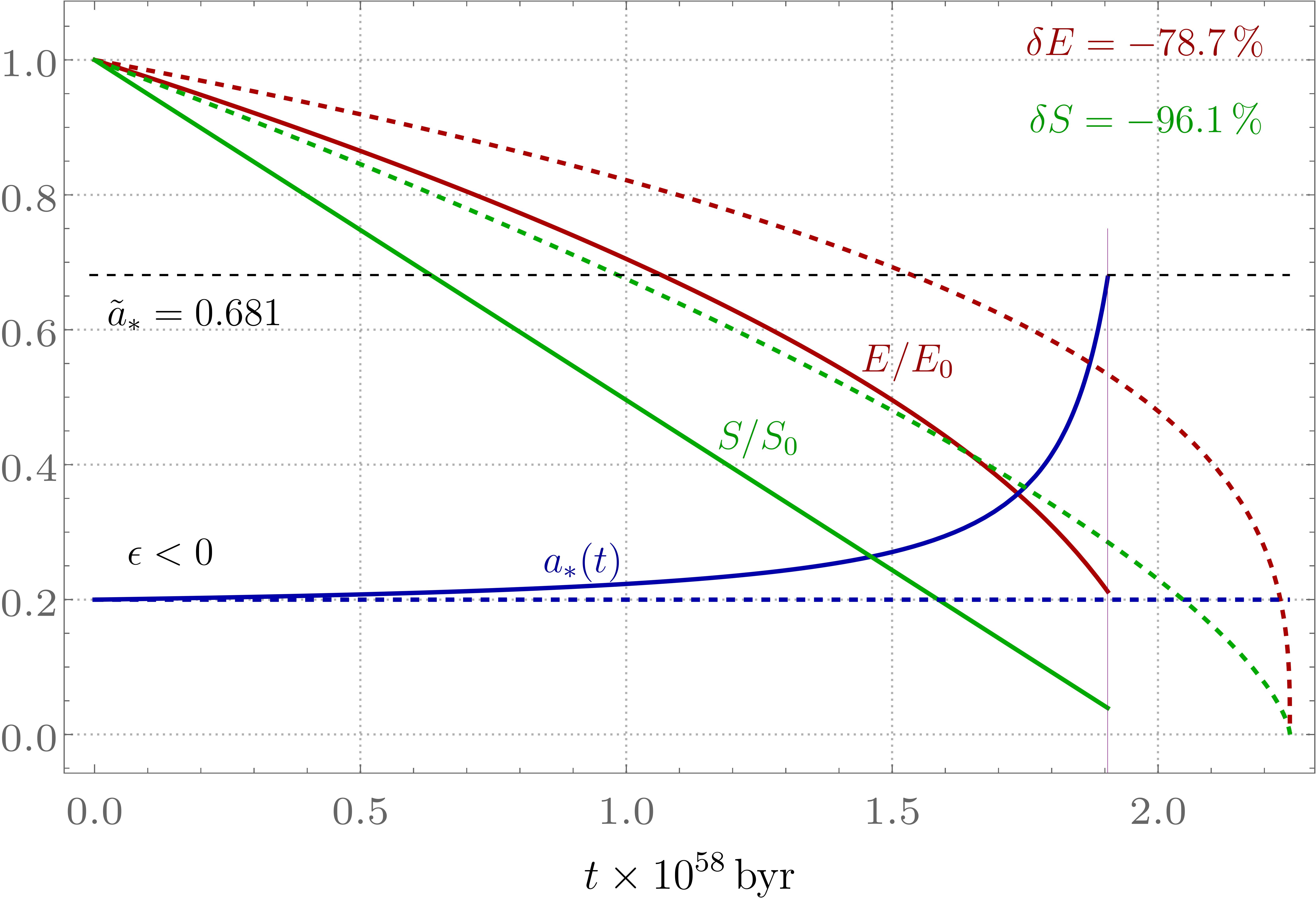}
		\caption{\scriptsize{$\dot E_0=1.7 \,\dot E_{H,0}$, $\dot J_0=1.5\, \dot J_{H,0}$, $\dot S_0=1.7 \,\dot S_{H,0}$.}\vspace{10pt}}\label{KerrSJ_02_a}
	\end{subfigure}
	\hspace{0.8 cm}
	\begin{subfigure}{0.45\textwidth}
		\includegraphics[width=1.0\linewidth]{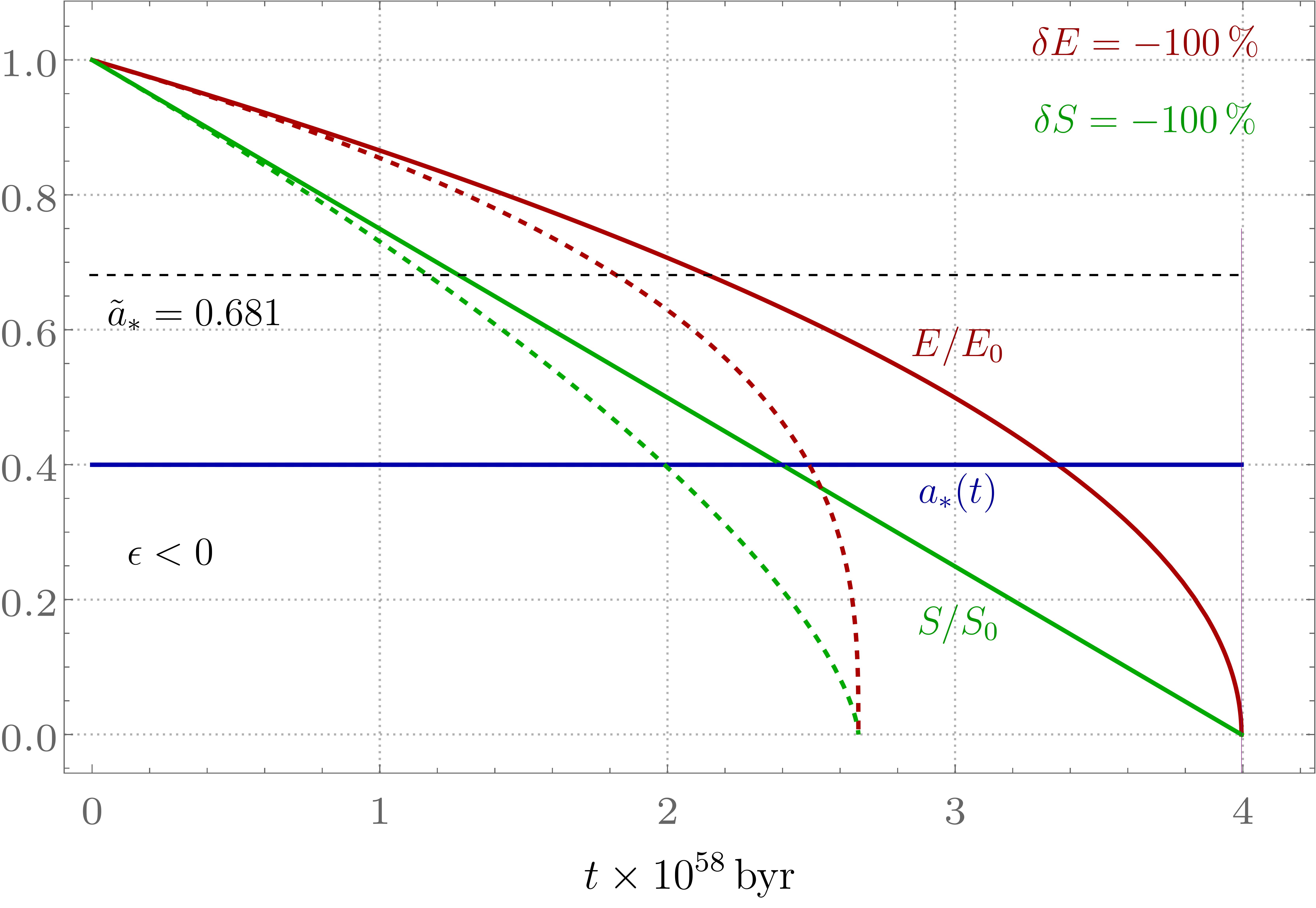}
		\caption{\scriptsize{$\dot E_0=1.0 \,\dot E_{H,0}$, $\dot J_0=1.0\, \dot J_{H,0}$, $\dot S_0=1.0 \,\dot S_{H,0}$.}\vspace{10pt}}\label{KerrSJ_04_a}
	\end{subfigure}
	\hspace{0.8 cm}
	\begin{subfigure}{0.45\textwidth}
		\includegraphics[width=1.0\linewidth]{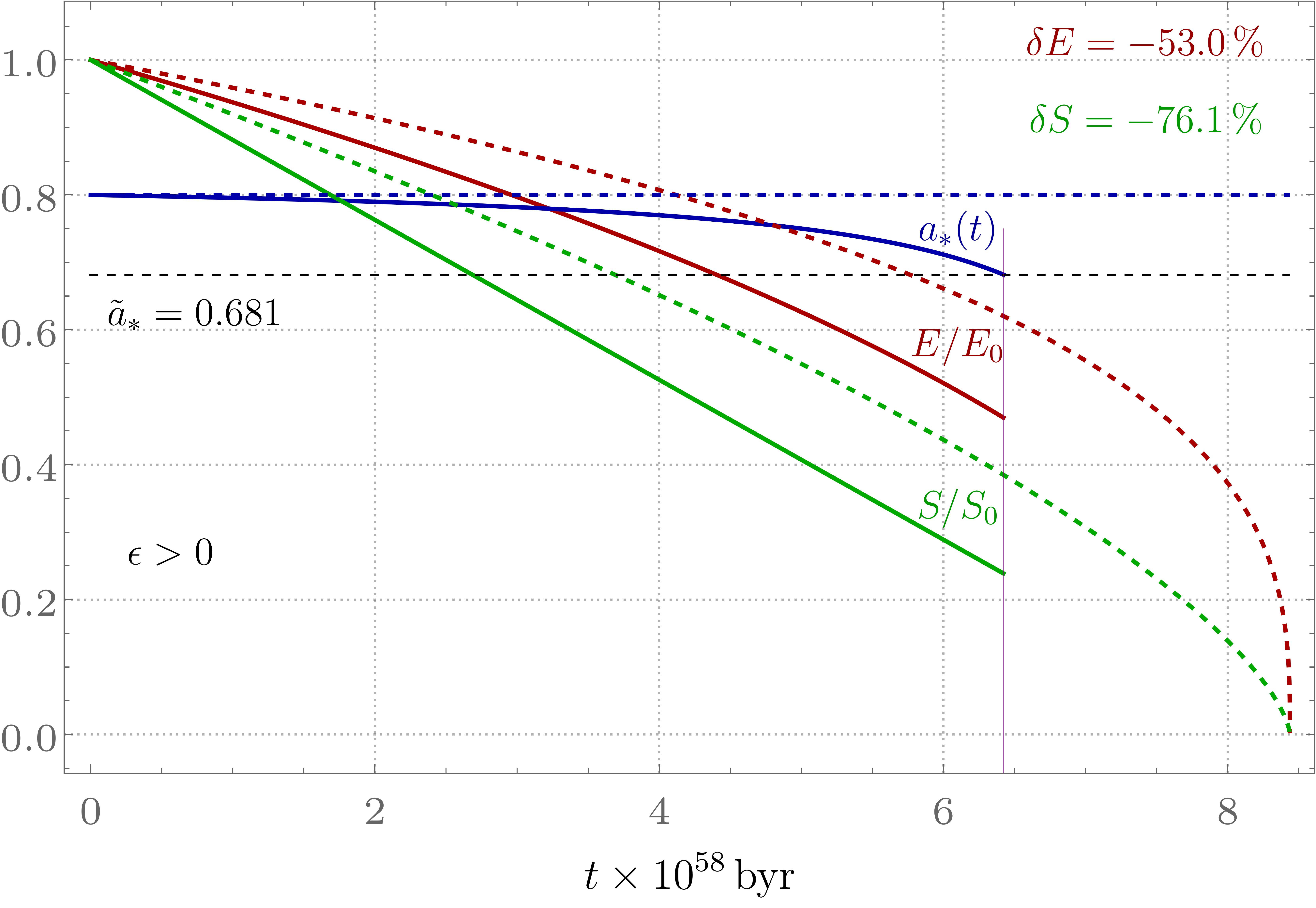}
		\caption{\scriptsize{$\dot E_0=1.54 \,\dot E_{H,0}$, $\dot J_0=1.6\, \dot J_{H,0}$, $\dot S_0=1.5 \,\dot S_{H,0}$.}\vspace{10pt}}\label{KerrSJ_08_a}
	\end{subfigure}
	\hspace{0.8 cm}
	\begin{subfigure}{0.45\textwidth}\includegraphics[width=1.0\linewidth]{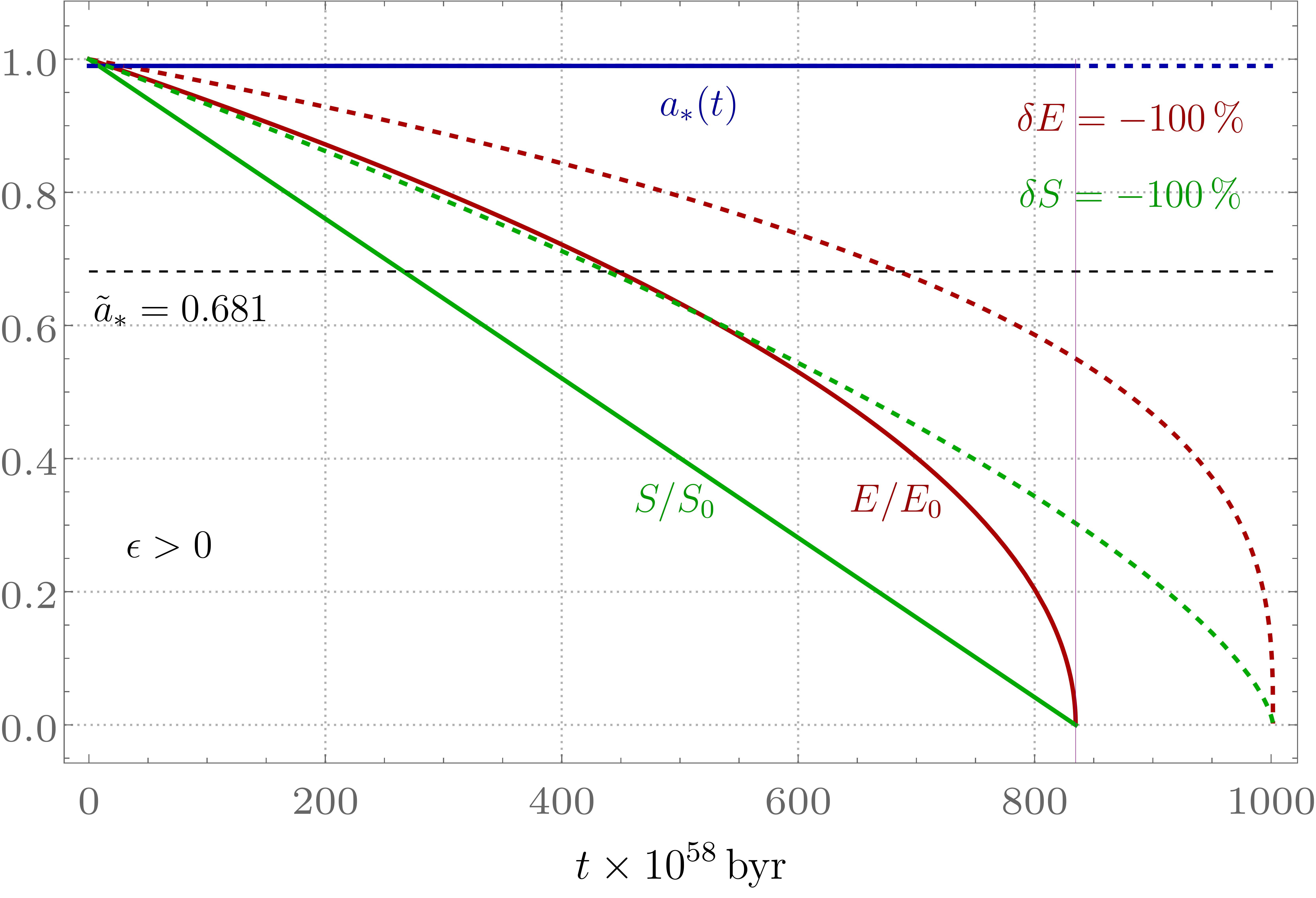}
		\caption{\scriptsize{$\dot E_0=1.8 \,\dot E_{H,0}$, $\dot J_0=1.8\, \dot J_{H,0}$, $\dot S_0=1.8 \,\dot S_{H,0}$.}\vspace{10pt}}\label{KerrSJ_099_a}
	\end{subfigure}
	\hspace{1.5 cm}
	\caption{The optimal fluctuation profiles of energy (solid red curves), entropy (solid green curves), and angular momentum (solid blue curves) for a solar-mass Kerr black hole are displayed in energy representation. The corresponding Hawking evaporation profiles are indicated by dashed curves, with the Hawking evaporation rates all negative and evaluated according to (\ref{eqHERESJ}). Their values for $a_*=0.2$, $a_*=0.4$, $a_*=0.8$, and $a_*=0.99$ are listed in Appendix \ref{appfigKerr0208}. The time scale is measured in $10^{58}$ billions of years. The information geometry is elliptic ($\epsilon > 0$) for initial spins above the Davies point ($a_* = 0.681$) and hyperbolic ($\epsilon < 0$) for initial spins below this threshold. The thin purple vertical line denotes the end time of the numerical solutions to the thermodynamic geodesic equations.}\label{fignSnJ}
\end{figure}

In Figure \ref{KerrSJ_02_a}, the spin of the black hole starts at $ a_{*0} = 0.2 $. After some time, the spin profile $a_*(t)$ crosses the Davies point at $\tilde a_*=0.681$. At this point the thermodynamic length becomes a complex number and all fluctuations terminate. The relative energy and  entropy changes are $\delta E\approx -79\%$ and $\delta S\approx -96\%$. 

Figure \ref{KerrSJ_04_a} depicts an optimal fluctuation profile starting at $a_{*0} = 0.4$, which ultimately leads to the complete evaporation of the Kerr black hole. The initial rates of change are chosen such that the optimal fluctuation-induced evaporation occurs after the Hawking evaporation time. Notably, throughout this process, the spin of the black hole does not change, until it stops suddenly at the end.

In Figure \ref{KerrSJ_08_a}, we illustrate a scenario starting at $ a_* = 0.8 $. At late times, the spin profile crosses a Davies critical point at $ \tilde{a}_* = 0.681 $. Although we prefer to stop the profiles here on physical grounds, there is no any indication by the TGO procedure that there is a phase transition at that point\footnote{The thermodynamic length stays real and positive passing through the Davies point.}. This reflects the fact that the thermodynamic information geometry in energy representation is flat, and as a result, the thermodynamic length is not expected to detect any defects in the state space, except  the extremal cases.

Finally, Figure \ref{KerrSJ_099_a} depicts a near-extremal case with an initial value of the spin at $ a_* = 0.99 $. Here, the initial rates of change of the parameters are sufficient to cause the Kerr black hole to evaporate earlier than the Hawking evaporation time. The spin of the black hole stays constant during this process, until  it suddenly stops at the end.

In energy representation, we identified optimal fluctuations that can directly lead to the evaporation of the Kerr black hole, which is in contrast to the results from the entropy representation. Additionally, the sign of the metric scale parameter $ \epsilon $ is reversed compared to its sign in the entropy representation when approaching the Davies critical point, hence we can still use it to identify the location of the phase transition. However, in the next subsection, we show that this no longer holds for accretion-driven processes.

\subsubsection{Accretion-driven profiles}

\begin{figure}[H]
	\centering
	\begin{subfigure}{0.45\textwidth}
		\includegraphics[width=1.0\linewidth]{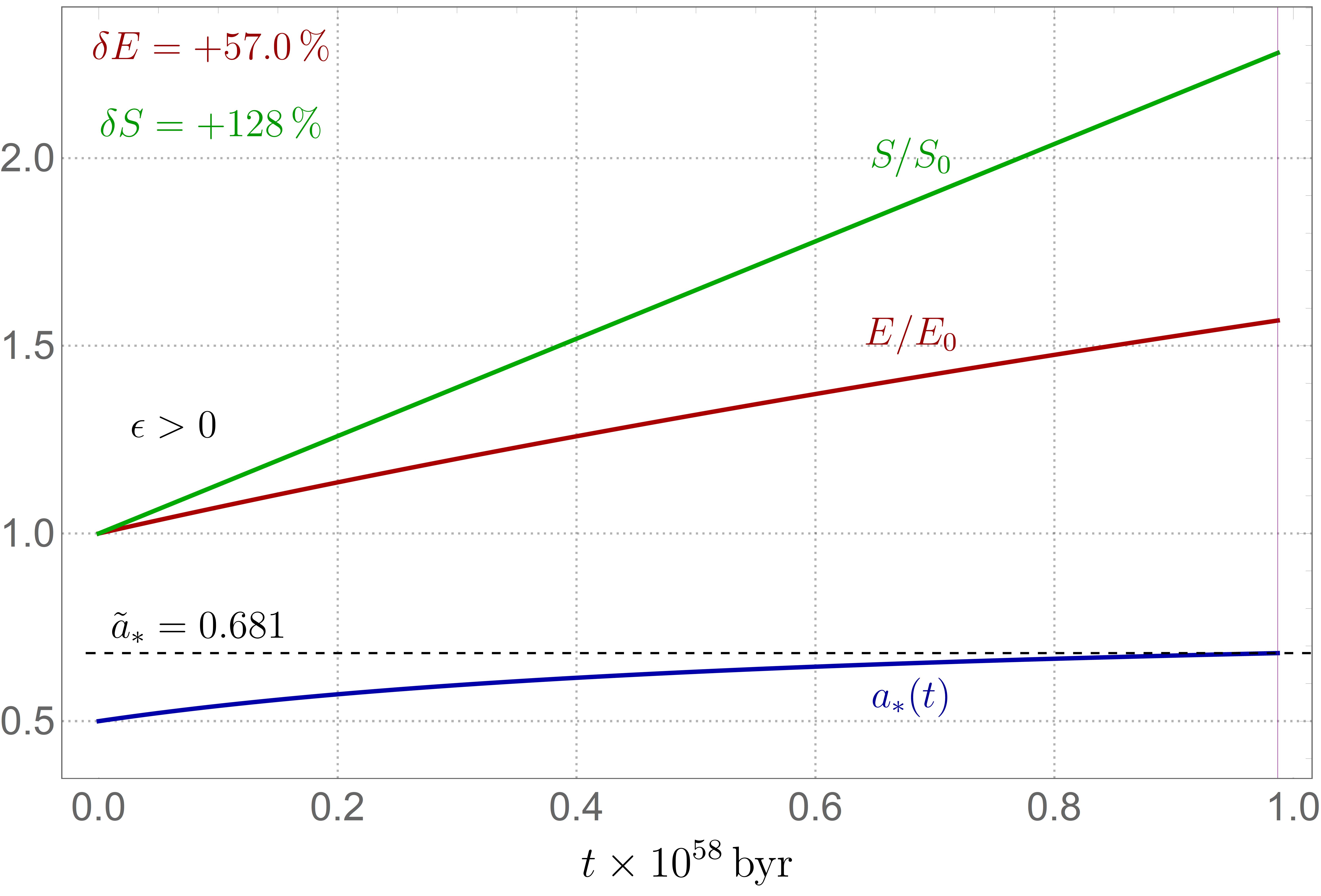}
		\caption{Initial spin at $a_{*0}=0.5$.}\label{fig6a}
	\end{subfigure}
	\hspace{0.8 cm}
	\begin{subfigure}{0.45\textwidth}\includegraphics[width=1.0\linewidth]{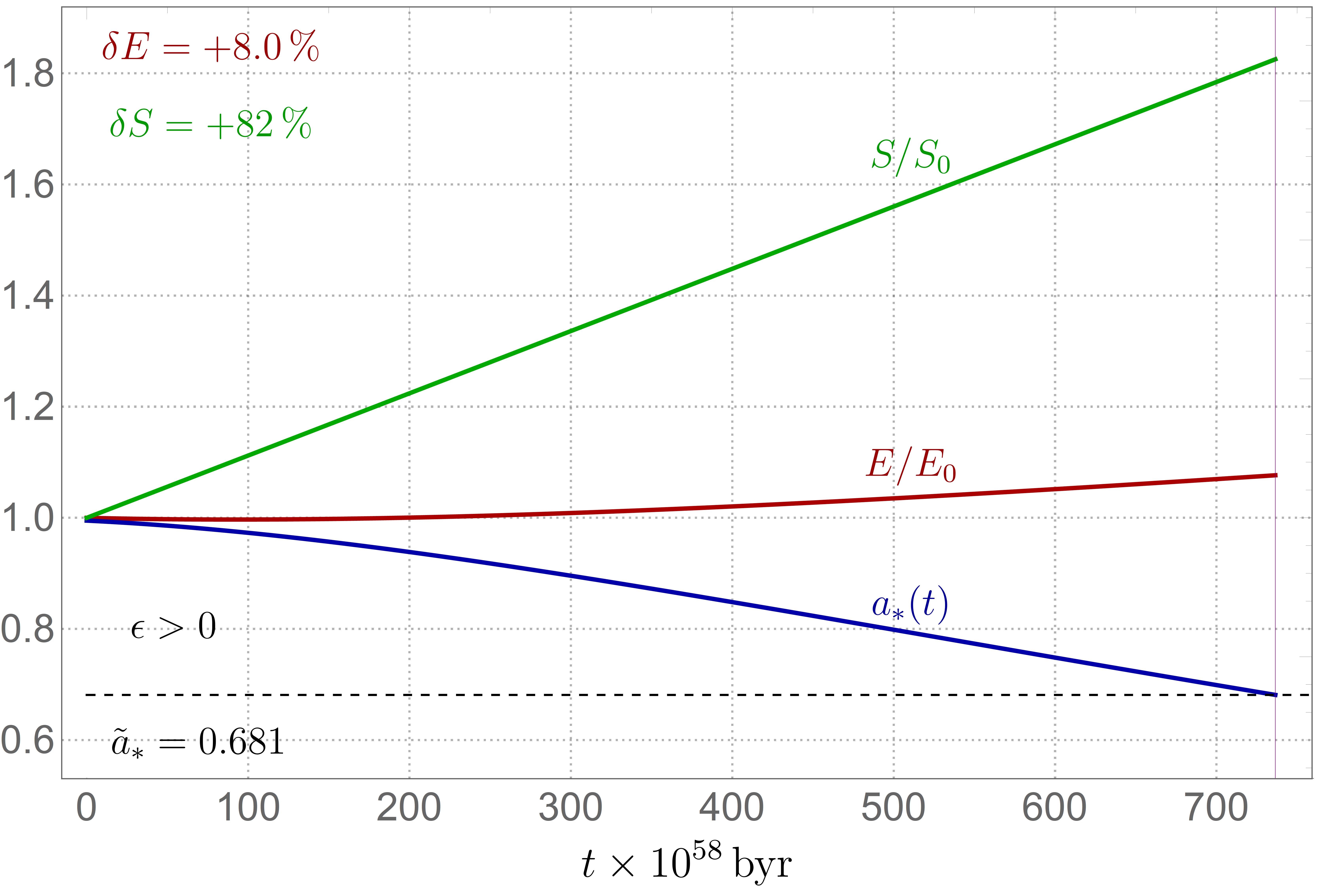}
		\caption{Initial spin at $a_{*0}=0.995$.}\label{fig6b}
	\end{subfigure}
	\hspace{1.5 cm}
	\caption{The data for these profiles is: \textbf{(a)} $\dot E_0=-6.7 \,\dot E_{H,0}$, $\dot J_0=-11 \,\dot J_{H,0}$, $\dot S_0=-6.0 \,\dot S_{H,0}$; \textbf{(b)} $\dot E_0=-0.75 \,\dot E_{H,0}$, $\dot J_0=-1.5 \,\dot J_{H,0}$, $\dot S_0=-6.0 \,\dot S_{H,0}$. The Hawking evaporation rates are negative and evaluated according to (\ref{eqHERESJ}). Their values for $a_*=0.5$ and $a_*=0.995$ are listed in Appendix \ref{appfigKerr0208}.}\label{figKerrposSJ}
\end{figure}

Figure \ref{figKerrposSJ}  shows two optimal accretion processes of matter or energy into the black hole.  
Figure \ref{fig6a} illustrates an optimal accretion process for a Kerr black hole initially spinning at $a_{*0}=0.5$. The spin gradually increases until it reaches the Davies critical value. At this point, the optimal process is terminated by an event location imposed by our numerical procedure\footnote{ Since the thermodynamic length failed to detect it.}.

Figure \ref{fig6b} presents another optimal accretion process, starting with a positive angular momentum rate of change, $\dot{J}_0 = -1.5\,\dot{J}_{H,0} > 0$, and initial spin at $a_{*0}=0.995$. In this scenario, the spin profile $a_*(t)$ falls toward the Davies critical value. At this point the process is terminated by an event location in our numerical procedure.

In both scenarios the thermodynamic length fails to detect the Davies phase transition and the processes have to be forcefully terminated by our numerical procedure. Furthermore, one notes that the metric scale $\epsilon$ also failed to detect the location of the critical points, since it did not change its sign relative to the Davies point. This is expected, since the thermodynamic curvature is zero and  $\epsilon$ cannot be associated with the type of the information geometry on the space of states.

\subsection{Short summary}

In energy representation, we identified optimal fluctuation profiles that can directly lead to the evaporation of the Kerr black hole. The latter is in contrast to the results from the entropy representation. Additionally, the sign of the metric scale $ \epsilon $ managed to identify the location of the Davies phase transition points in both representations, except for accretion-driven processes in energy representation. This outcome is anticipated, given that the thermodynamic curvature is zero and $\epsilon$ is not associated with the underlying information geometry of the state space in energy representation.

\section{Conclusion}\label{secConcl}

To quantify the optimal time evolution of black holes, we developed a finite-time geometric framework called Thermogeometric Optimization (TGO). This framework is rooted in the concept of the shortest path between thermodynamic states, as measured by the thermodynamic length. Calculating this length requires the definition of a suitable metric structure on the system's state space. In our work, we utilized the Hessians of entropy and energy as the simplest metric choices. However, the TGO framework is highly adaptable and can incorporate alternative metric definitions.

Applying the TGO method to Schwarzschild and Kerr black hole solutions, we successfully captured the key aspects of their finite-time optimal thermodynamic behavior. For example, our analysis revealed that Schwarzschild black holes can fully evaporate through optimal fluctuations across all relevant thermodynamic representations. A similar result was observed for Kerr black holes in the energy ensemble, but it did not hold in the entropy representation. Furthermore, we compared these optimal processes to standard evaporation via Hawking radiation, highlighting their differences and potential implications.

Remarkably, in the entropy representation of the Kerr black hole, the TGO method successfully identified the Davies phase transition points, corresponding to the specific spin at \(\tilde{a}_* = \pm 0.681\). Our analysis demonstrated that, in all cases, the sign of the metric scale \(\epsilon\) consistently changes depending on whether the initial spin \(a_{*0}\) of the black hole is smaller or larger than \(\tilde{a}_*\). This shift of sign indicates that the information geometries, responsible for the existence of optimal protocols, differ below and above the critical Davies phase transition points, offering valuable insights into their nature. The success of the thermogeometric method in detecting the Davies points within the entropy representation partially resolves a long-standing challenge in Hessian thermodynamic geometry. Conversely, in energy representation, the sign of \(\epsilon\) did not reliably identify the phase transition points, since the thermodynamic information space is flat, preventing \(\epsilon\) from being associated with the type of the underlying information geometry on the state space. This limitation suggests construction and utilization of other techniques or a generalization of the TGO method.

We demonstrated that the mass-energy content of a black hole can undergo significant changes during an optimal process. Building on this, an interesting challenge is to apply our approach to model black holes as heat engines and assess their efficiency. Such thermodynamic cycles could enable the extraction of useful work from black holes in an optimal manner, presenting a novel alternative to the classical Penrose process. 

Furthermore, we raise the possibility of applying the thermogeometric optimization method within the framework of Geometrothermodynamics (GTD), which utilizes Legendre-invariant metrics. Due to the robustness of the geometric approach, it can be extended to explore optimal processes in other systems, including AdS black holes and their strongly correlated holographic dual quark-gluon models. Additionally, we hypothesize that TGO may prove valuable in modeling cosmological fluctuations and phase transitions. 

{Finally, in the pursuit of solving thermal problems, numerous theories and methods have been developed, most of which remain confined to specific fields and problems. Since TGO is rooted in rigid geometric structures it offers broader applicability that extends beyond gravitational systems to a wide range of real-world thermal phenomena. A natural next step would be to compare this geometric framework with conventional non-geometric optimization schemes in finite-time thermodynamics and finite-size systems -- particularly in contexts involving entropy generation minimization, energy utilization efficiency, minimum economic cost, maximum output work, and maximum thermal efficiency.}

These topics, along with other open questions, will be addressed in future research.

\section*{Acknowledgments}

The authors would like to express their gratitude to  S. Yazadjiev, P. Nedkova,  K. Hristov,  G. Gyulchev, K. Staykov, P. Ivanov, D. Marvakov, A. Isaev and S. Krivonos for their invaluable comments and discussions. V. A. is grateful to the support by Sofia University Grant 80-10-22/08.04.2024 and the Bulgarian NSF grant K$\Pi$-06-H88/3. H. D. thankfully acknowledges the support by the program “JINR-Bulgaria” of the Bulgarian Nuclear Regulatory Agency. M. R., R. R., and T. V. were fully financed by the European Union-NextGeneration EU, through the National Recovery and Resilience Plan of the Republic of Bulgaria, project BG-RRP-2.004-0008-C01.

\appendix

\section{Nambu brackets and heat capacities of Kerr black hole}\label{appA}

Nambu brackets  generalize the Poisson bracket for three or more variables \cite{Mansoori:2014oia, avramov2023thermodynamic}, accounting for the determinant of the Jacobian when working in certain coordinates, i.e.

\begin{equation}
\{f,x^1,...,x^{ n-1}\}_{y^1,y^2,...,y^n}=\left|
\begin{array}{cccc}
	\frac{\partial f}{\partial y^1}\big|_{y^2,y^3,...,y^n}& \frac{\partial f}{\partial y^2}\big|_{{y^1,y^3,...,y^n}} & \cdots  & \frac{\partial f}{\partial y^n}\big|_{{y^1,y^2,..., y^{n-1}}} \\[5pt]
	\frac{\partial x^1}{\partial y^1}\big|_{{y^2,y^3,...,y^n}}	& 	\frac{\partial x^1}{\partial y^2}\big|_{{y^1,y^3,...,y^n}} &  \cdots & \frac{\partial x^1}{\partial y^n}\big|_{{y^1,y^2,...,y^{n-1}}} \\[5pt]
	\vdots	& \vdots  &   & \vdots\\[5pt]
	\frac{\partial x^{ n-1}}{\partial y^1}\big|_{{y^2,y^3,...,y^n}}	&  	\frac{\partial x^{ n-1}}{\partial y^2}\big|_{{y^1,y^3,...,y^n}}&  \cdots & 	\frac{\partial x^{ n-1}}{\partial y^n}\big|_{{y^1,y^2,...,y^{n-1}}}
	\end{array}
	\right|.
\end{equation}
For instance, when $n=2$ recovers the standard Poisson bracket: 
\begin{equation}
    \{f,x\}_{u,v}=   \left|
\begin{array}{cc}
\frac{\partial f}{\partial u}\big|_v& \frac{\partial f}{\partial v}\big|_u\\[5pt]
\frac{\partial x}{\partial u}\big|_v & \frac{\partial x}{\partial v}\big|_u\end{array}\right|
=\frac{\partial f}{\partial u}\bigg|_v 
 \frac{\partial x}{\partial v}\bigg|_u-\frac{\partial f}{\partial v}\bigg|_u \frac{\partial x}{\partial u}\bigg|_v,
\end{equation}
where $(u,v)$ are variables and $f=f(u,v)$ and $x=x(u,v)$ are functions expressed in those variables. Using the Nambu bracket formalism one can easily calculate the three heat capacities of the Kerr black hole in entropy and energy representations respectively\footnote{There are also $C_S=0$ and $C_T=\infty$.}: 
\begin{align}
    &C_E(E,J)=T\frac{\partial S}{\partial T}\bigg|_E=T\frac{\{S,E\}_{E,J}}{\{T,E\}_{E,J}},\quad C_E(S,J)=T\frac{\partial S}{\partial T}\bigg|_E=T\frac{\{S,E\}_{S,J}}{\{T,E\}_{S,J}},
    \\
    &C_J(E,J)=T\frac{\partial S}{\partial T}\bigg|_J=T\frac{\{S,J\}_{E,J}}{\{T,J\}_{E,J}},\quad\,\, C_J(S,J)=T\frac{\partial S}{\partial T}\bigg|_J=T\frac{\{S,J\}_{S,J}}{\{T,J\}_{S,J}}
    \\
    &C_\Omega(E,J)=T\frac{\partial S}{\partial T}\bigg|_\Omega=T\frac{\{S,\Omega\}_{E,J}}{\{T,\Omega\}_{E,J}},\quad\, C_\Omega(S,J)=T\frac{\partial S}{\partial T}\bigg|_\Omega=T\frac{\{S,\Omega\}_{S,J}}{\{T,\Omega\}_{S,J}}.
\end{align}
For example, let us calculate the $C_J$ heat capacity in entropy representation with variables ($E,J$):
\begin{align}
    C_J(E,J)=T\frac{\partial S}{\partial T}\bigg|_J=T\frac{\{S,J\}_{E,J}}{\{T,J\}_{E,J}}
    =T\frac{\left|
\begin{array}{cc}
\frac{\partial S}{\partial E}\big|_J& \frac{\partial S}{\partial J}\big|_E\\[5pt]
\frac{\partial J}{\partial J}\big|_E & \frac{\partial J}{\partial E}\big|_J\end{array}\right|}{\left|
\begin{array}{cc}
\frac{\partial T}{\partial E}\big|_J& \frac{\partial T}{\partial J}\big|_E\\[5pt]
\frac{\partial J}{\partial J}\big|_E & \frac{\partial J}{\partial E}\big|_J\end{array}\right|}
=T \frac{\left|
\begin{array}{cc}
\frac{\partial S}{\partial E}\big|_J& \frac{\partial S}{\partial J}\big|_E\\[5pt]
1 & 0\end{array}\right|}{\left|
\begin{array}{cc}
\frac{\partial T}{\partial E}\big|_J& \frac{\partial T}{\partial J}\big|_E\\[5pt]
1 & 0\end{array}\right|}
=T \frac{\frac{\partial S}{\partial J}\big|_E}{\frac{\partial T}{\partial J}\big|_E}.
\end{align}
This simply tells us that we need $S=S(E,J)$ and $T=T(E,J)$, which are given by (\ref{eqKerrEntr}) and (\ref{eqKerrTemp}), hence:
\begin{equation}
       C_J(E,J) = \lambda E^2\frac{ E^2\big(E^2+ \sqrt{E^4-\zeta ^2 J^2}\big)-\zeta ^2 J^2}{E^2\big(E^2-2 \sqrt{E^4-\zeta ^2 J^2}\big)+\zeta ^2 J^2}.
\end{equation}

Similar calculations lead to the expression for $C_J$ in energy representation with variables $(S,J)$:
\begin{equation}
    C_J(S,J)=T\frac{\partial S}{\partial T}\bigg|_J=T\frac{\{S,J\}_{S,J}}{\{T,J\}_{S,J}}= -\frac{2 S \left(4 S^2-\zeta ^2 J^2 \lambda ^2\right)\left(4 S^2+\zeta ^2 J^2 \lambda ^2\right)}{16 S^4-3 \zeta ^4 J^4 \lambda ^4-24 \zeta ^2 J^2 \lambda ^2 S^2}.
\end{equation}
\section{Kerr geodesic equations in energy representation}\label{appKTDGEER}

The system of thermodynamic geodesic equations in $(S,J)$ space is written by:

\begin{equation}\label{eqGeodEnerKerr}
    \ddot S+A_1 \dot S^2+A_2 \dot J \dot S+A_3 \dot J^2=0,\quad \ddot J+B_1 \dot S^2+B_2 \dot J \dot S+B_3 \dot J^2=0,
\end{equation}
where the explicit form of the coefficients yield:
\begin{align*}
   & A_1=\frac{3 (2 S-\zeta  J \lambda )^2 (2 S+\zeta  J \lambda )^2 }{4 S V^4},\quad V=\sqrt{\zeta ^2 \lambda ^2 J^2+4 S^2},
    \\\\
    &A_2=\frac{12 \zeta ^2 J \lambda ^2 S^2 \big(192 S^6-\zeta ^6 J^6 \lambda ^6-\zeta ^4 J^4 \lambda ^4 \left(12 S^2+V^2\right)-8 \zeta ^2 J^2 \lambda ^2 S^2 \left(V^2-2 S^2\right)-80 S^4 V^2\big)}{V^{10}},
\\\\
  &A_3=  \frac{12 \zeta ^2 \lambda ^2 S^3 \big(\zeta ^6 J^6 \lambda ^6+8 \zeta ^4 J^4 \lambda ^4 S^2+4 \zeta ^2 J^2 \lambda ^2 S^2 \left(12 S^2+V^2\right)-16 S^4 \left(V^2-8 S^2\right)\big)}{V^{10}},
\\\\
&B_1=\frac{3}{8 S^2 V^{10}} \big[3 \zeta ^8 J^9 \lambda ^8 \left(4 S^2+V^2\right)+48 \zeta ^6 J^7 \lambda ^6 S^2 \left(2 S^2+V^2\right)-\zeta ^{10} J^{11} \lambda ^{10}
    \\
    &\quad\,\,+32 \zeta ^4 J^5 \lambda ^4 S^4 \left(13 V^2-20 S^2\right)+256 \zeta ^2 J^3 \lambda ^2 S^6 \left(7 V^2-5 S^2\right)+256 J S^8 \left(28 S^2-5 V^2\right)\big],
\\\\
&B_2=-\frac{3}{2 S V^{10}} \big[24 \zeta ^8 J^8 \lambda ^8 S^2+4 \zeta ^6 J^6 \lambda ^6 S^2 \left(64 S^2+3 V^2\right)+\zeta ^{10} J^{10} \lambda ^{10}
  \\
  &\quad\,\,+16 \zeta ^4 J^4 \lambda ^4 S^4 \left(104 S^2+3 V^2\right)+64 \zeta ^2 J^2 \lambda ^2 S^6 \left(60 S^2-7 V^2\right)+256 S^8 V^2\big],
\\\\
  &B_3=  \frac{12 \zeta ^2 J \lambda ^2 S^2 \left(\zeta ^2 J^2 \lambda ^2+4 S^2\right)^2}{V^8}.
\end{align*}
These equations are highly nonlinear and can be treated numerically as shown on Figures \ref{fignSnJ} and \ref{figKerrposSJ}.

\section{Conversion between SI and Planck systems of units}

To convert between different systems of units, we only need the appropriate Planck parameters, which can be derived through a straightforward dimensional analysis of the following product of fundamental constants:
\begin{equation}
P = {\hbar ^x}{G^y}{c^z}\varepsilon _0^u{e^v}k^w = [{M^{x - y - u + w}}{L^{2x + 3y + z - 3u + 2w}}{T^{ - x - 2y - z + 2u - 2w}}{C^{2u + v}}{K^{ - w}}].
\end{equation}
Notable solutions are the Planck length $l_p$, mass $m_p$, time $t_p$, energy $E_p$ and temperature $T_p$:
\begin{align}
{l_p} = \sqrt {\frac{{\hbar G}}{{{c^3}}}},\quad
{m_p} = \sqrt {\frac{{\hbar c}}{G}},\quad
{t_p} = \sqrt {\frac{{\hbar G}}{{{c^5}}}},\quad 
{E_p} =  \sqrt {\frac{{\hbar c^5}}{G}},\quad
{T_p} = \sqrt {\frac{{\hbar {c^5}}}{{Gk^2}}}.
\end{align}

The conversion between Planck units $\hbar=c=G=k=1$ and SI units is defined by the following relations:
\begin{align}
{l^{(P)}} = \frac{{{l^{(SI)}}[\text{m}]}}{{{l_p}[\text{m}]}},\,\,\, 
 {m^{(P)}} = \frac{{{m^{(SI)}}[\text{kg}]}}{{{m_p}[\text{kg}]}},\,\,\, {t^{(P)}} = \frac{{{t^{(SI)}}[\text{s}]}}{{{t_p}[\text{s}]}},\,\,\, {E^{(P)}} = \frac{{{E^{(SI)}}[\text{J}]}}{{{E_p}[\text{J}]}},\,\,\, {T^{(P)}} = \frac{{{T^{(SI)}}[\text{K}]}}{{{T_p}[\text{K}]}}.
\end{align}
Basically, the Planck quantity, e.g. $l^{(P)}$, counts how many Planck lengths fit into the length of $l^{(SI)}$ meters. 

\section{Black hole thermodynamics in Planck units}\label{appPlUnits}

In Planck units, $\hbar=c=G=k=1$, traditionally employed in black hole physics, one has:
\begin{equation}
    \lambda = 4\pi,\quad \eta=\frac{1}{2 \sqrt \pi},\quad \zeta= 1, \quad l_p= 1,\quad E=M,\quad a = \frac{J}{M},\quad a_*=\frac{a}{M}=\frac{J}{M^2}.
\end{equation}
Therefore, the Schwarzschild black hole thermodynamics becomes:
\begin{equation}
    dM=T dS,\quad S=\frac{A}{4 }=4 \pi M^2,\quad R_S=2 M,\quad A=16 \pi M^2,\quad T=\frac{1 }{8 \pi  M}.
\end{equation}

Since $R_S=2 M$, then $r_{\pm}=M\pm \sqrt{M^2-a^2}$, and Kerr black hole thermodynamics in entropy representation (\ref{eqKerrEntr}) takes the form $a_*=J/M^2$ and:
\begin{align}
    &S = 2\pi \big( M^2+\sqrt{ M^4 - J^2} \big),\,\,\,
    T =\frac{\sqrt{M^4-J^2}}{4 \pi M \big(M^2+\sqrt{M^4-J^2}\big)},\,\,\, \Omega= \frac{J}{2 M\big( M^2+ \sqrt{M^4-J^2}\big)}.
\end{align}
Similarly, the energy representation (\ref{eqEnRepKerr}) becomes $dM=T dS+\Omega dJ$ and:
\begin{align}
M=\frac{\sqrt{4\pi^2 J^2+ S^2}}{2\sqrt{\pi S}},\,\,\, T=\frac{ S^2- 4 \pi^2 J^2}{4 \sqrt{\pi S^3 \left(4 \pi ^2 J^2+S^2\right)}},\,\,\, \Omega=\frac{2 \pi ^{3/2} J}{\sqrt{S \left(4 \pi ^2 J^2+S^2\right)}},\,\,\,a_*=\frac{\pi J S}{4\pi^2 J^2+ S^2}.
\end{align}
Finally, the two non-extremality conditions (\ref{eqKerrExsistS}) and (\ref{eqKerrExsistE}) yield:
\begin{equation}
    M^2> |J|,\quad  S>  2\pi |J| .
\end{equation}

\section{Initial data for Hawking and fluctuation profiles}

\subsection{Numeric data for Figure \ref{figEScomSchwz}}\label{appNDF1}

In Fig. \ref{figEHcomp} the energy profiles are normalized to the energy of a solar-mass Schwarzschild black hole: $ E_0 \equiv E_\odot = 1.8 \times 10^{47}\, \text{J} $.
The initial rates of change $\dot E_0$ of the energy for the optimal processes (\ref{eqGeodEprofSchw}) are multiples of the Hawking energy rate of change (\ref{eqHawkEvapESchw}):
\begin{equation}
\dot E_{H,0}=\dot E_H(0)=-\frac{\alpha }{E_0^2}=-8.9 \times 10^{-29} \,\text{J/s},\quad \alpha=2.9 \times 10^{66} \,\text{J$^3$/s}.
\end{equation}
They are as follows:  (Black curve) $ \dot{E}_0 = \dot E_{H,0} $. 
 (Red curve) $ \dot{E}_0 = 3\,\dot E_{H,0}$.  
 (Blue curve) $ \dot{E}_0 =4.5 \,\dot E_{H,0}$.  
(Green curve) $ \dot{E}_0 = 2.3 2\,\dot E_{H,0}$ .

In Fig. \ref{figSHcomp} the entropy profiles are normalized to the entropy of a solar-mass Schwarzschild black hole: $ S_0 \equiv S_\odot = 1.46 \times 10^{54} $.
The initial rates of change $\dot S_0$ of the entropy for the optimal processes (\ref{eqSprofHessSchw}) are multiples of the Hawking entropy  rate of change (\ref{eqHEPSSchwz}):
\begin{equation}
\dot S_{H,0}=\dot S_H(0)=-\frac{2 \alpha }{\eta^3\sqrt{S_0}}=-1.45\times 10^{-21} \,\text{J/(Ks)},\quad \eta=1.49\times 10^{20}\, \text{(JK)}^{1/2}.
\end{equation}
They are as follows:  (Black curve) $ \dot{S}_0 = \dot S_{H,0} $. 
 (Red curve) $ \dot{S}_0 = 3\,\dot S_{H,0}$.  
 (Blue curve) $ \dot{S}_0 =4.5 \,\dot S_{H,0}$.  
(Green curve) $ \dot{S}_0 = 2.32\,\dot S_{H,0}$.

\subsection{Numeric data for Figure \ref{figKerr0208} and later}\label{appfigKerr0208}

The Hawking evaporation rates are evaluated according to (\ref{eqHERESJ}). The other relevant constants are: $b=2.31\times 10^{67} \,\text{J}^3\text{/s}$, $\zeta= 3.64\times 10^{52} \,\text{J/s}$ and $\lambda= 4.52\times 10^{-41}\, (\text{JK})^{-1}$.
\\

\noindent For $a_*=0.2$, Hawking evaporation rates are given by:
\begin{align}
&E_{H,0}=E_\odot = 1.8 \times 10^{47}\, \text{J},\quad J_{H,0}=1.78 \times 10^{41} \,\text{Js},\quad S_{H,0} = 1.45 \times 10^{54}\, \text{J/K},
\\
&\dot E_{H,0}=-8.46\times 10^{-29} \,\text{J/s},\quad {\dot J}_{H,0}=-1.67\times 10^{-34} \,\text{J},\quad \dot S_{H,0}=-1.36\times 10^{-21} \,\text{J/(Ks)},
\\
&\gamma_1=\frac{b(1-a_*^2)^2}{(1+\sqrt{1-a_*^2})^3}=2.74\times 10^{66}\,\text{J}^3\text{/s},\quad \gamma_2=\frac{2\gamma_1 \zeta J_{H,0}}{a_* E_{H,0}^2}=5.48\times 10^{66} \,\text{J}^3\text{/s}.
\end{align}

\noindent For $a_*=0.4$, Hawking evaporation rates are given by:
\begin{align}
&E_{H,0}=E_\odot = 1.8 \times 10^{47}\, \text{J},\quad J_{H,0}=3.56 \times 10^{41} \,\text{Js},\quad S_{H,0}= 1.40 \times 10^{54}\, \text{J/K},
\\
&\dot E_{H,0}=-7.14\times 10^{-29} \,\text{J/s},\quad {\dot J}_{H,0}=-2.82\times 10^{-34} \,\text{J},\quad \dot S_{H,0}=-1.11\times 10^{-21} \,\text{J/(Ks)},
\\
&\gamma_1=2.31\times 10^{66}\,\text{J}^3\text{/s},\quad \gamma_2=4.63\times 10^{66} \,\text{J}^3\text{/s}.
\end{align}

\noindent For $a_*=0.5$, Hawking evaporation rates are given by:
\begin{align}
&E_{H,0}=E_\odot = 1.8 \times 10^{47}\, \text{J},\quad J_{H,0}=4.45 \times 10^{41} \,\text{Js},\quad S_{H,0}= 1.37\times 10^{54}\, \text{J/K},
\\
&\dot E_{H,0}=-6.17\times 10^{-29} \,\text{J/s},\quad {\dot J}_{H,0}=-3.05\times 10^{-34} \,\text{J},\quad \dot S_{H,0}=-9.36\times 10^{-22} \,\text{J/(Ks)},
\\
&\gamma_1=2.0\times 10^{66}\,\text{J}^3\text{/s},\quad \gamma_2=4.0\times 10^{66} \,\text{J}^3\text{/s}.
\end{align}

\noindent For $a_*=0.8$, Hawking evaporation rates are given by:
\begin{align}
&E_{H,0}=E_\odot = 1.8 \times 10^{47}\, \text{J},\quad J_{H,0}=7.12 \times 10^{41} \,\text{Js},\quad S_{H,0}= 1.17 \times 10^{54}\, \text{J/K},
\\
&\dot E_{H,0}=-2.25\times 10^{-29} \,\text{J/s},\quad {\dot J}_{H,0}=-1.78\times 10^{-34} \,\text{J},\quad \dot S_{H,0}=-2.93\times 10^{-22} \,\text{J/(Ks)},
\\
&\gamma_1=7.3\times 10^{65}\,\text{J}^3\text{/s},\quad \gamma_2=1.46\times 10^{66} \,\text{J}^3\text{/s}.
\end{align}

\noindent For $a_*=0.99$, Hawking evaporation rates are given by:
\begin{align}
&E_{H,0}=E_\odot = 1.8 \times 10^{47}\, \text{J},\,\,\, J_{H,0}=8.81 \times 10^{41} \,\text{Js},\,\,\, S_{H,0}= 8.36 \times 10^{53}\, \text{J/K},
\\
&\dot E_{H,0}=-1.9\times 10^{-31} \,\text{J/s},\quad {\dot J}_{H,0}=-1.86\times 10^{-36} \,\text{J},\quad \dot S_{H,0}=-1.76\times 10^{-24} \,\text{J/(Ks)},
\\
&\gamma_1=6.15\times 10^{63}\,\text{J}^3\text{/s},\quad \gamma_2=1.23\times 10^{64} \,\text{J}^3\text{/s}.
\end{align}

\noindent For $a_*=0.995$, Hawking evaporation rates are given by:
\begin{align}
&E_{H,0}=E_\odot = 1.8 \times 10^{47}\, \text{J},\,\,\, J_{H,0}=8.85 \times 10^{41} \,\text{Js},\,\,\, S_{H,0}= 8.05 \times 10^{53}\, \text{J/K},
\\
&\dot E_{H,0}=-5.33\times 10^{-32} \,\text{J/s},\quad {\dot J}_{H,0}=-5.24\times 10^{-37} \,\text{J},\quad \dot S_{H,0}=-7.77\times 10^{-25} \,\text{J/(Ks)},
\\
&\gamma_1=1.73\times 10^{63}\,\text{J}^3\text{/s},\quad \gamma_2=3.45\times 10^{63} \,\text{J}^3\text{/s}.
\end{align}

\bibliographystyle{utphys}

\begin{thebibliography}{10}
	
	\bibitem{Bekenstein:1972tm}
	J.~D. Bekenstein, ``{Black holes and the second law},''
	\href{http://dx.doi.org/10.1007/BF02757029}{{\em Lett. Nuovo Cim.} {\bfseries
			4} (1972) 737--740}.
	
	\bibitem{Bekenstein:1973ur}
	J.~D. Bekenstein, ``{Black holes and entropy},''
	\href{http://dx.doi.org/10.1103/PhysRevD.7.2333}{{\em Phys. Rev. D}
		{\bfseries 7} (1973) 2333--2346}.
	
	\bibitem{Bekenstein:1974ax}
	J.~D. Bekenstein, ``{Generalized second law of thermodynamics in black hole
		physics},'' \href{http://dx.doi.org/10.1103/PhysRevD.9.3292}{{\em Phys. Rev.
			D} {\bfseries 9} (1974) 3292--3300}.
	
	\bibitem{Bekenstein:1975tw}
	J.~D. Bekenstein, ``{Statistical Black Hole Thermodynamics},''
	\href{http://dx.doi.org/10.1103/PhysRevD.12.3077}{{\em Phys. Rev. D}
		{\bfseries 12} (1975) 3077--3085}.
	
	\bibitem{Hawking:1971tu}
	S.~W. Hawking, ``{Gravitational radiation from colliding black holes},''
	\href{http://dx.doi.org/10.1103/PhysRevLett.26.1344}{{\em Phys. Rev. Lett.}
		{\bfseries 26} (1971) 1344--1346}.
	
	\bibitem{Hawking:1975vcx}
	S.~W. Hawking, ``{Particle Creation by Black Holes},''
	\href{http://dx.doi.org/10.1007/BF02345020}{{\em Commun. Math. Phys.}
		{\bfseries 43} (1975) 199--220}. [Erratum: Commun.Math.Phys. 46, 206 (1976)].
	
	\bibitem{Hawking:1976de}
	S.~W. Hawking, ``{Black Holes and Thermodynamics},''
	\href{http://dx.doi.org/10.1103/PhysRevD.13.191}{{\em Phys. Rev. D}
		{\bfseries 13} (1976) 191--197}.
	
	\bibitem{Bardeen:1973gs}
	J.~M. Bardeen, B.~Carter, and S.~W. Hawking, ``{The Four laws of black hole
		mechanics},'' \href{http://dx.doi.org/10.1007/BF01645742}{{\em Commun. Math.
			Phys.} {\bfseries 31} (1973) 161--170}.
	
	\bibitem{Davies:1977bgr}
	P.~C.~W. Davies, ``{Thermodynamics of Black Holes},''
	\href{http://dx.doi.org/10.1098/rspa.1977.0047}{{\em Proc. Roy. Soc. Lond. A}
		{\bfseries 353} (1977) 499--521}.

	\bibitem{Davies_1978}
	P.~C.~W. Davies, ``Thermodynamics of black holes,''
	\href{http://dx.doi.org/10.1088/0034-4885/41/8/004}{{\em Reports on Progress
			in Physics} {\bfseries 41} no.~8, (Aug, 1978) 1313}.
	\url{https://dx.doi.org/10.1088/0034-4885/41/8/004}.
	
	\bibitem{EventHorizonTelescope:2019dse}
	{\bfseries Event Horizon Telescope} Collaboration, K.~Akiyama {\em et~al.},
	``{First M87 Event Horizon Telescope Results. I. The Shadow of the
		Supermassive Black Hole},''
	\href{http://dx.doi.org/10.3847/2041-8213/ab0ec7}{{\em Astrophys. J. Lett.}
		{\bfseries 875} (2019) L1}, \href{http://arxiv.org/abs/1906.11238}{{\ttfamily
			arXiv:1906.11238 [astro-ph.GA]}}.
	
	\bibitem{EventHorizonTelescope:2021dqv}
	{\bfseries Event Horizon Telescope} Collaboration, P.~Kocherlakota {\em
		et~al.}, ``{Constraints on black-hole charges with the 2017 EHT observations
		of M87*},'' \href{http://dx.doi.org/10.1103/PhysRevD.103.104047}{{\em Phys.
			Rev. D} {\bfseries 103} no.~10, (2021) 104047},
	\href{http://arxiv.org/abs/2105.09343}{{\ttfamily arXiv:2105.09343 [gr-qc]}}.
	
	\bibitem{EventHorizonTelescope:2022wkp}
	{\bfseries Event Horizon Telescope} Collaboration, K.~Akiyama {\em et~al.},
	``{First Sagittarius A* Event Horizon Telescope Results. I. The Shadow of the
		Supermassive Black Hole in the Center of the Milky Way},''
	\href{http://dx.doi.org/10.3847/2041-8213/ac6674}{{\em Astrophys. J. Lett.}
		{\bfseries 930} no.~2, (2022) L12}.
	
	\bibitem{LIGOScientific:2016aoc}
	{\bfseries LIGO Scientific, Virgo} Collaboration, B.~P. Abbott {\em et~al.},
	``{Observation of Gravitational Waves from a Binary Black Hole Merger},''
	\href{http://dx.doi.org/10.1103/PhysRevLett.116.061102}{{\em Phys. Rev.
			Lett.} {\bfseries 116} no.~6, (2016) 061102},
	\href{http://arxiv.org/abs/1602.03837}{{\ttfamily arXiv:1602.03837 [gr-qc]}}.
	
	\bibitem{LIGOScientific:2016vlm}
	{\bfseries LIGO Scientific, Virgo} Collaboration, B.~P. Abbott {\em et~al.},
	``{Properties of the Binary Black Hole Merger GW150914},''
	\href{http://dx.doi.org/10.1103/PhysRevLett.116.241102}{{\em Phys. Rev.
			Lett.} {\bfseries 116} no.~24, (2016) 241102},
	\href{http://arxiv.org/abs/1602.03840}{{\ttfamily arXiv:1602.03840 [gr-qc]}}.
	
	\bibitem{Luminet:1979nyg}
	J.~P. Luminet, ``{Image of a spherical black hole with thin accretion disk},''
	{\em Astron. Astrophys.} {\bfseries 75} (1979) 228--235.
	
	\bibitem{Page:1974he}
	D.~N. Page and K.~S. Thorne, ``{Disk-Accretion onto a Black Hole. Time-Averaged
		Structure of Accretion Disk},'' \href{http://dx.doi.org/10.1086/152990}{{\em
			Astrophys. J.} {\bfseries 191} (1974) 499--506}.
	
	\bibitem{Thorne:1974ve}
	K.~S. Thorne, ``{Disk accretion onto a black hole. 2. Evolution of the
		hole.},'' \href{http://dx.doi.org/10.1086/152991}{{\em Astrophys. J.}
		{\bfseries 191} (1974) 507--520}.
	
	\bibitem{Gyulchev:2019tvk}
	G.~Gyulchev, P.~Nedkova, T.~Vetsov, and S.~Yazadjiev, ``{Image of the
		Janis-Newman-Winicour naked singularity with a thin accretion disk},''
	\href{http://dx.doi.org/10.1103/PhysRevD.100.024055}{{\em Phys. Rev. D}
		{\bfseries 100} no.~2, (2019) 024055},
	\href{http://arxiv.org/abs/1905.05273}{{\ttfamily arXiv:1905.05273 [gr-qc]}}.
	
	\bibitem{Gyulchev:2021dvt}
	G.~Gyulchev, P.~Nedkova, T.~Vetsov, and S.~Yazadjiev, ``{Image of the thin
		accretion disk around compact objects in the
		Einstein\textendash{}Gauss\textendash{}Bonnet gravity},''
	\href{http://dx.doi.org/10.1140/epjc/s10052-021-09624-5}{{\em Eur. Phys. J.
			C} {\bfseries 81} no.~10, (2021) 885},
	\href{http://arxiv.org/abs/2106.14697}{{\ttfamily arXiv:2106.14697 [gr-qc]}}.
	
	\bibitem{1982TaylorPulse}
	J.~H. {Taylor} and J.~M. {Weisberg}, ``{A new test of general relativity -
		Gravitational radiation and the binary pulsar PSR 1913+16},''
	\href{http://dx.doi.org/10.1086/159690}{{\em APJ} {\bfseries 253} (Feb.,
		1982) 908--920}.
	
	\bibitem{penrose1971extraction}
	R.~Penrose and R.~Floyd, ``Extraction of rotational energy from a black hole,''
	\href{http://dx.doi.org/10.1038/physci229177a0}{{\em Nature Physical Science}
		{\bfseries 229} no.~6, (1971) 177--179}.
	
	\bibitem{Weinhold:1975a}
	F.~Weinhold, ``{Metric geometry of equilibrium thermodynamics},''
	\href{http://dx.doi.org/10.1063/1.431689}{{\em J. Chem. Phys.} {\bfseries 63}
		no.~6, (1975) 2479--2483}. \url{http://dx.doi.org/10.1063/1.431689}.
	
	\bibitem{Ruppeiner:1983zz}
	G.~Ruppeiner, ``{Thermodynamic Critical Fluctuation Theory?},''
	\href{http://dx.doi.org/10.1103/PhysRevLett.50.287}{{\em Phys. Rev. Lett.}
		{\bfseries 50} (1983) 287--290}.
	
	\bibitem{Ruppeiner:1995ss}
	G.~Ruppeiner, ``{Riemannian geometry in thermodynamic fluctuation theory},''
	\href{http://dx.doi.org/10.1103/RevModPhys.67.605}{{\em Rev. Mod. Phys.}
		{\bfseries 67} (1995) 605--659}.
	\url{http://dx.doi.org/10.1103/RevModPhys.67.605}.
	
	\bibitem{salamon1984relation}
	P.~Salamon, J.~Nulton, and E.~Ihrig, ``{On the relation between entropy and
		energy versions of thermodynamic length},''
	\href{http://dx.doi.org/10.1063/1.446468}{{\em Journal of Chemical Physics}
		{\bfseries 80} no.~1, (January, 1984) 436--437}.
	
	\bibitem{mrugala1984equivalence}
	R.~Mrugala, ``{On equivalence of two metrics in classical thermodynamics},''
	\href{http://dx.doi.org/10.1016/0378-4371(84)90064-1}{{\em Physica A:
			Statistical Mechanics and its Applications} {\bfseries 125} no.~2--3, (1984)
		631--639}.
	
	\bibitem{Quevedo:2007ws}
	H.~Quevedo and A.~Vazquez, ``{The Geometry of thermodynamics},''
	\href{http://dx.doi.org/10.1063/1.2902782}{{\em AIP Conf. Proc.} {\bfseries
			977} no.~1, (2008) 165--172},
	\href{http://arxiv.org/abs/0712.0868}{{\ttfamily arXiv:0712.0868 [math-ph]}}.
	
	\bibitem{Quevedo:2007mj}
	H.~Quevedo, ``{Geometrothermodynamics of black holes},''
	\href{http://dx.doi.org/10.1007/s10714-007-0586-0}{{\em Gen. Rel. Grav.}
		{\bfseries 40} (2008) 971--984},
	\href{http://arxiv.org/abs/0704.3102}{{\ttfamily arXiv:0704.3102 [gr-qc]}}.
	
	\bibitem{Quevedo:2017tgz}
	H.~Quevedo, M.~N. Quevedo, and A.~Sanchez, ``{Homogeneity and thermodynamic
		identities in geometrothermodynamics},''
	\href{http://dx.doi.org/10.1140/epjc/s10052-017-4739-3}{{\em Eur. Phys. J. C}
		{\bfseries 77} no.~3, (2017) 158},
	\href{http://arxiv.org/abs/1701.06702}{{\ttfamily arXiv:1701.06702 [gr-qc]}}.
	\url{http://arxiv.org/abs/1701.06702}.
	
	\bibitem{pineda2019physical}
	V.~Pineda, H.~Quevedo, M.~N. Quevedo, A.~Sanchez, and E.~Valdes, ``The physical
	significance of geometrothermodynamic metrics,''
	\href{http://dx.doi.org/10.1142/S0219887819501688}{{\em International Journal
			of Geometric Methods in Modern Physics} {\bfseries 16} no.~11, (2019)
		1950168}.

\bibitem{Quevedo:2023vip}
H.~Quevedo and M.~N.~Quevedo,
{``Unified representation of homogeneous and quasi-homogenous systems in geometrothermodynamics''},
\href{doi:10.1016/j.physletb.2023.137678}{{\em Phys. Lett. B}
	{\bfseries 838} (2023) 137678}

	\bibitem{Ruppeiner:2007hr}
	G.~Ruppeiner, ``{Stability and fluctuations in black hole thermodynamics},''
	\href{http://dx.doi.org/10.1103/PhysRevD.75.024037}{{\em Phys. Rev. D}
		{\bfseries 75} (2007) 024037}.
	
	\bibitem{Ruppeiner:2008kd}
	G.~Ruppeiner, ``{Thermodynamic curvature and phase transitions in Kerr-Newman
		black holes},'' \href{http://dx.doi.org/10.1103/PhysRevD.78.024016}{{\em
			Phys. Rev. D} {\bfseries 78} (2008) 024016},
	\href{http://arxiv.org/abs/0802.1326}{{\ttfamily arXiv:0802.1326 [gr-qc]}}.
	
	\bibitem{Ruppeiner:2013yca}
	G.~Ruppeiner, ``{Thermodynamic curvature and black holes},''
	\href{http://dx.doi.org/10.1007/978-3-319-03774-5_10}{{\em Springer Proc.
			Phys.} {\bfseries 153} (2014) 179--203},
	\href{http://arxiv.org/abs/1309.0901}{{\ttfamily arXiv:1309.0901 [gr-qc]}}.
	
	\bibitem{Ruppeiner:2018pgn}
	G.~Ruppeiner, ``{Thermodynamic Black Holes},''
	\href{http://dx.doi.org/10.3390/e20060460}{{\em Entropy} {\bfseries 20}
		no.~6, (2018) 460}, \href{http://arxiv.org/abs/1803.08990}{{\ttfamily
			arXiv:1803.08990 [gr-qc]}}.
	
	\bibitem{Mansoori:2013pna}
	S.~A.~H. Mansoori and B.~Mirza, ``{Correspondence of phase transition points
		and singularities of thermodynamic geometry of black holes},''
	\href{http://dx.doi.org/10.1140/epjc/s10052-013-2681-6}{{\em Eur. Phys. J. C}
		{\bfseries 74} no.~99, (2014) 2681},
	\href{http://arxiv.org/abs/1308.1543}{{\ttfamily arXiv:1308.1543 [gr-qc]}}.
	
	\bibitem{Mansoori:2014oia}
	S.~A.~H. Mansoori, B.~Mirza, and M.~Fazel, ``{Hessian matrix, specific heats,
		Nambu brackets, and thermodynamic geometry},''
	\href{http://dx.doi.org/10.1007/JHEP04(2015)115}{{\em JHEP} {\bfseries 04}
		(2015) 115}, \href{http://arxiv.org/abs/1411.2582}{{\ttfamily arXiv:1411.2582
			[gr-qc]}}.
	
	\bibitem{Mansoori:2016jer}
	S.~A.~H. Mansoori, B.~Mirza, and E.~Sharifian, ``{Extrinsic and intrinsic
		curvatures in thermodynamic geometry},''
	\href{http://dx.doi.org/10.1016/j.physletb.2016.05.096}{{\em Phys. Lett. B}
		{\bfseries 759} (2016) 298--305},
	\href{http://arxiv.org/abs/1602.03066}{{\ttfamily arXiv:1602.03066 [gr-qc]}}.
	
	\bibitem{HosseiniMansoori:2019jcs}
	S.~A. Hosseini~Mansoori and B.~Mirza, ``{Geometrothermodynamics as a singular
		conformal thermodynamic geometry},''
	\href{http://dx.doi.org/10.1016/j.physletb.2019.135040}{{\em Phys. Lett. B}
		{\bfseries 799} (2019) 135040},
	\href{http://arxiv.org/abs/1905.01733}{{\ttfamily arXiv:1905.01733 [gr-qc]}}.
	
	\bibitem{Mahmoudi:2023uxr}
	S.~Mahmoudi, K.~Jafarzade, and S.~H. Hendi, ``{A comprehensive review of
		geometrical thermodynamics: From fluctuations to black holes},''
	\href{http://dx.doi.org/10.55730/1300-0101.2748}{{\em Turk. J. Phys.}
		{\bfseries 47} no.~5, (2023) 214--278},
	\href{http://arxiv.org/abs/2307.00010}{{\ttfamily arXiv:2307.00010 [gr-qc]}}.
	
	\bibitem{Dimov:2019pkk}
	H.~Dimov, R.~C. Rashkov, and T.~Vetsov, ``{Thermodynamic Information Geometry
		and Applications in Holography},''
	\href{http://dx.doi.org/10.1007/978-981-15-7775-8_19}{{\em Springer Proc.
			Math. Stat.} {\bfseries 335} (2019) 285--298}.
	
	\bibitem{Dimov:2019fxp}
	H.~Dimov, R.~C. Rashkov, and T.~Vetsov, ``{Thermodynamic information geometry
		and complexity growth of a warped AdS black hole and the warped
		AdS$_3$/CFT$_2$ correspondence},''
	\href{http://dx.doi.org/10.1103/PhysRevD.99.126007}{{\em Phys. Rev. D}
		{\bfseries 99} no.~12, (2019) 126007},
	\href{http://arxiv.org/abs/1902.02433}{{\ttfamily arXiv:1902.02433
			[hep-th]}}.
	
	\bibitem{Vetsov:2018dte}
	T.~Vetsov, ``{Information Geometry on the Space of Equilibrium States of Black
		Holes in Higher Derivative Theories},''
	\href{http://dx.doi.org/10.1140/epjc/s10052-019-6553-6}{{\em Eur. Phys. J. C}
		{\bfseries 79} no.~1, (2019) 71},
	\href{http://arxiv.org/abs/1806.05011}{{\ttfamily arXiv:1806.05011 [gr-qc]}}.
	
	\bibitem{salamon1977finite}
	P.~Salamon, B.~Andresen, and R.~S. Berry, ``{Thermodynamics in finite time. II.
		Potentials for finite-time processes},'' {\em Physical Review A} {\bfseries
		15} no.~5, (1977) 2094--2102.
	
	\bibitem{andresen1977extremals}
	B.~Andresen, P.~Salamon, and R.~S. Berry, ``{Thermodynamics in finite time.
		Extremals for imperfect heat engines},'' {\em Journal of Chemical Physics}
	{\bfseries 66} no.~4, (1977) 1571--1577.
	
	\bibitem{andresen1977optimization}
	B.~Andresen, R.~S. Berry, and P.~Salamon, ``{Optimization of processes with
		finite-time thermodynamics},'' in {\em International Conference on Energy Use
		Management}, R.~Fazzolare and C.~B. Smith, eds., vol.~II, pp.~1--9.
	\newblock Pergamon Press, New York, NY, USA, 1977.
	
	\bibitem{salamon1980significance}
	P.~Salamon, B.~Andresen, P.~D. Gait, and R.~S. Berry, ``{The significance of
		Weinhold’s length},'' {\em Journal of Chemical Physics} {\bfseries 73}
	no.~2, (1980) 1001--1002.
	
	\bibitem{salamon1980minimum}
	P.~Salamon, A.~Nitzan, B.~Andresen, and R.~S. Berry, ``{Minimum entropy
		production and the optimization of heat engines},'' {\em Physical Review A}
	{\bfseries 21} no.~6, (1980) 2115--2129.
	
	\bibitem{salamon1983thermodynamic}
	P.~Salamon and R.~S. Berry, ``{Thermodynamic length and dissipated
		availability},'' {\em Physical Review Letters} {\bfseries 51} no.~13, (1983)
	1127.
	
	\bibitem{andresen1984thermodynamics}
	B.~Andresen, P.~Salamon, and R.~S. Berry, ``{Thermodynamics in finite time},''
	{\em Physics Today} {\bfseries 37} no.~9, (1984) 62.
	
	\bibitem{andresen1983availability}
	B.~Andresen, M.~H. Rubin, and R.~S. Berry, ``{Availability for finite-time
		processes. General theory and a model},'' {\em Journal of Physical Chemistry}
	{\bfseries 87} no.~15, (1983) 2704--2713.
	
	\bibitem{salamon1985length}
	P.~Salamon, J.~D. Nulton, and R.~S. Berry, ``{Length in Statistical
		Thermodynamics},'' {\em Journal of Chemical Physics} {\bfseries 82} no.~1,
	(1985) 2433--2436.
	
	\bibitem{andresen2011current}
	B.~Andresen, ``{Current Trends in Finite-Time Thermodynamics},'' {\em
		Angewandte Chemie International Edition} {\bfseries 50} no.~12, (2011)
	2690--2704.
	
	\bibitem{andresen1996finite}
	B.~Andresen, ``{Finite-time thermodynamics and thermodynamic length},''
	\href{http://dx.doi.org/10.1016/S0035-3159(96)80060-2}{{\em Revue Générale
			de Thermique} {\bfseries 35} no.~418--419, (1996) 621--626}.
	
	\bibitem{berry2022finite}
	R.~S. Berry, P.~Salamon, and B.~Andresen, eds.,
	\href{http://dx.doi.org/10.3390/books978-3-0365-4950-7}{{\em {Finite-Time
				Thermodynamics}}}.
	\newblock MDPI, 2022.
	
		\bibitem{CurzonAhlborn1975}
	F.~L.~Curzon and B.~Ahlborn,
	``Efficiency of a Carnot engine at maximum power output,''
	\href{https://doi.org/10.1119/1.10023}{{\em American Journal of Physics}
		{\bfseries 43, 22} (1975).}
			
	\bibitem{Bejan1995}
	A.~Bejan, 
	\emph{Entropy Generation Minimization: The Method of Thermodynamic Optimization of Finite-Size Systems and Finite-Time Processes}, CRC Press, 1995, ISBN: 0849396514.
	
	
	\bibitem{Cheng2019}
	X.~T.~Cheng, 
	\emph{A Critical Perspective of Entropy Generation Minimization in Thermal Analyses and Optimizations}, 
	Cambridge Scholars Publishing, 2019. ISBN: 1527518175.
	
	\bibitem{2007PhRvL99j0602C}
	G.~E. Crooks, ``{Measuring Thermodynamic Length},''
	\href{http://dx.doi.org/10.1103/PhysRevLett.99.100602}{{\em Physical Review
			Letters} {\bfseries 99} no.~10, (September, 2007) 100602},
	\href{http://arxiv.org/abs/0706.0559}{{\ttfamily arXiv:0706.0559
			[cond-mat.stat-mech]}}.
	
	\bibitem{cafaro2022thermodynamic}
	C.~Cafaro, O.~Luongo, S.~Mancini, and H.~Quevedo, ``{Thermodynamic length,
		geometric efficiency and Legendre invariance},''
	\href{http://dx.doi.org/10.1016/j.physa.2021.126740}{{\em Physica A:
			Statistical Mechanics and its Applications} {\bfseries 590} (2022) 126740}.
	
	\bibitem{avramov2023thermodynamic}
	V.~Avramov, H.~Dimov, M.~Radomirov, R.~C. Rashkov, and T.~Vetsov,
	``{Thermodynamic stability of ACGL Chern-Simons Black Hole and Optimal
		Processes},'' {\em Annals of the University of Craiova, Physics} {\bfseries
		33} (2023) 78--97.
	
	\bibitem{Bravetti:2015xsp}
	A.~Bravetti, C.~Gruber, and C.~S. Lopez-Monsalvo, ``{Thermodynamic optimization
		of a Penrose process: An engineers' approach to black hole thermodynamics},''
	\href{http://dx.doi.org/10.1103/PhysRevD.93.064070}{{\em Physical Review D}
		{\bfseries 93} no.~6, (2016) 064070},
	\href{http://arxiv.org/abs/1511.06801}{{\ttfamily arXiv:1511.06801 [gr-qc]}}.
	
	\bibitem{2020Entrp221076A}
	P.~Abiuso, H.~J.~D. Miller, M.~Perarnau-Llobet, and M.~Scandi, ``{Geometric
		Optimisation of Quantum Thermodynamic Processes},''
	\href{http://dx.doi.org/10.3390/e22101076}{{\em Entropy} {\bfseries 22}
		no.~10, (2020) 1076}, \href{http://arxiv.org/abs/2008.13593}{{\ttfamily
			arXiv:2008.13593 [quant-ph]}}.
	
	\bibitem{ANDRESEN1996647}
	B.~Andresen, ``{Finite-time thermodynamics and thermodynamic length},''
	\href{http://dx.doi.org/10.1016/S0035-3159(96)80060-2}{{\em Revue Générale
			de Thermique} {\bfseries 35} no.~418, (1996) 647--650}.
	
	\bibitem{2019Quant3197S}
	M.~Scandi and M.~Perarnau-Llobet, ``{Thermodynamic length in open quantum
		systems},'' \href{http://dx.doi.org/10.22331/q-2019-10-24-197}{{\em Quantum}
		{\bfseries 3} (2019) 197}, \href{http://arxiv.org/abs/1810.05583}{{\ttfamily
			arXiv:1810.05583 [quant-ph]}}.
	
	\bibitem{Gruber:2016xui}
	C.~Gruber, \href{http://dx.doi.org/10.1142/9789813226609_0164}{``Black hole
		thermodynamics in finite time,''} in {\em Proceedings of the Thirteenth
		Marcel Grossmann Meeting on General Relativity}.
	\newblock World Scientific, 2015.
	\newblock \url{https://doi.org/10.1142/9789813226609_0164}.
	
	\bibitem{Gruber:2016mqb}
	C.~Gruber, O.~Luongo, and H.~Quevedo,
	\href{http://dx.doi.org/10.1142/9789813226609_0025}{``{Geometric approaches
			to the thermodynamics of black holes},''} in {\em {14th Marcel Grossmann
			Meeting on Recent Developments in Theoretical and Experimental General
			Relativity, Astrophysics, and Relativistic Field Theories}}, vol.~1,
	pp.~453--466.
	\newblock 2017.
	\newblock \href{http://arxiv.org/abs/1603.09443}{{\ttfamily arXiv:1603.09443
			[gr-qc]}}.
	
	\bibitem{page1976thermal}
	D.~N. Page, ``{Thermal emission of energy from black holes},''
	\href{http://dx.doi.org/10.1103/PhysRevD.14.3260}{{\em Physical Review D}
		{\bfseries 14} no.~12, (1976) 3260--3273}.
	
	\bibitem{page2013jcap}
	D.~N. Page, ``{Time Dependence of Hawking Radiation Entropy},''
	\href{http://dx.doi.org/10.1088/1475-7516/2013/09/028}{{\em Journal of
			Cosmology and Astroparticle Physics} {\bfseries 2013} no.~09, (2013) 028}.
	
	\bibitem{Nian:2019buz}
	J.~Nian, ``{Kerr black hole evaporation and Page curve},''
	\href{http://dx.doi.org/10.1142/S0218271824500305}{{\em Int. J. Mod. Phys. D}
		{\bfseries 33} no.~07n08, (2024) 2450030},
	\href{http://arxiv.org/abs/1912.13474}{{\ttfamily arXiv:1912.13474
			[hep-th]}}.
	
	\bibitem{Arevalo:2024kmo}
	V.~A. Ar\'evalo, D.~Andrade, and C.~Rojas, ``{Time evolution of the Von Neumann
		entropy for a Kerr\textendash{}Taub\textendash{}NUT black hole},''
	\href{http://dx.doi.org/10.1140/epjc/s10052-024-13290-8}{{\em Eur. Phys. J.
			C} {\bfseries 84} no.~9, (2024) 932},
	\href{http://arxiv.org/abs/2406.19224}{{\ttfamily arXiv:2406.19224 [gr-qc]}}.
	
	\bibitem{Belgiorno:2002}
	F.~{Belgiorno}, ``{Notes on Quasi-Homogeneous Functions in Thermodynamics},''
	\href{http://dx.doi.org/10.48550/arXiv.physics/0210031}{{\em arXiv e-prints}
		(Oct., 2002) physics/0210031},
	\href{http://arxiv.org/abs/physics/0210031}{{\ttfamily arXiv:physics/0210031
			[physics.class-ph]}}.
	
	\bibitem{smarr1973mass}
	L.~B. Smarr, ``{Mass Formula for Kerr Black Holes},''
	\href{http://dx.doi.org/10.1103/PhysRevLett.30.71}{{\em Physical Review
			Letters} {\bfseries 30} no.~2, (1973) 71--73}.
	
	\bibitem{Avramov:2023eif}
	V.~Avramov, H.~Dimov, M.~Radomirov, R.~C. Rashkov, and T.~Vetsov, ``{On
		thermodynamic stability of black holes. Part I: classical stability},''
	\href{http://dx.doi.org/10.1140/epjc/s10052-024-12639-3}{{\em Eur. Phys. J.
			C} {\bfseries 84} no.~3, (2024) 281},
	\href{http://arxiv.org/abs/2302.11998}{{\ttfamily arXiv:2302.11998 [gr-qc]}}.
	
	\bibitem{Avramov:2024hys}
	V.~Avramov, H.~Dimov, M.~Radomirov, R.~C. Rashkov, and T.~Vetsov, ``{On
		Thermodynamic Stability of Black Holes. Part II: AdS Family of Solutions},''
	\href{http://arxiv.org/abs/2402.07272}{{\ttfamily arXiv:2402.07272 [gr-qc]}}.
	
	\bibitem{bazarov1964thermodynamics}
	I.~Bazarov, F.~Immirzi, and A.~Hayes, {\em Thermodynamics}.
	\newblock Pergamon Press; [distributed in the Western Hemisphere by Macmillan,
	New York], 1964.
	
	\bibitem{callen2006thermodynamics}
	H.~Callen, {\em Thermodynamics and an Introduction to Thermostatistics}.
	\newblock Student Edition. Wiley India Pvt. Limited, 2006.
	
	\bibitem{greiner2012thermodynamics}
	W.~Greiner, D.~Rischke, L.~Neise, and H.~St{\"o}cker, {\em Thermodynamics and
		Statistical Mechanics}.
	\newblock Classical Theoretical Physics. Springer New York, 2012.
	
	\bibitem{swendsen2020introduction}
	R.~Swendsen, {\em An Introduction to Statistical Mechanics and Thermodynamics:
		Second Edition}.
	\newblock Oxford Graduate Texts. Oxford University Press, 2020.
	
	\bibitem{blundell2010concepts}
	S.~Blundell and K.~Blundell, {\em Concepts in Thermal Physics}.
	\newblock OUP Oxford, 2010.
	
	\bibitem{Ruppeiner:2010}
	G.~{Ruppeiner}, ``{Thermodynamic curvature measures interactions},''
	\href{http://dx.doi.org/10.1119/1.3459936}{{\em American Journal of Physics}
		{\bfseries 78} no.~11, (Nov., 2010) 1170--1180},
	\href{http://arxiv.org/abs/1007.2160}{{\ttfamily arXiv:1007.2160
			[cond-mat.stat-mech]}}.
	
	\bibitem{Quevedo:2023ypd}
	H.~Quevedo, ``{Extended Black Hole Geometrothermodynamics},''
	\href{http://dx.doi.org/10.1134/S1063772923140160}{{\em Astron. Rep.}
		{\bfseries 67} no.~Suppl 2, (2023) S214--S218},
	\href{http://arxiv.org/abs/2312.01535}{{\ttfamily arXiv:2312.01535 [gr-qc]}}.
	
	\bibitem{Johnson:2014yja}
	C.~V. Johnson, ``{Holographic Heat Engines},''
	\href{http://dx.doi.org/10.1088/0264-9381/31/20/205002}{{\em Class. Quant.
			Grav.} {\bfseries 31} (2014) 205002},
	\href{http://arxiv.org/abs/1404.5982}{{\ttfamily arXiv:1404.5982 [hep-th]}}.
	
	\bibitem{Setare:2015yra}
	M.~R. Setare and H.~Adami, ``{Polytropic black hole as a heat engine},''
	\href{http://dx.doi.org/10.1007/s10714-015-1979-0}{{\em Gen. Rel. Grav.}
		{\bfseries 47} no.~11, (2015) 133}.
	
	\bibitem{Johnson:2015fva}
	C.~V. Johnson, ``{Born\textendash{}Infeld AdS black holes as heat engines},''
	\href{http://dx.doi.org/10.1088/0264-9381/33/13/135001}{{\em Class. Quant.
			Grav.} {\bfseries 33} no.~13, (2016) 135001},
	\href{http://arxiv.org/abs/1512.01746}{{\ttfamily arXiv:1512.01746
			[hep-th]}}.
	
	\bibitem{Johnson:2016pfa}
	C.~V. Johnson, ``{An Exact Efficiency Formula for Holographic Heat Engines},''
	\href{http://dx.doi.org/10.3390/e18040120}{{\em Entropy} {\bfseries 18}
		(2016) 120}, \href{http://arxiv.org/abs/1602.02838}{{\ttfamily
			arXiv:1602.02838 [hep-th]}}.
	
	\bibitem{Chakraborty:2016ssb}
	A.~Chakraborty and C.~V. Johnson, ``{Benchmarking black hole heat engines,
		I},'' \href{http://dx.doi.org/10.1142/S0218271819500123}{{\em Int. J. Mod.
			Phys. D} {\bfseries 27} no.~16, (2018) 1950012},
	\href{http://arxiv.org/abs/1612.09272}{{\ttfamily arXiv:1612.09272
			[hep-th]}}.
	
	\bibitem{Hennigar:2017apu}
	R.~A. Hennigar, F.~McCarthy, A.~Ballon, and R.~B. Mann, ``{Holographic heat
		engines: general considerations and rotating black holes},''
	\href{http://dx.doi.org/10.1088/1361-6382/aa7f0f}{{\em Class. Quant. Grav.}
		{\bfseries 34} no.~17, (2017) 175005},
	\href{http://arxiv.org/abs/1704.02314}{{\ttfamily arXiv:1704.02314
			[hep-th]}}.
	
	\bibitem{Chakraborty:2017weq}
	A.~Chakraborty and C.~V. Johnson, ``{Benchmarking Black Hole Heat Engines,
		II},'' \href{http://dx.doi.org/10.1142/S0218271819500068}{{\em Int. J. Mod.
			Phys. D} {\bfseries 27} no.~16, (2018) 1950006},
	\href{http://arxiv.org/abs/1709.00088}{{\ttfamily arXiv:1709.00088
			[hep-th]}}.
	
	\bibitem{Johnson:2018amj}
	C.~V. Johnson and F.~Rosso, ``{Holographic Heat Engines, Entanglement Entropy,
		and Renormalization Group Flow},''
	\href{http://dx.doi.org/10.1088/1361-6382/aaf1f1}{{\em Class. Quant. Grav.}
		{\bfseries 36} no.~1, (2019) 015019},
	\href{http://arxiv.org/abs/1806.05170}{{\ttfamily arXiv:1806.05170
			[hep-th]}}.
	
	\bibitem{Johnson:2019olt}
	C.~V. Johnson, ``{Holographic Heat Engines as Quantum Heat Engines},''
	\href{http://dx.doi.org/10.1088/1361-6382/ab5ba9}{{\em Class. Quant. Grav.}
		{\bfseries 37} no.~3, (2020) 034001},
	\href{http://arxiv.org/abs/1905.09399}{{\ttfamily arXiv:1905.09399
			[hep-th]}}.
	
	\bibitem{DiMarco:2022yhp}
	M.~C. DiMarco, S.~L. Jess, R.~A. Hennigar, and R.~B. Mann, ``{Universality for
		black hole heat engines near critical points},''
	\href{http://dx.doi.org/10.1103/PhysRevD.107.044001}{{\em Phys. Rev. D}
		{\bfseries 107} no.~4, (2023) 044001},
	\href{http://arxiv.org/abs/2211.14856}{{\ttfamily arXiv:2211.14856 [gr-qc]}}.
	
\end{thebibliography}

\providecommand{\href}[2]{#2}\begingroup\raggedright\endgroup
\end{document}